\documentclass[iop]{emulateapj}

\usepackage{wasysym}
\usepackage{natbib}
\usepackage{longtable}
\usepackage{graphicx}
\usepackage{fancyref}
\usepackage{gensymb}
\usepackage{adjustbox}
\usepackage{booktabs}
\usepackage{threeparttable}

\shortauthors{Fischer et al.}

\begin{document}

\title{Gemini Near Infrared Field Spectrograph Observations of the Seyfert 2 Galaxy Mrk 573: In Situ Acceleration of Ionized and Molecular Gas off Fueling Flows.}

\author{Travis C. Fischer\altaffilmark{1}\altaffilmark{\textdagger},
C. Machuca\altaffilmark{2},
M. R. Diniz\altaffilmark{3},
D. M. Crenshaw\altaffilmark{2},
S. B. Kraemer\altaffilmark{4},
R. A. Riffel\altaffilmark{3},
H. R. Schmitt\altaffilmark{5},
F. Baron\altaffilmark{2},
T. Storchi-Bergmann\altaffilmark{6},
A. N. Straughn\altaffilmark{1},
M. Revalski\altaffilmark{2},
C. L. Pope\altaffilmark{2}}

\altaffiltext{1}{Astrophysics Science Division, Goddard Space Flight Center,
Code 665, Greenbelt, MD 20771, USA}

\altaffiltext{2}{Department of Physics and Astronomy, Georgia State 
University, Astronomy Offices, 25 Park Place, Suite 605,
Atlanta, GA 30303, USA}

\altaffiltext{3}{Departamento de F\'isica, Centro de Ci\^encias Naturais e Exatas, 
Universidade Federal de Santa Maria, 97105-900 Santa Maria, RS, Brazil}

\altaffiltext{4}{Institute for Astrophysics and Computational Sciences,
Department of Physics, The Catholic University of America, Washington, DC
20064, USA}

\altaffiltext{5}{Naval Research Laboratory, Washington, DC 20375, USA}

\altaffiltext{6}{Departamento de Astronomia, Universidade Federal do Rio Grande 
do Sul, IF, CP 15051, 91501-970 Porto Alegre, RS, Brazil}

\altaffiltext{\textdagger}{James Webb Space Telescope NASA Postdoctoral Program Fellow; travis.c.fischer@nasa.gov}

\begin{abstract}

We present near-infrared and optical emission-line and stellar kinematics of the Seyfert 2 galaxy Mrk 573 using the Near-Infrared Field 
Spectrograph (NIFS) at {\it Gemini North} and Dual Imaging Spectrograph (DIS) at Apache Point Observatory, respectively. 
By obtaining full kinematic maps of the infrared ionized and molecular gas and stellar kinematics in a $\sim 700 \times 2100$ pc$^2$ 
circumnuclear region of Mrk~573, we find that kinematics within the Narrow-Line Region (NLR) are largely due to a 
combination of both rotation and in situ acceleration of material originating in the host disk. Combining these observations with large-scale, 
optical long-slit spectroscopy that traces ionized gas emission out to several kpcs, we find that rotation kinematics dominate the majority 
of the gas. We find that outflowing gas extends to distances less than 1 kpc, suggesting that outflows in Seyfert galaxies may not 
be powerful enough to evacuate their entire bulges.

\end{abstract}


~~~~~

\section{Introduction}
\label{sec1}

Supermassive black holes (SMBHs) reside within the centers of all massive galaxies possessing bulges, with a small percentage actively 
gaining mass from the surrounding accretion disk, 
which we define as Active Galactic Nuclei (AGN). The fueling of AGN and subsequent feedback is thought to play a critical role in the 
formation of large-scale structure in the early Universe \citep{Sca04}, chemical enrichment of the intergalactic medium \citep{Kha08}, and 
self-regulation of super-massive black hole (SMBH) and galactic bulge growth \citep{Hop05}. 

In optical and infrared spectroscopy, AGN feedback can be observed as high-velocity gas ($\sim$200 - 1000 km s$^{-1}$) inside the 
Narrow-Line Region (NLR), a region of relatively low density ionized gas extending from the torus to distances between 10 - 1000 pc from their 
central nuclei. In recent studies using resolved spectroscopy of nearby AGN with the Space Telescope Imaging Spectrograph (STIS) aboard 
the {\it Hubble Space Telescope} ({\it HST}) \citep{Cre09,Cre15} and ground-based 
Integral Field Units (IFUs; \citealt{Sto10, Mul11}), these high-velocity clouds have been attributed to nuclear mass outflows, with mass 
outflow rates on the order of 1 - 10 M$_{\odot}$ yr$^{-1}$. Associated accretion rates would be 10-100 times greater than what is 
required for the AGN bolometric luminosities, which suggests that most of the fueling flow does not make it to the central AGN, but is
instead blown out by radiation pressure and/or highly-ionized winds \citep{Eve07}. Our group has previously modeled these outflow 
kinematics in nearby AGN, where the observed velocity pattern often possessed a signature of radial acceleration followed by deceleration 
\citep{Das05,Das06,Cre10b,Sto10,Fis10,Fis11,Fis13,Bar14}, credited to either radiative driving or entrainment of clouds in a highly-ionized 
wind \citep{Kra07} followed by mass loading or interaction with an ambient medium \citep{Cre10b}. However, the mechanisms and geometries 
of mass outflows and fueling flows remain poorly understood.  

\begin{figure}
\centering
\includegraphics[width=0.48\textwidth]{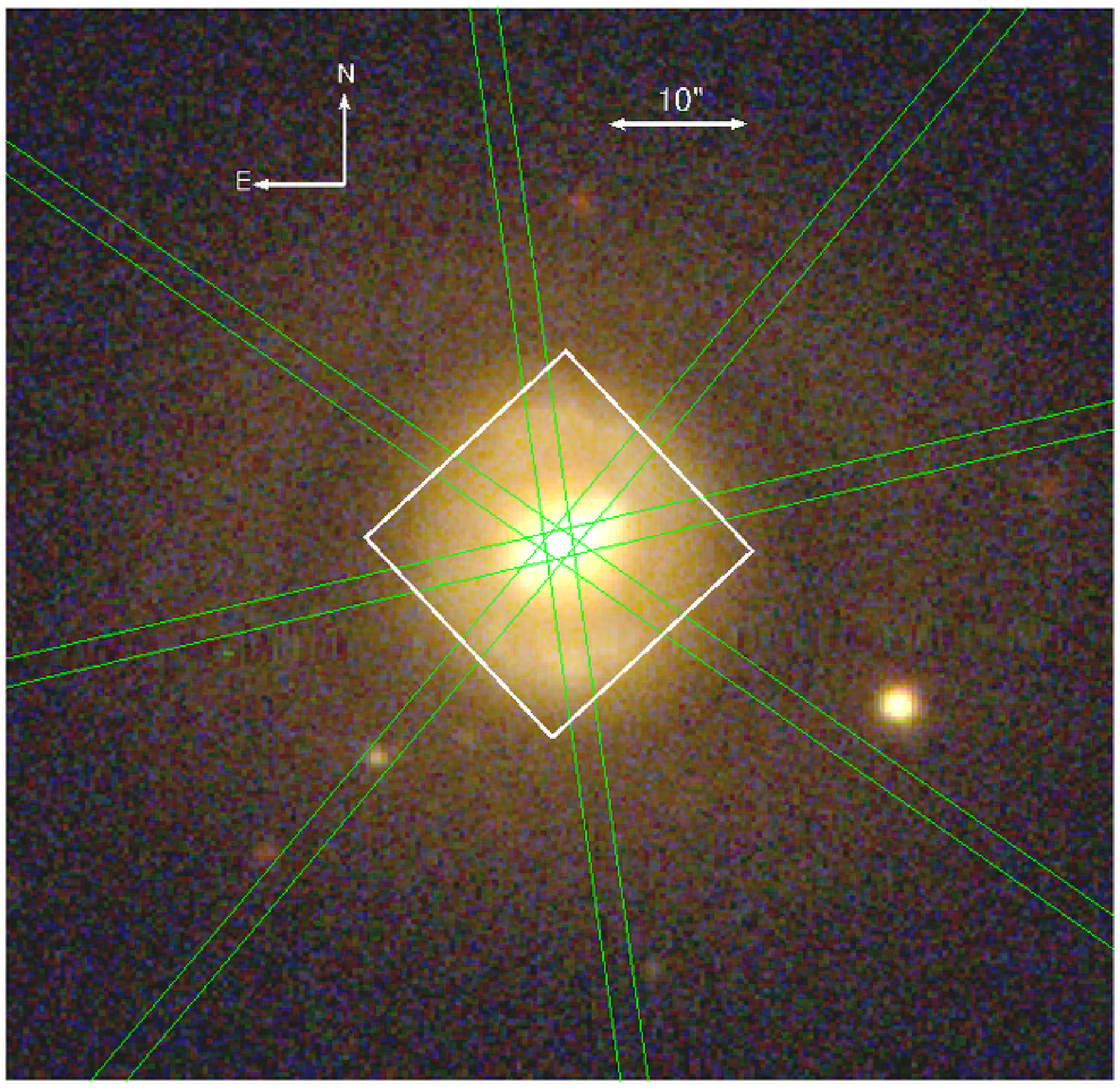}\\
\includegraphics[width=0.49\textwidth,viewport=0 0 590 550,clip]{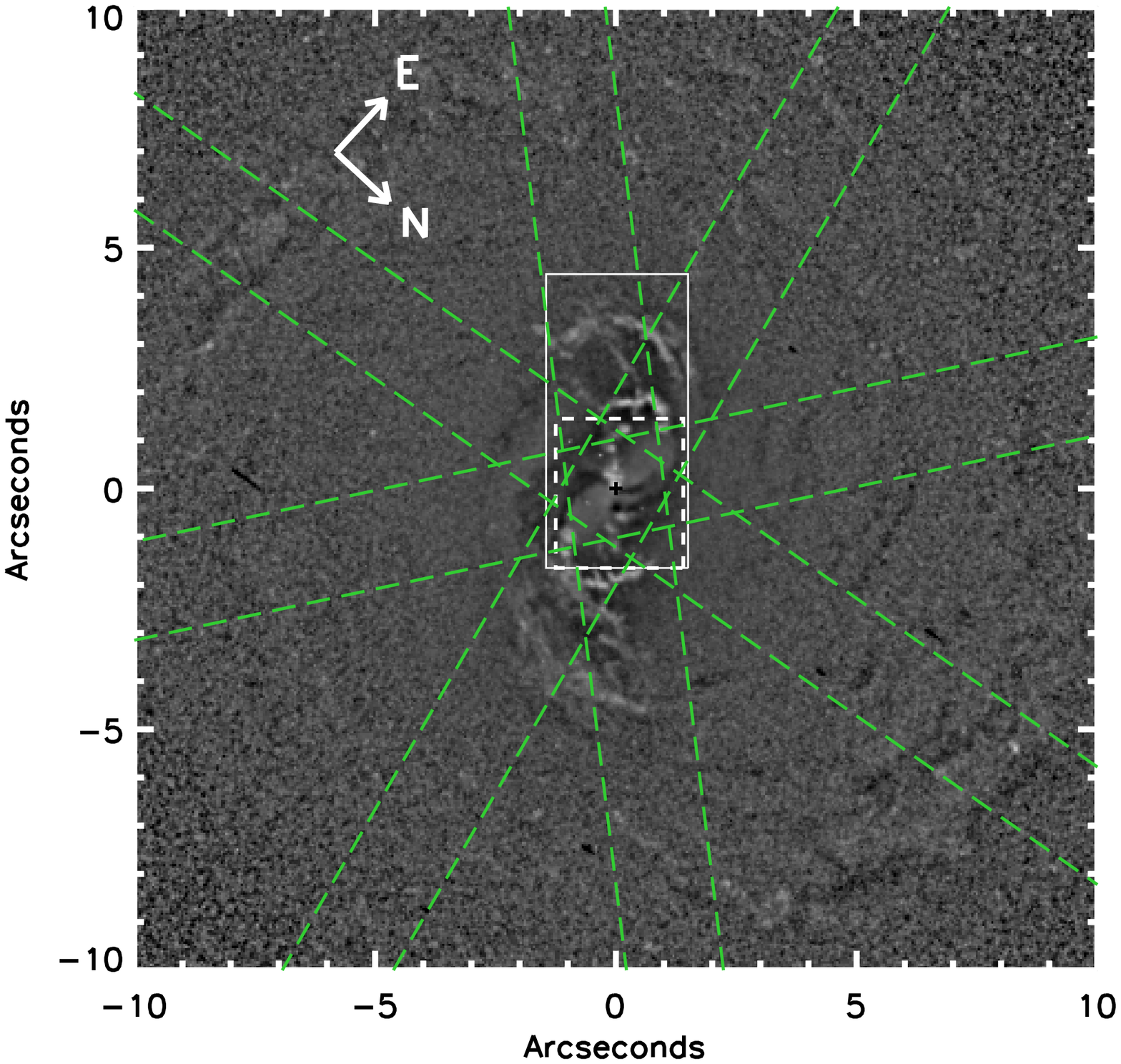}  
\caption{Top:  Large-scale combined $gri$ band image of Mrk 573 from SDSS DR10. APO DIS observations shown in green. 
20.0$" \times$20.0$"$ structure map blow-up outlined in white. Bottom: Enhanced contrast 20.0$" \times$20.0$"$ structure 
map \citep{Pog02, Fis10} of the {\it HST} WFPC2/F606W image of Mrk 573. Bright areas correspond to line emission and dark areas correspond 
to dust absorption. Fields of view for Gemini/NIFS Z- and K-band observations used in our analysis are indicated by the 
white solid and dashed line boxes. APO DIS long-slit observations are shown in green. The central cross marks the 
continuum centroid of the image.}

\label{fig:structure}
\end{figure}

Seyfert 2 AGN Mrk 573 was chosen as an ideal candidate to investigate the connection between fueling flows, 
in the form of dust spirals \citep{Pog02,Sim07}, and outflows of ionized gas in AGN, as the central engine is radiating near the Eddington limit 
(log($M_{BH}/M_{\astrosun}$) = 7.28, L$_{bol} = 10^{45.5}$, \citealt{Bia07,Kra09}). 
As an (R)SAB0+(rs) spiral galaxy, Mrk 573 has an outer ring, 
inner ring and host disk generally aligned along a PA of $\sim$ 90$\degree$, with weak arms forming a bar along a PA $\sim 0\degree$, visible 
in SDSS imaging as shown in Figure \ref{fig:structure}. A close up of the inner host disk, observed via {\it HST} imaging (also Figure \ref{fig:structure}), 
shows arcs of ionized gas that are a result of nuclear dust spirals crossing into the ionizing bicone from the central AGN. {\it HST}/STIS long-slit spectra 
along a position angle of 108.8$^{\circ}$ previously revealed that the ionized gas is in outflow \citep{Sch09,Fis10}, with the observed radial velocity 
pattern at projected distances $< 1"$ from the nucleus following the classic acceleration/deceleration found in our kinematic studies of other Seyfert galaxies. 

Using kinematic and geometric models (\citet{Fis10}, hereafter Paper I)
we had previously concluded that the ionized arcs in the extended NLR were due to an intersection between the inner disk, with an axis inclined 
30$^{\circ}$ from our line of sight, and a wide bicone of ionizing radiation, with an axis inclined by 60$^{\circ}$ from our line of sight. 
However, our previous study found a lack of agreement between our kinematic model and radial velocities at projected distances $> 2"$ 
($\sim$700pc) from the nucleus. Large velocities at this distance were hypothesized to be either A) rotation of a high velocity 
host disk or B) in situ acceleration of gas off spiral arms by the intersecting radiation field or an outflowing wind. The large 
velocities at this distance could be due to rotation, as suggested by \citet{Sch09}, but the unprojected rotation velocities from our previous 
model in Paper I extend up to $\sim$400 km s$^{-1}$, which are rare in spiral galaxies but not unheard of in spiral galaxies \citep{Spa00}. In addition, the 
model geometry required the SW side of the inner disk to be closer to us, which, given the velocity curve, would require the disk to be 
rotating in the sense that the spiral arms are unwinding. Again, this is a very rare but not impossible occurrence \citep{But03}. Alternatively, 
the velocities at distances $> 2"$ from the nucleus could be primarily due to in situ acceleration, where the radiation field or an outflowing 
wind is impinging on the spiral arms. This is an intriguing possibility, because it suggests that at least a portion of the previous fueling flow 
to the nucleus was ionized after the AGN turned on, and is now being driven out by the AGN. 

To address these issues, we have obtained kinematic maps of Mrk 573 in the infrared and optical using Gemini NIFS and APO DIS observations, 
respectively. By analyzing the interaction between the inner host disk, AGN ionization bicone, and fueling flow, we have determined the source of 
the kinematics in both the ionized gas arcs and the molecular gas close to the nucleus and their relationship to dust spirals and fueling flows. 

\begin{figure*}[htp]
\begin{center}
\begin{minipage}{.47\textwidth}
\vspace{0.5cm}
\hspace{0.6cm}\includegraphics[width=0.48\textwidth,viewport=0 42 360 275,clip]{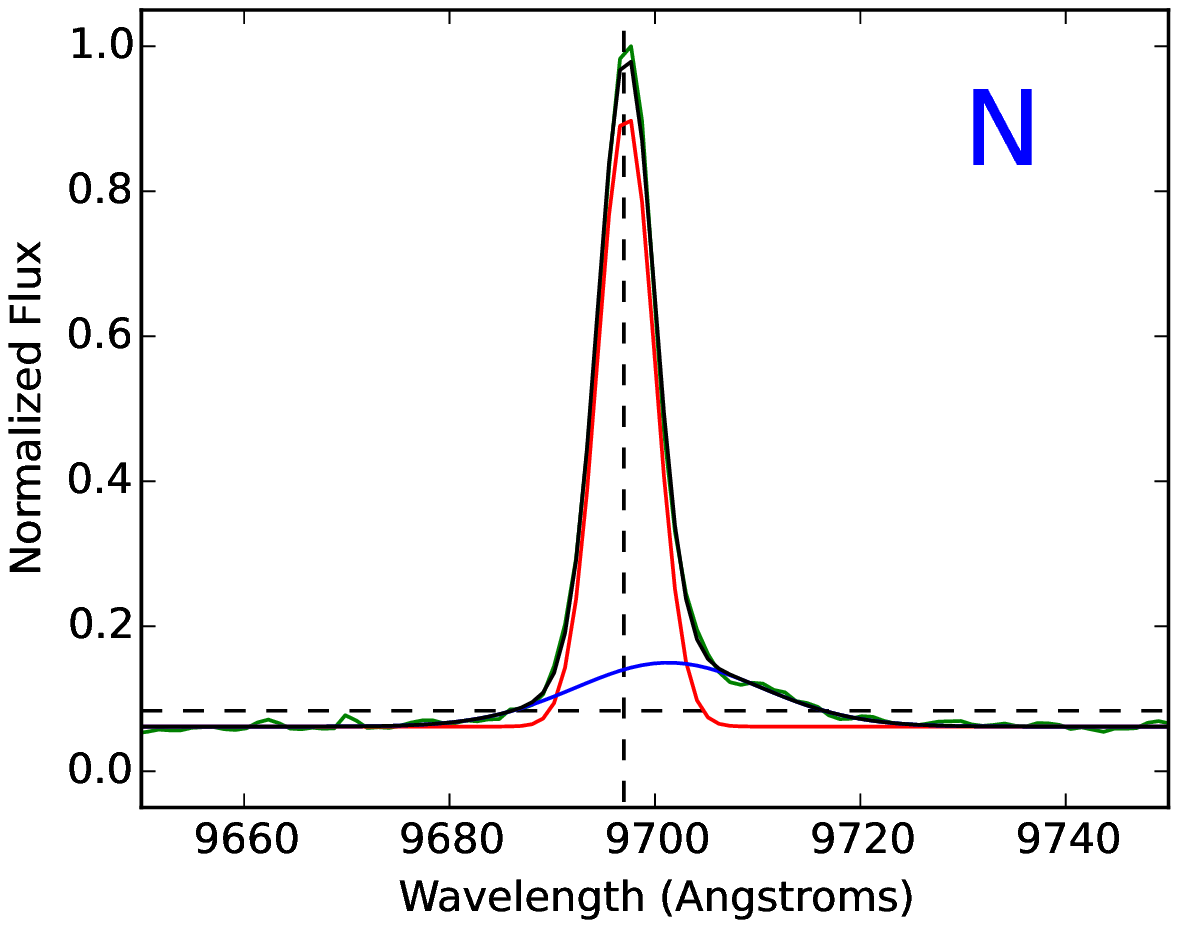} \hspace{-0.3cm}
\includegraphics[width=0.4125\textwidth,viewport=50 42 360 275,clip]{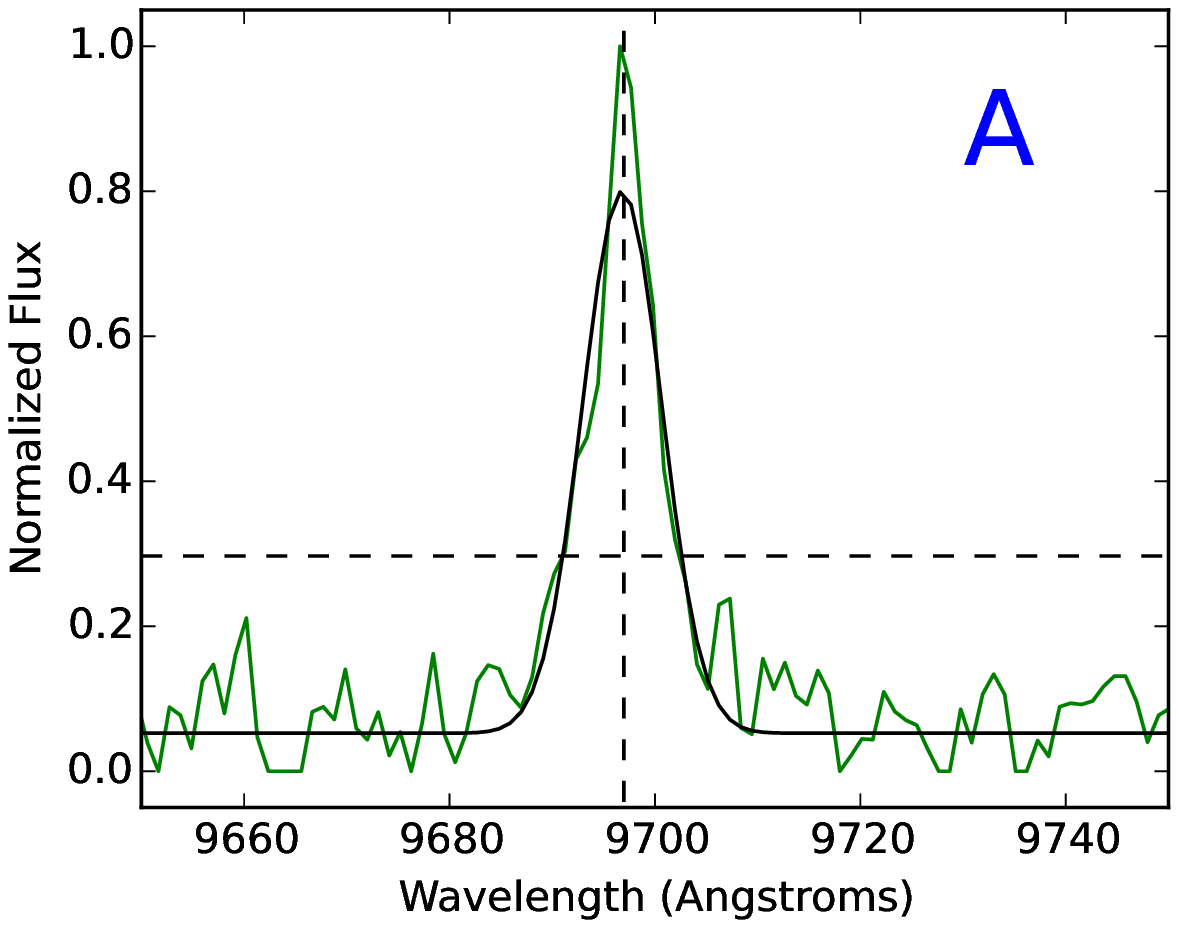}\\ 
\vspace{-0.05cm}
\hspace{0.5cm}\includegraphics[width=0.48\textwidth,viewport=0 42 360 275,clip]{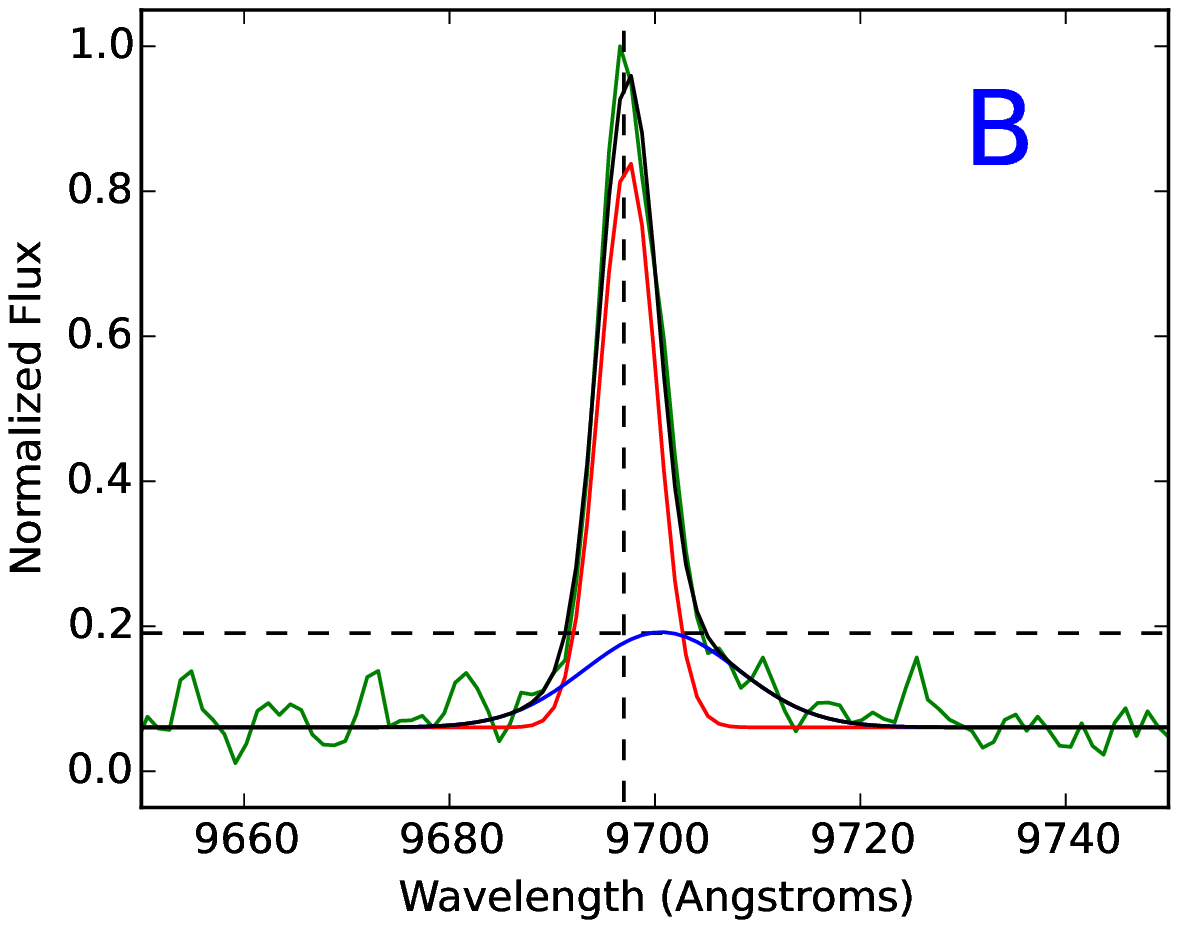} \hspace{-0.3cm}
\includegraphics[width=0.4125\textwidth,viewport=50 42 360 275,clip]{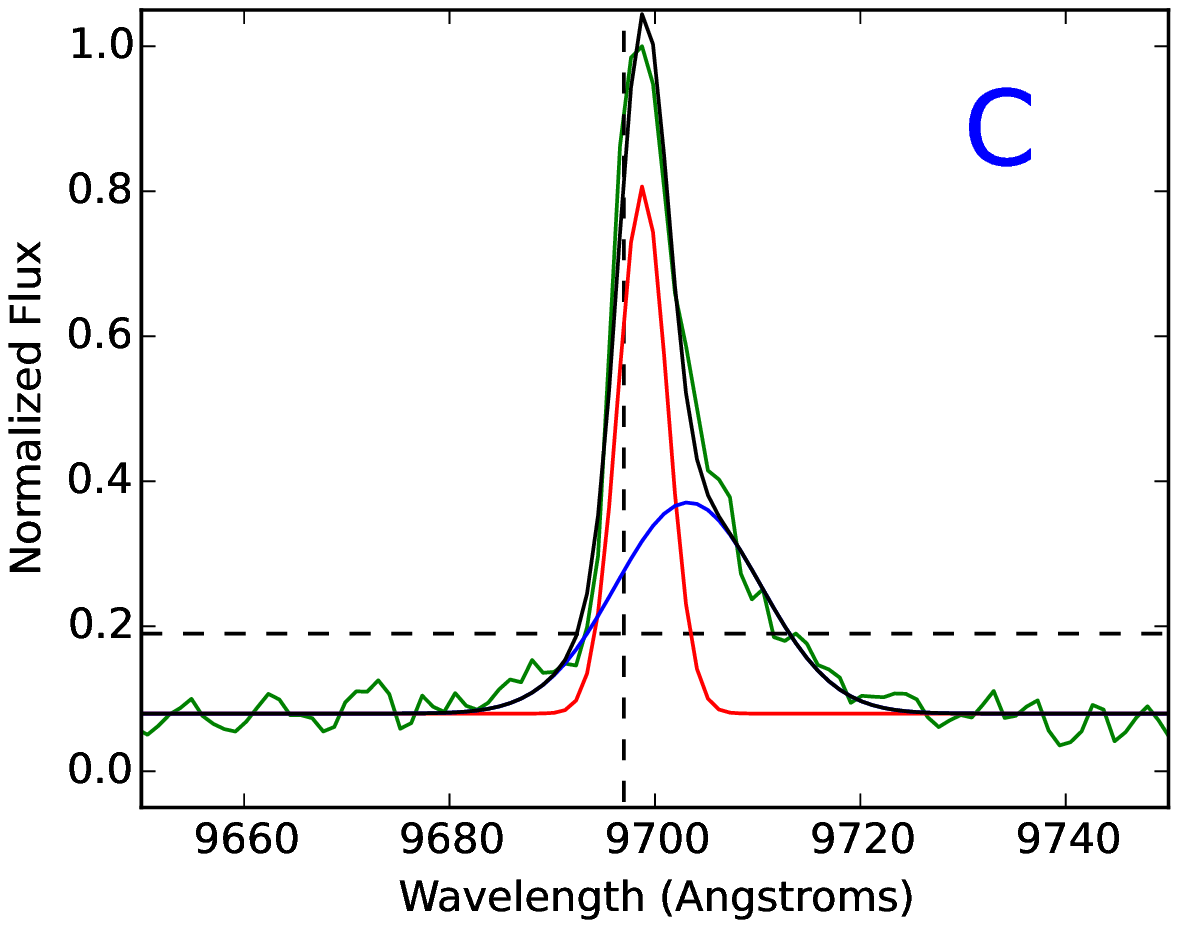}\\
\vspace{-0.05cm}
\hspace{0.5cm}\includegraphics[width=0.48\textwidth,viewport=0 42 360 275,clip]{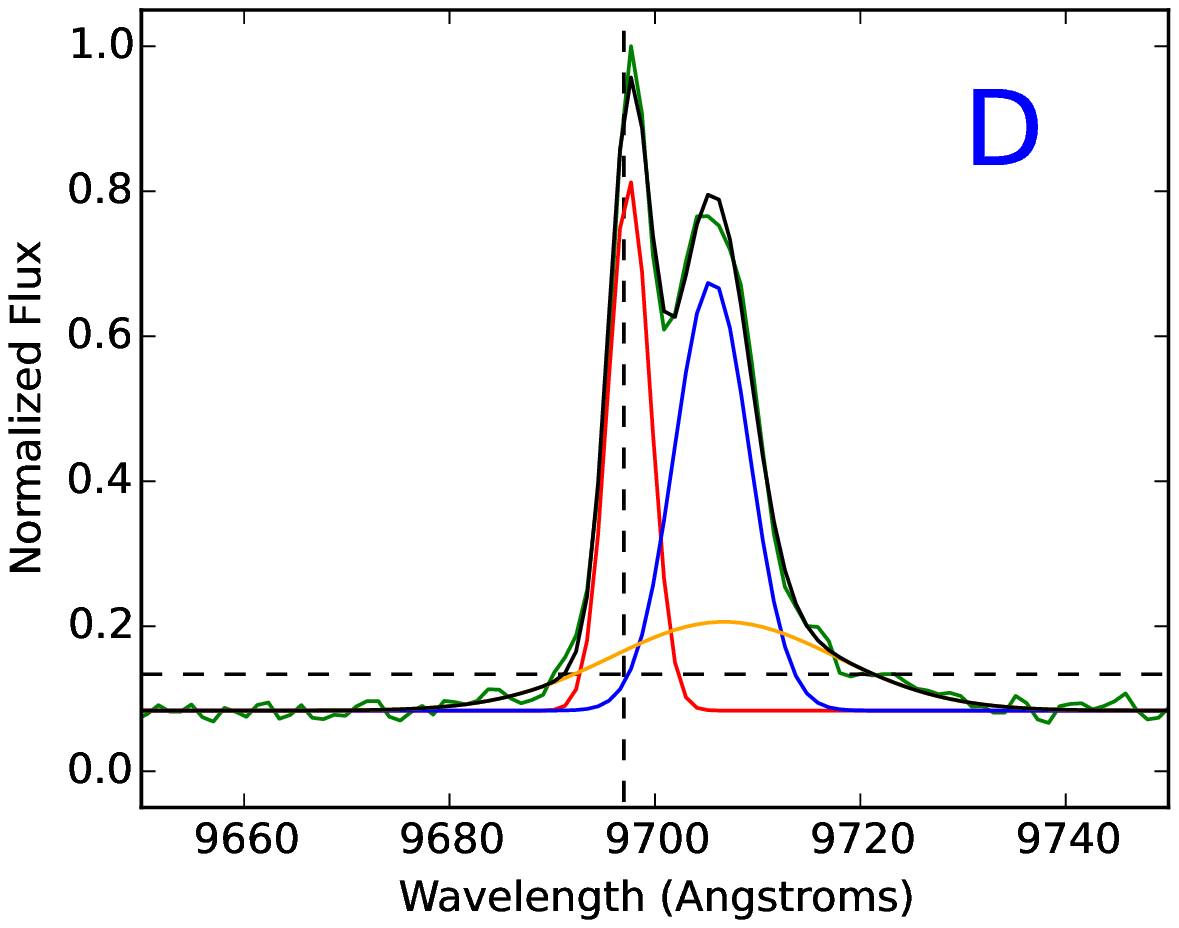} \hspace{-0.3cm}
\includegraphics[width=0.4125\textwidth,viewport=50 42 360 275,clip]{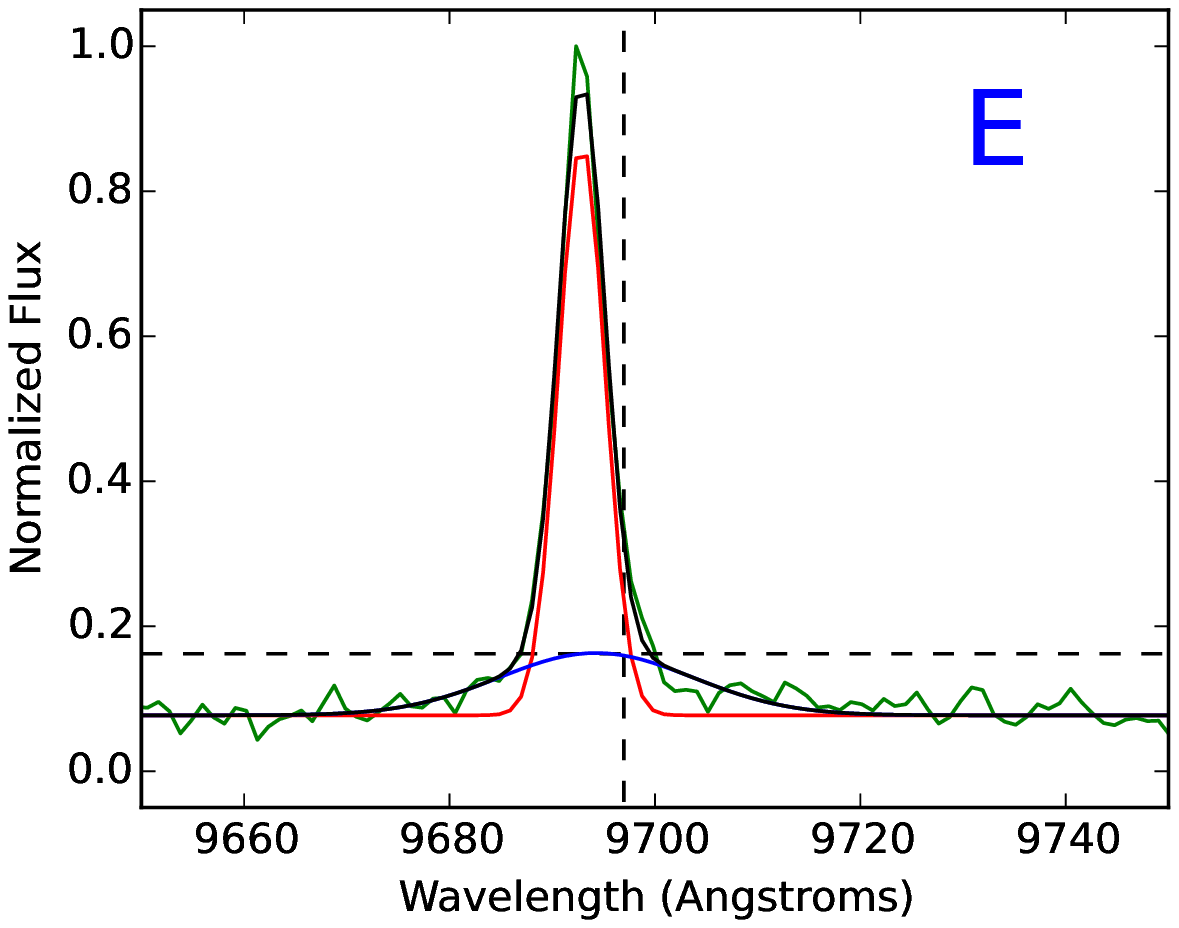}\\
\vspace{-0.05cm}
\hspace{0.5cm}\includegraphics[width=0.48\textwidth,viewport=0 42 360 275,clip]{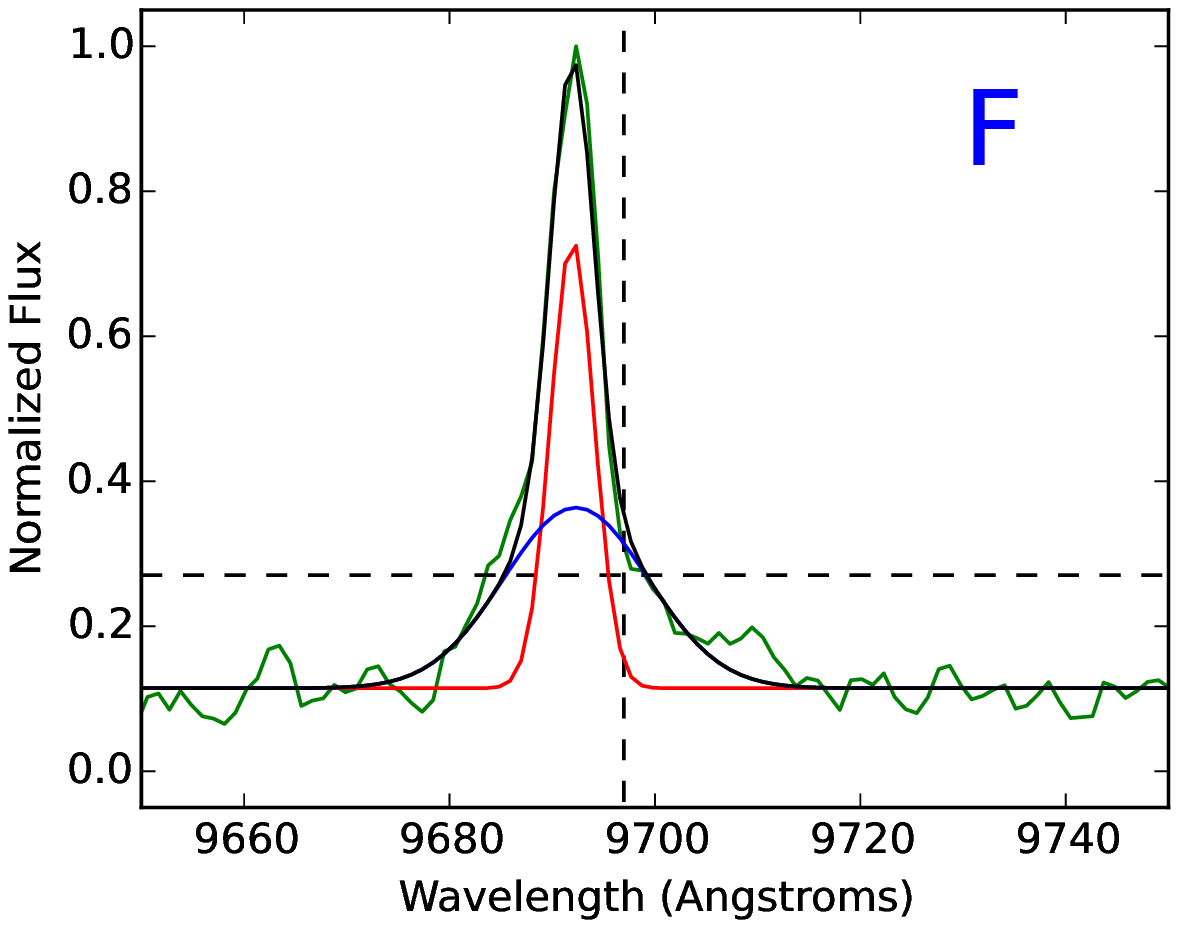} \hspace{-0.3cm}
\includegraphics[width=0.4125\textwidth,viewport=50 42 360 275,clip]{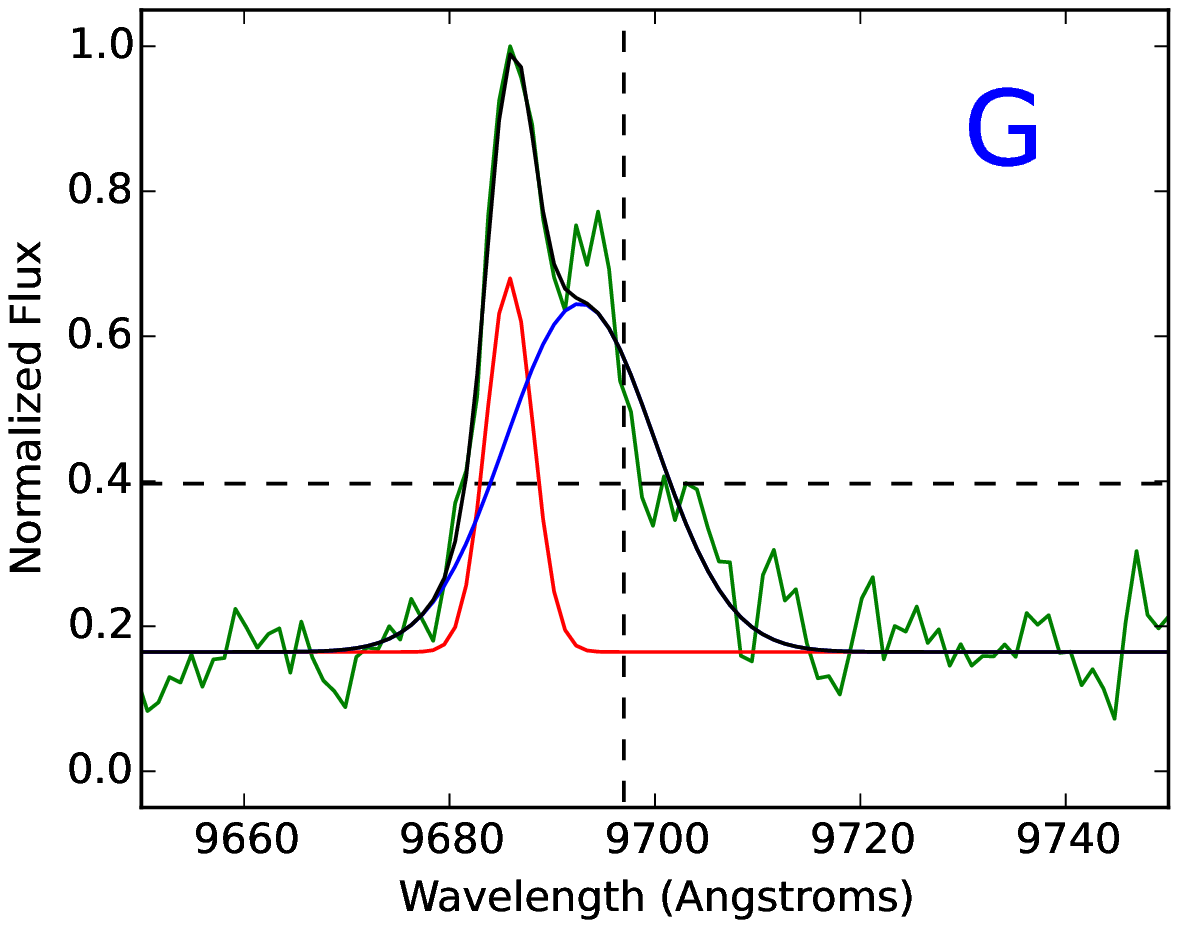}\\
\vspace{-0.05cm}
\hspace{0.5cm}\includegraphics[width=0.48\textwidth,viewport=0 0 360 275,clip]{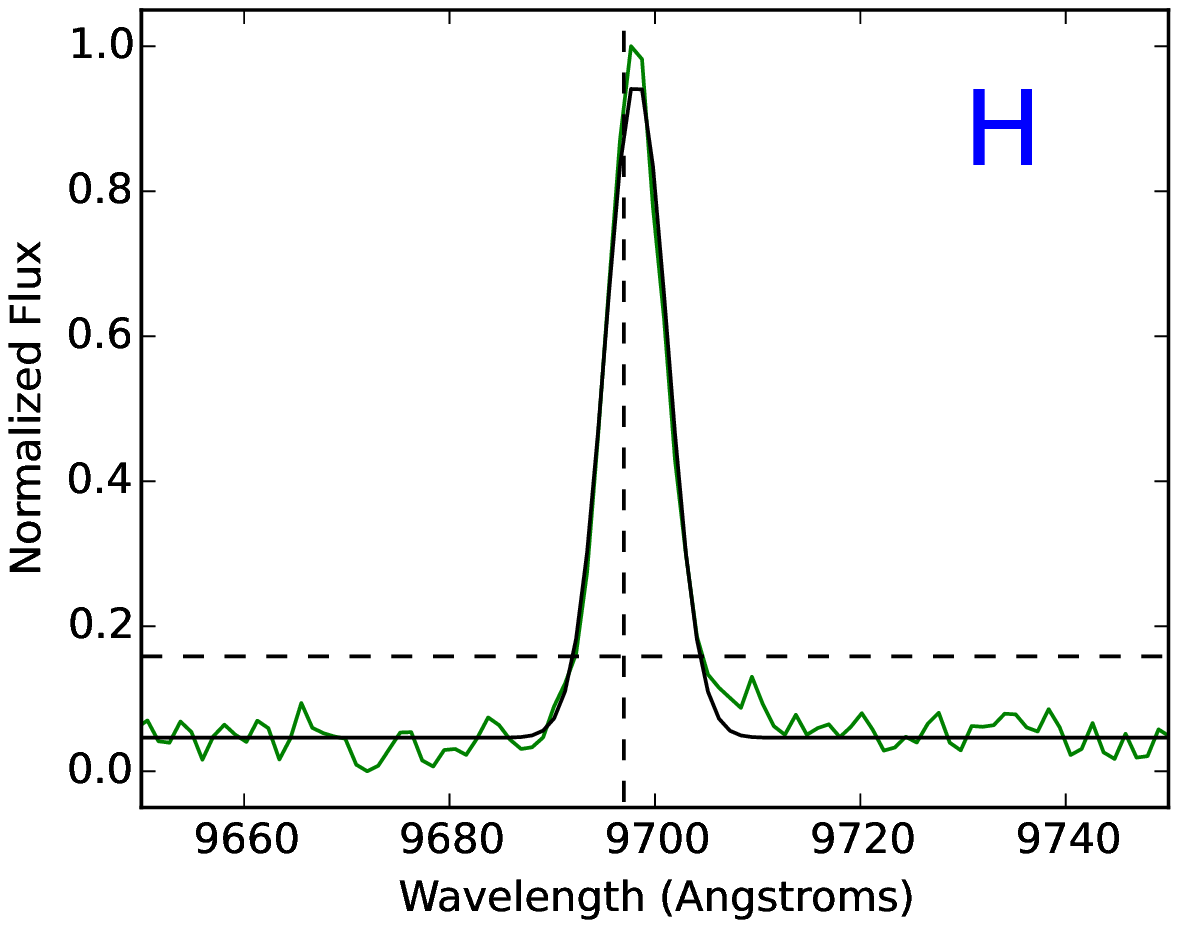} \hspace{-0.3cm}
\includegraphics[width=0.4125\textwidth,viewport=50 0 360 275,clip]{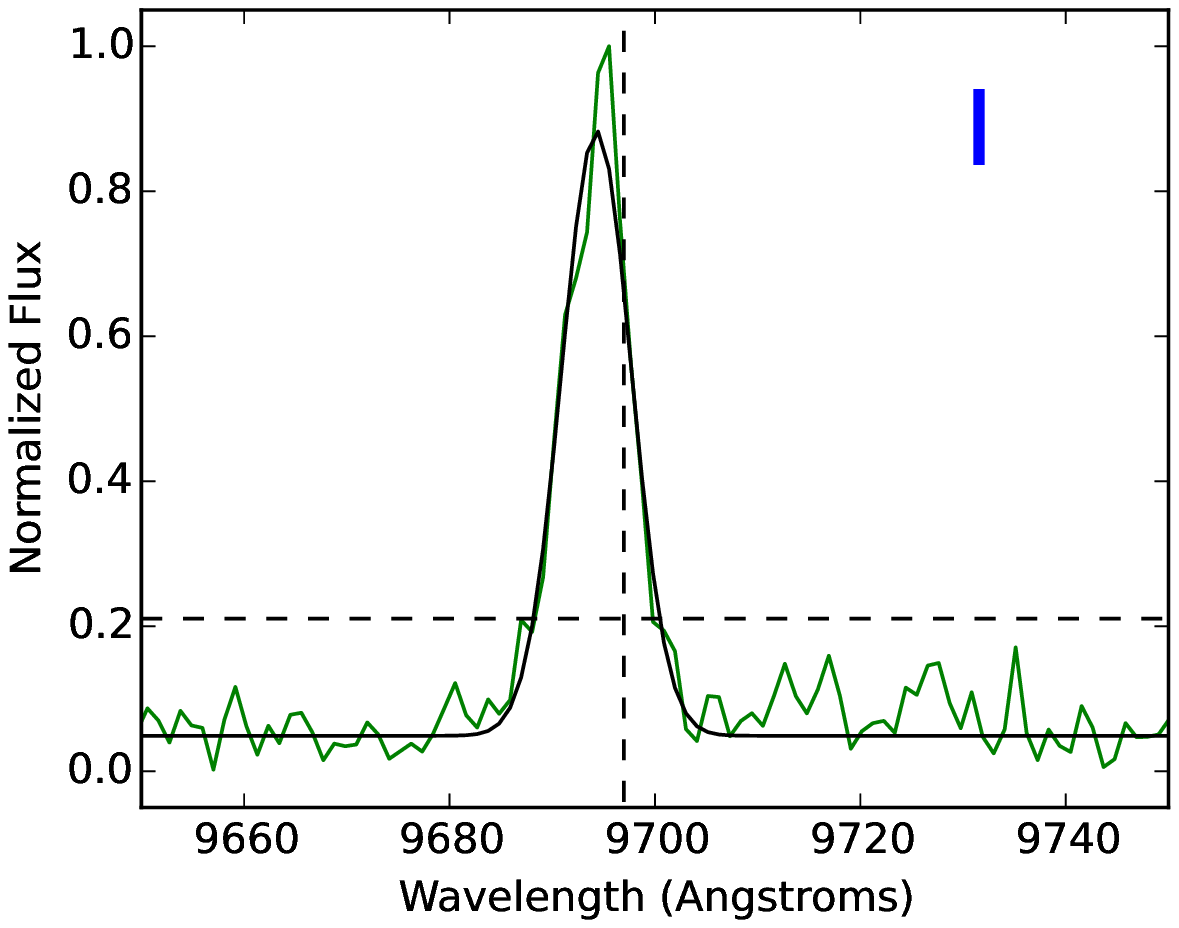}\\
\vskip -0.15in
\end{minipage}
\hfill
\begin{minipage}{.51\textwidth}

\includegraphics[width=\textwidth,viewport=12 20 475 830,clip]{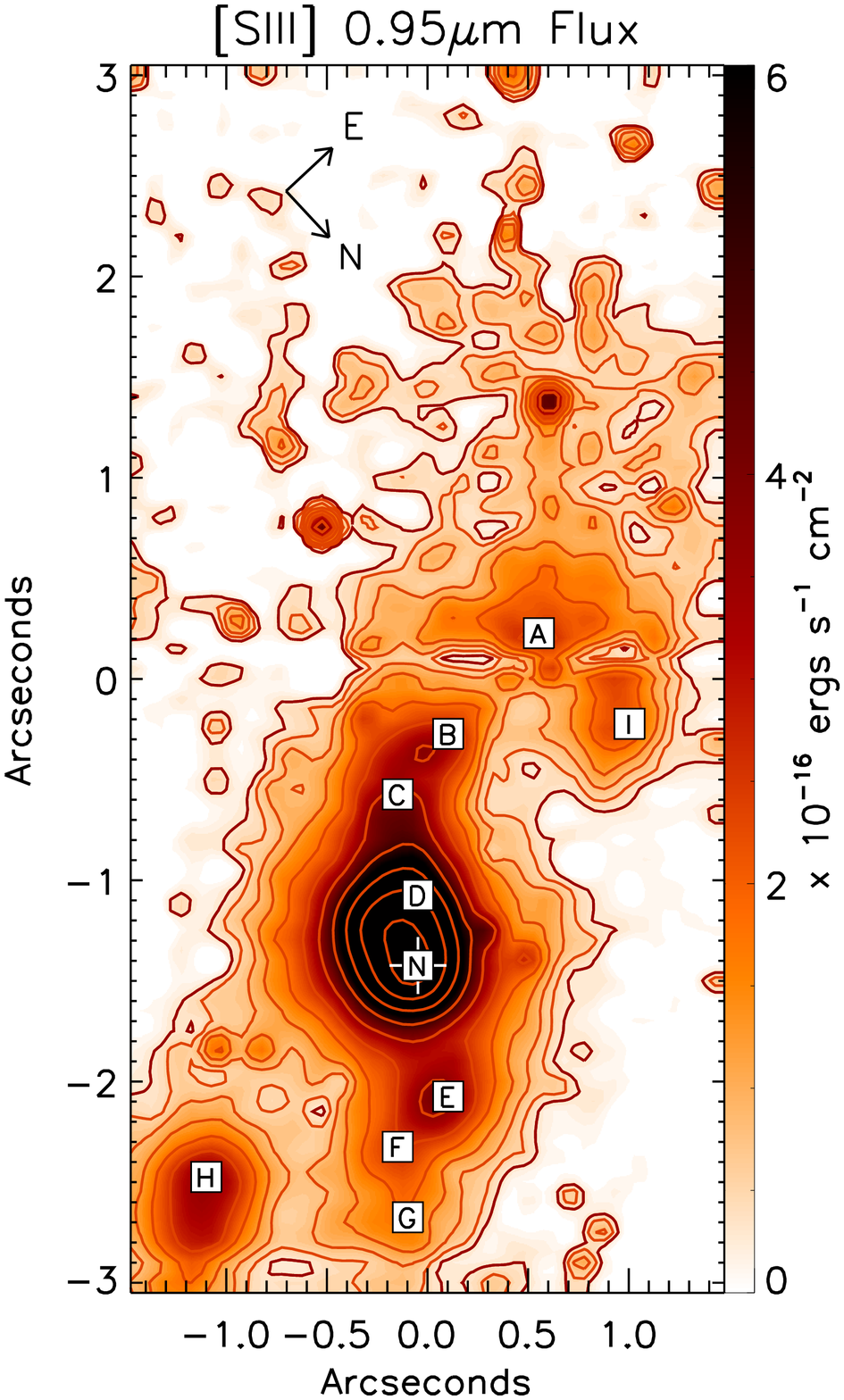}

\end{minipage}
\caption{Continuum-subtracted [SIII] flux distribution obtained from the combined $\sim 3" \times 6"$ Z-band datacube. Individual NLR knots are labeled. 
Spectra of each knot are shown to the left, each overplotted with their best fitting model. The continuum centroid of Mrk 573 is represented by a white 
cross and letter N. Outer, dark-red contours represent a 10$\sigma$ S/N lower flux limit. The horizontal decrease in flux along $y \sim +0.1"$ is an 
artifact from combining the central and offset positioned datacubes.}  

\label{fig:SIIIflux}
\end{center}
\end{figure*}

\begin{figure*}[htp]
\begin{center}
\begin{minipage}{.47\textwidth}
\vspace{0.3cm}
\hspace{0.6cm}\includegraphics[width=0.48\textwidth,viewport=0 42 360 275,clip]{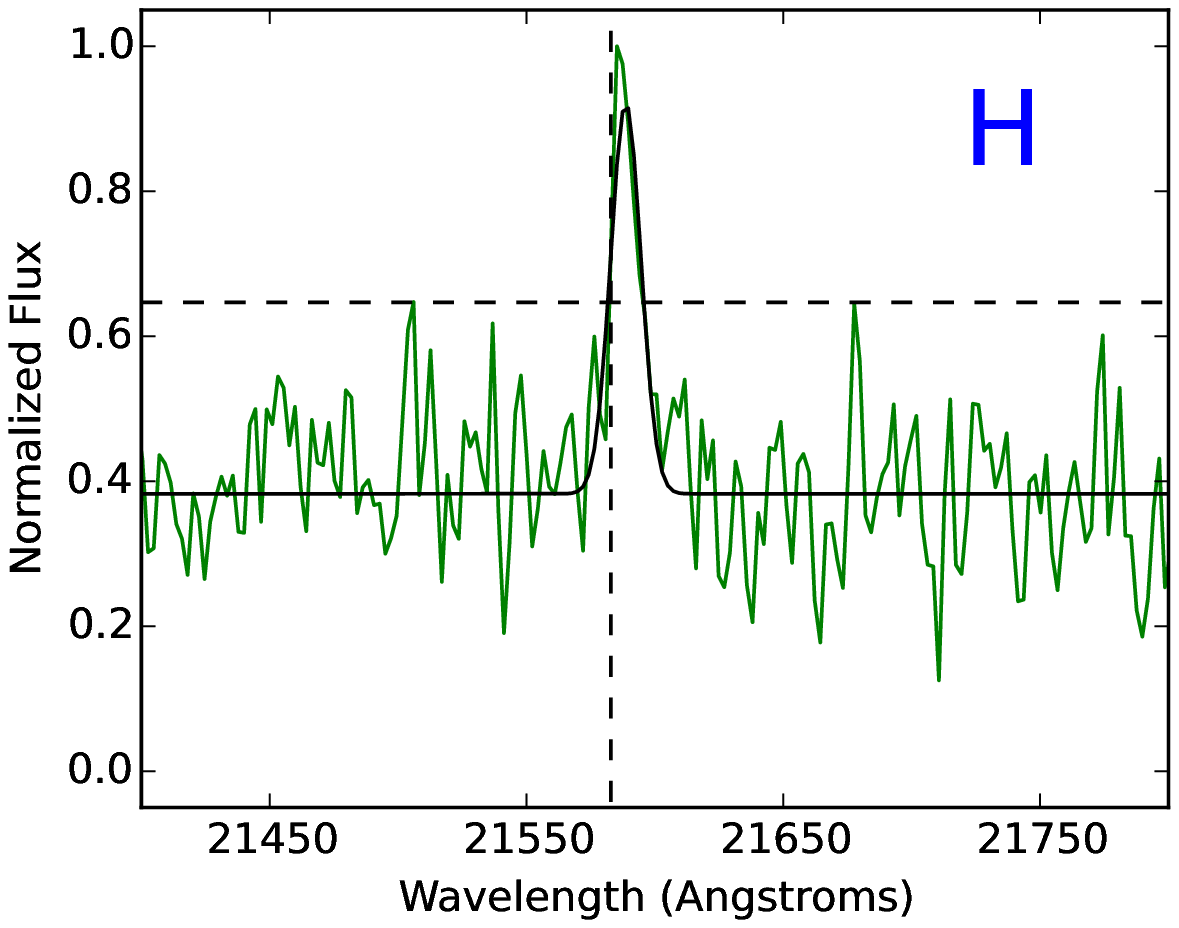} \hspace{-0.3cm}
\includegraphics[width=0.4125\textwidth,viewport=50 42 360 275,clip]{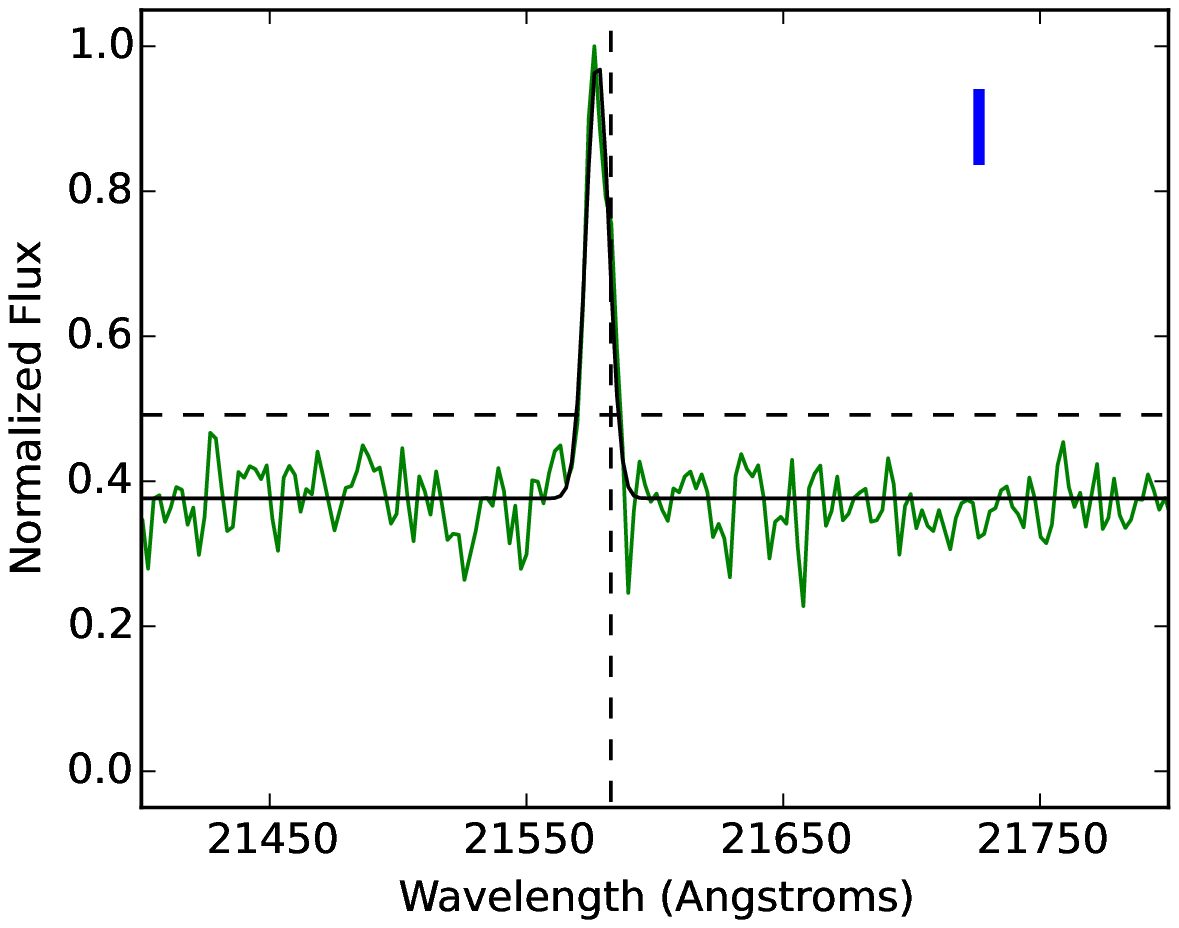}\\ 
\vspace{-0.05cm}
\hspace{0.5cm}\includegraphics[width=0.48\textwidth,viewport=0 0 360 275,clip]{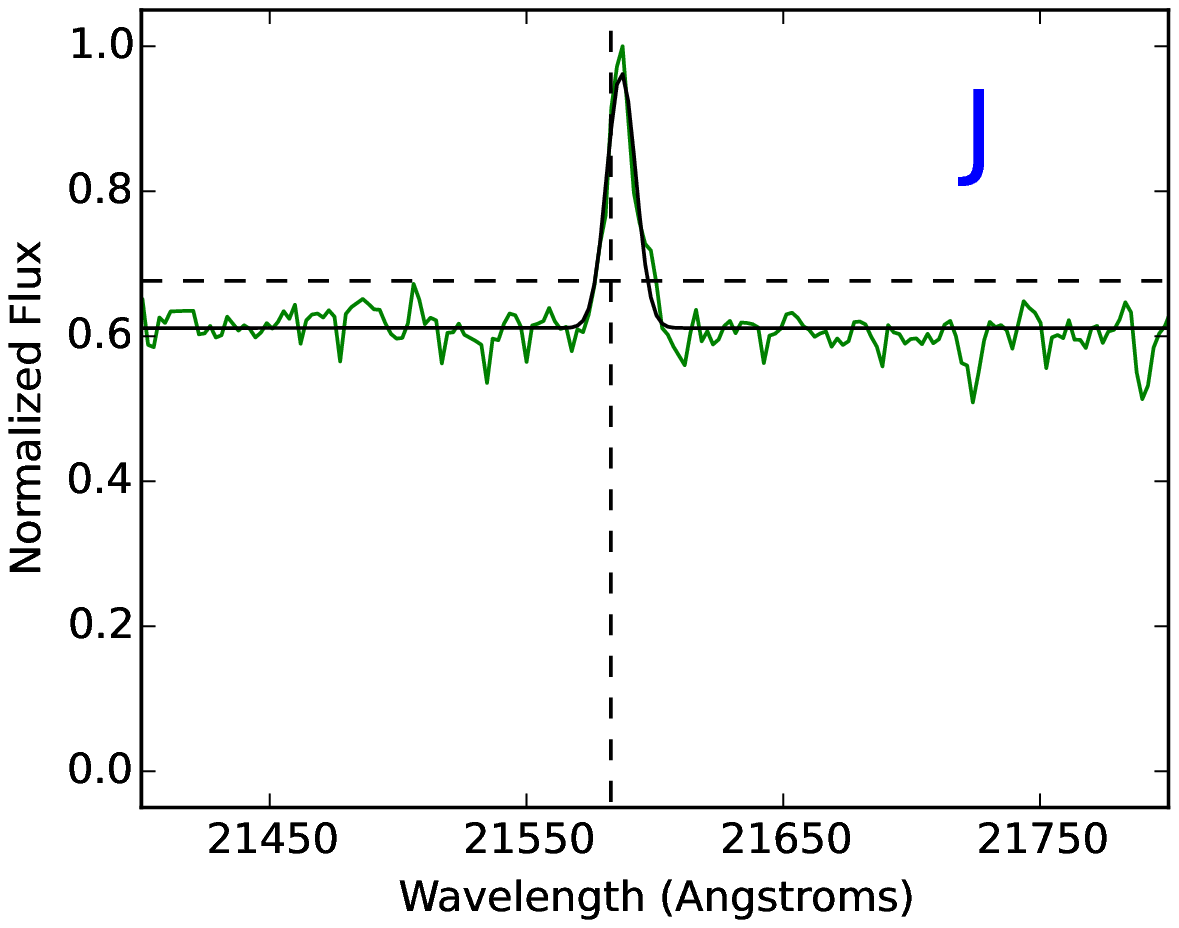} \hspace{-0.3cm}
\includegraphics[width=0.4125\textwidth,viewport=50 0 360 275,clip]{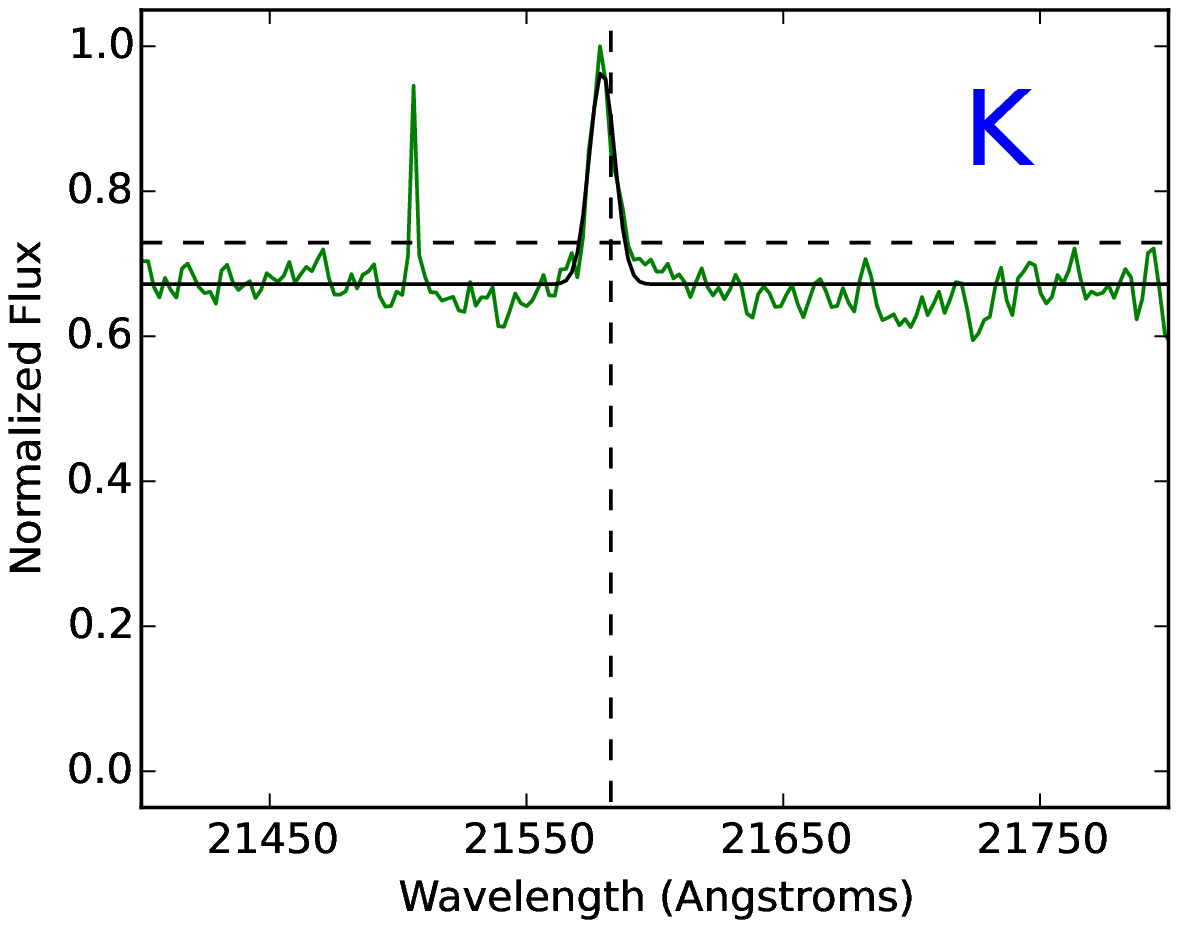}\\

\vskip -0.15in
\end{minipage}
\hfill
\begin{minipage}{.51\textwidth}
\includegraphics[width=\textwidth,viewport=18 00 500 430,clip]{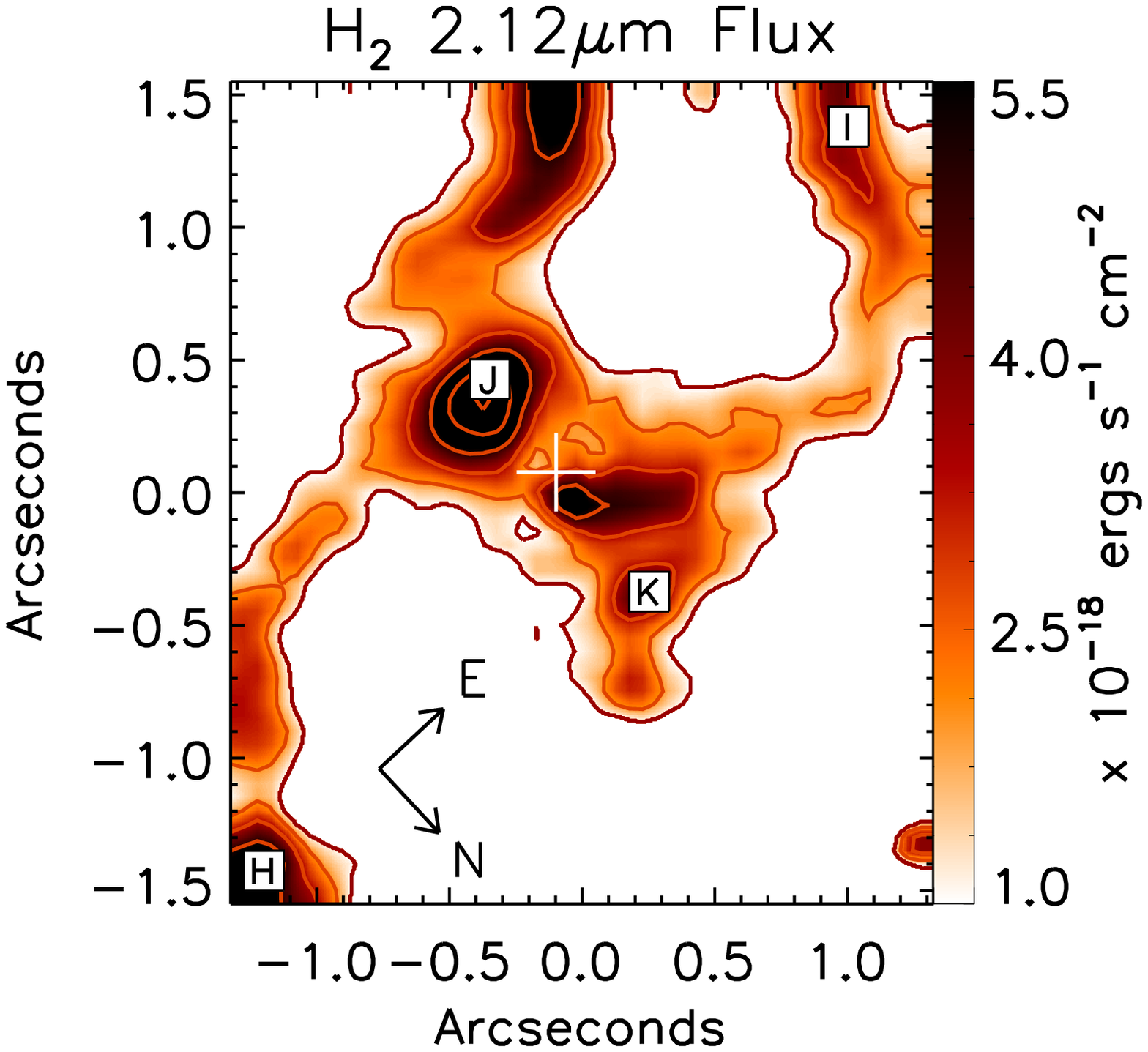}

\end{minipage}
\caption{Continuum-subtracted H$_{2}$ flux distribution obtained from the $\sim 3" \times 3"$ K-band datacube. Representative knots of H$_2$ 
are labeled, with spectra of each knot and overplotted best fitting models are shown to the left. The continuum centroid is plotted as a white cross. 
Outer, dark-red contours represent a 1$\sigma$ S/N lower flux limit.}  
\label{fig:H2flux}
\end{center}
\end{figure*}

\section{Observations and Data Reduction}
\label{sec2}

\subsection{Gemini NIFS}

We observed Mrk 573 using Gemini/NIFS employing the Gemini North Altitude Conjugate 
Adaptive Optics for the Infrared (ALTAIR) adaptive optics system. We obtained observations
in the Z-, J-, and K$_{l}$-bands, with spectral resolutions of $R = \lambda / \Delta\lambda = $ 4990, 6040, and 5290 respectively, and 
covering spectral regions between 0.94 - 1.33 and 2.1 - 2.54$\mu$m. Observation sequencing 
followed standard object-sky-object dithering \citep{Rif11,Rif13} with off-source sky positions common 
for extended targets. Six individual exposures of 600 s were obtained, for 3600 s in total. 
Data reduction was performed using tasks contained in the NIFS subpackage within the 
GEMINI IRAF package, in addition to standard IRAF tasks. The reduction process included 
image trimming, flat fielding, sky subtraction, and wavelength and s-distortion 
calibrations. Frames were corrected for telluric absorptions and then flux calibrated by 
interpolating a blackbody function to the spectrum of a telluric standard star. The 
resultant data cube was median combined into a single data cube 
via the gemcombine task of the GEMINI IRAF package. 

In order to map the kinematics of the ionized spirals into the entire ENLR $\sim8"$ in length, we obtained NIFS $3" \times 3"$ observations of 
the nucleus and adjacent offset positions roughly SE and NW of the nucleus at a position angle of 133$^{\circ}$ east of north. Due to poor S/N 
in the offset positions, H$_2$ and stellar 
kinematic measurements were not obtained outside the nuclear FOV, and [SIII] measurements were only obtained in the nuclear and adjacent 
SE FOVs. The final data cubes for Z- and K$_l$-bands contain approximately 8000 and 4500 spatial pixels respectively, with each pixel 
corresponding to an angular sampling of $0.05'' \times 0.05''$, allowing us to finely sample the inner regions of Mrk 573. At a redshift of $z = 
0.017179$ \citep{Nel95}, pixels cover an area of 17.1 pc $\times$ 17.1 pc, across a field of view of the inner $3'' \times 3''$ (1.026 kpc $\times
$ 1.026 kpc) of the AGN. We note that the observations oversample the point-spread function, which has a full-width at half-maximum (FWHM) of 2 
pixels ($\sim$ 0.1$"$).

Continuum-subtracted [SIII] and H$_{2}$ images, taken from Z- and K$_{l}$-band datacubes respectively, are shown in Figures \ref{fig:SIIIflux} 
and \ref{fig:H2flux}. Structures in both bands correspond closely to what is seen in the {\it HST} imaging of Figure \ref{fig:structure}. A near-linear 
feature is seen near the nucleus in [SIII], and the inner southeast arc and a portion of the northwest arc are seen further from the nucleus. Arcs 
in H$_2$ appear to extend from the nucleus to the north and south, extending out toward the ends of the [SIII] arcs.

\subsection{APO DIS}
To compare the nuclear kinematics to those observed in the extended host galaxy, we also obtained large-scale long-slit observations of Mrk 573 
on four separate occasions using the Dual Imaging Spectrograph (DIS) on the 3.5 m Astrophysical Research Consortium 
telescope at the Apache Point Observatory in Sunspot, New Mexico. Each observation set was taken with a 2.0$''$ slit rotated to various 
position angles, chosen strategically to cover as much of Mrk 573 as possible. The first observation set (total exposure time of 1800 seconds) 
was taken on 2013 December 3 and has a position angle of 103$^{\circ}$ - following the major axis of the inner host galaxy as determined through 
imaging. The second observation set (total exposure time of 1200 seconds) was taken on 2014 December 24 and has a position angle of 
140$^{\circ}$. Finally, the third and fourth sets (total exposure time of 2700 seconds for each set) were taken on 2015 August 14 and have 
position angles of 8$^{\circ}$ and 55$^{\circ}$, respectively. Each observation set results in two spectral images: a $``$blue$"$ image using the 
B1200 grating and a $``$red$"$ image using the R1200 grating. Observations were reduced using standard IRAF tasks plus LA-Cosmic \citep{Dok01} 
to detect and clean cosmic ray hits.  At the given redshift, the scale of the images is  $\sim$342 pc/arcsecond and we are able to characterize radial 
velocities at a distance up to 6 kpc from the center of the galaxy. 

\begin{table}[h!] 
  \centering
  \caption{NIFS / DIS observational parameters}
      \label{tab:obsparams}
  \begin{tabular}{llccc}
    \toprule
    Inst. & Grating & Spatial Res.       & Spectral Res. 				& Wave Range\\
    		     &		     & 	($''$/pixel) &  ($\lambda / \Delta\lambda$)		& ($\mu$m) \\
    \midrule
    NIFS          & Z			& 0.05  	      &	 	4990	  &  0.94 - 1.15 \\	
    NIFS          & K$_{long}$ 	& 0.05	      & 	5290	   & 2.10 - 2.54 \\	
    DIS            & B1200  	   	& 0.40	      & 	4000	   & 0.43 - 0.55 \\	
    DIS            & R1200  	   	& 0.42	      & 	5500	   & 0.60 - 0.72 \\	
    \bottomrule
  \end{tabular}
\end{table}

\begin{figure*}
\centering

\includegraphics[width=0.95\textwidth]{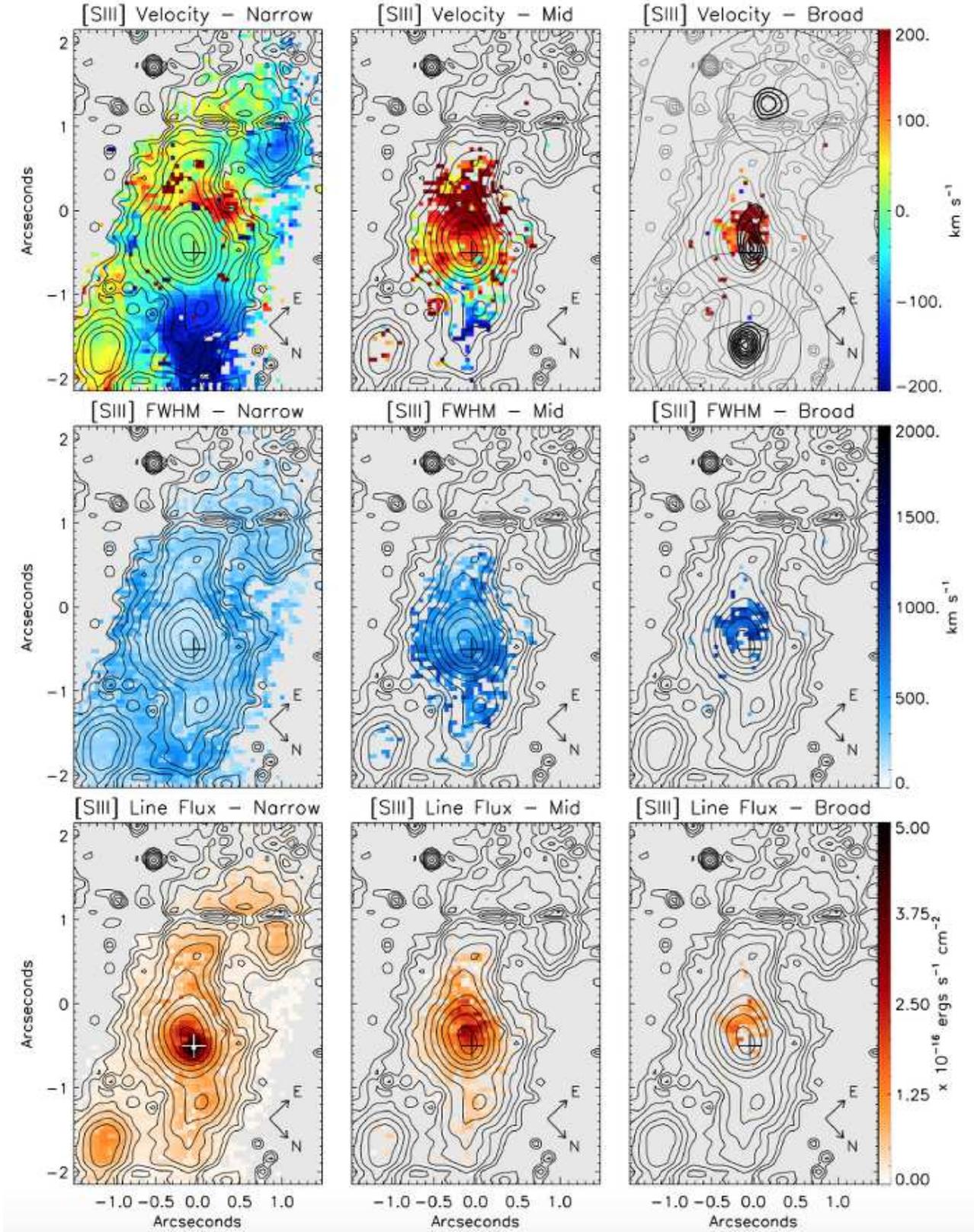} \\

\caption{ Top, middle, and bottom rows display centroid velocity, FWHM, and integrated flux maps of [S~III] $0.9533 ~ \mu$m 
emission-line profiles, respectively. Left, middle, and right columns represent the narrow-, mid-, and broad-component fits of the 
[S~III] $0.9533 ~ \mu$m emission-line profiles, respectively. Black contours represent integrated, continuum-subtracted [S~III] flux images. 
Continuum centroid is depicted by a cross. Radio contours of archival VLA observations at 1.43 GHz and 8.5 GHz are overlaid in 
the top right plot, where [SIII] flux contours are grey and 1.43 GHz and 8.5 GHz distributions are thin and thick black contours, respectively.} 
\label{fig:SIIImaps}
\end{figure*}

\begin{figure}
\centering

\includegraphics[width=0.49\textwidth]{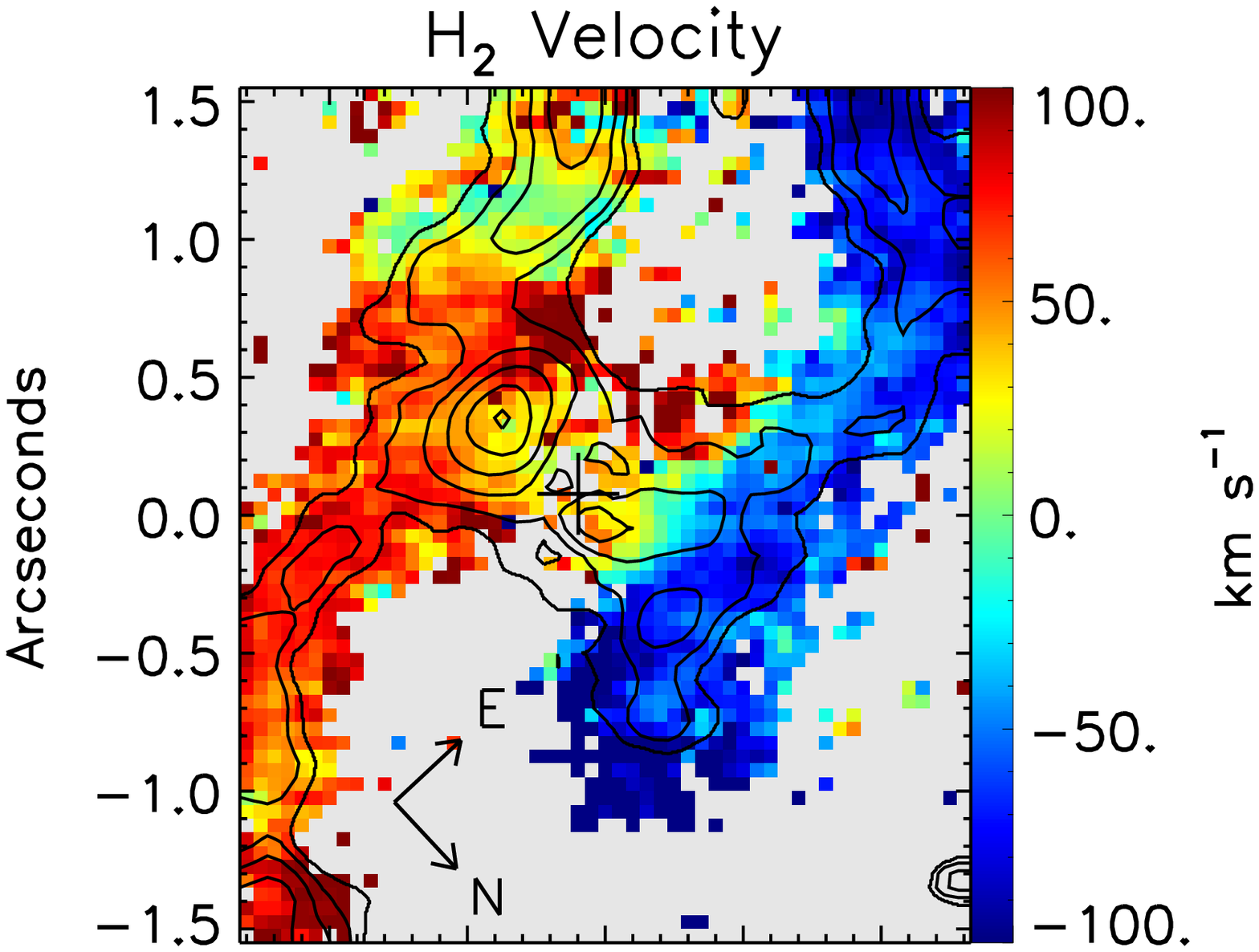} \\
\vspace{-1.3cm}
\includegraphics[width=0.49\textwidth]{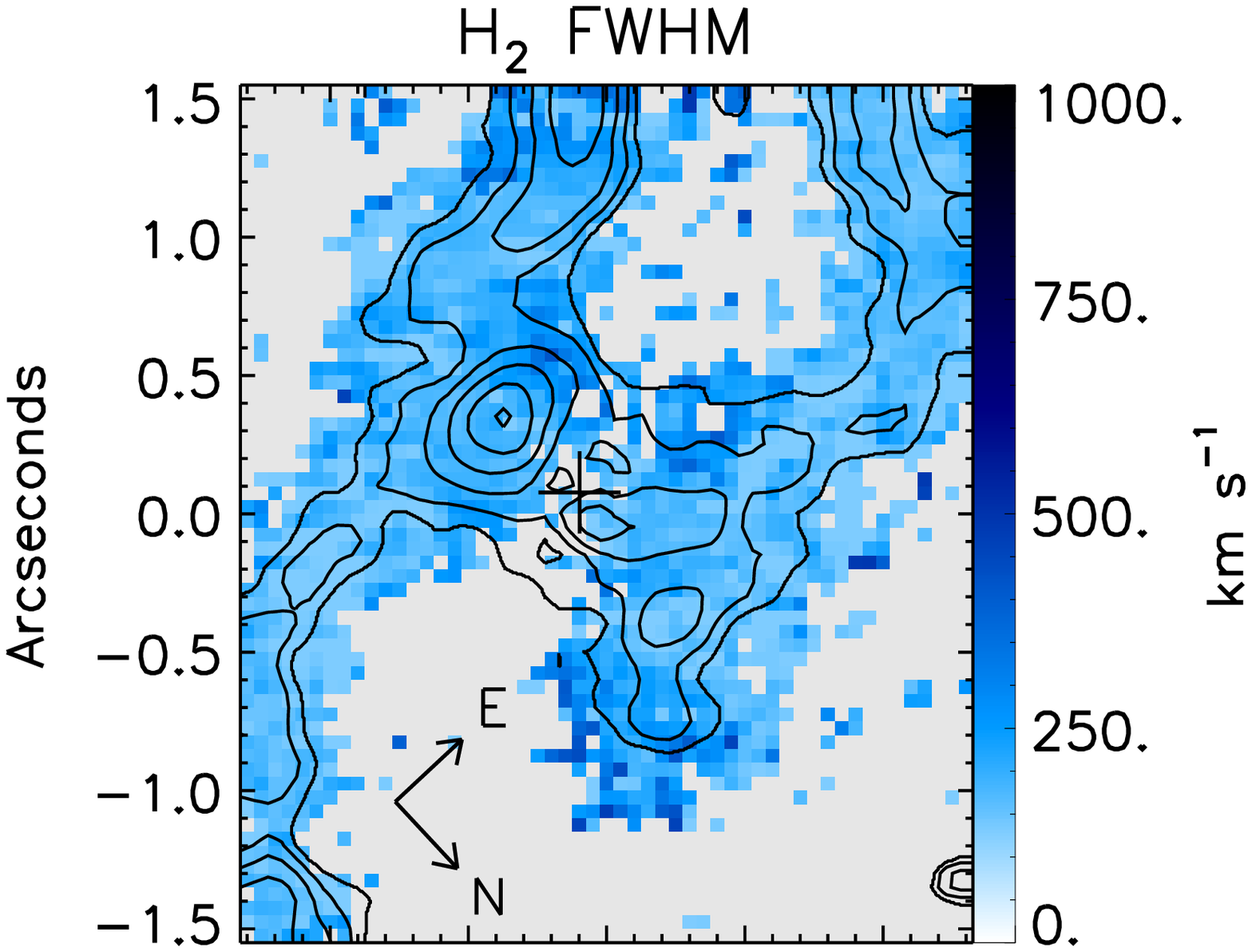} \\
\vspace{-1.3cm}
\includegraphics[width=0.49\textwidth]{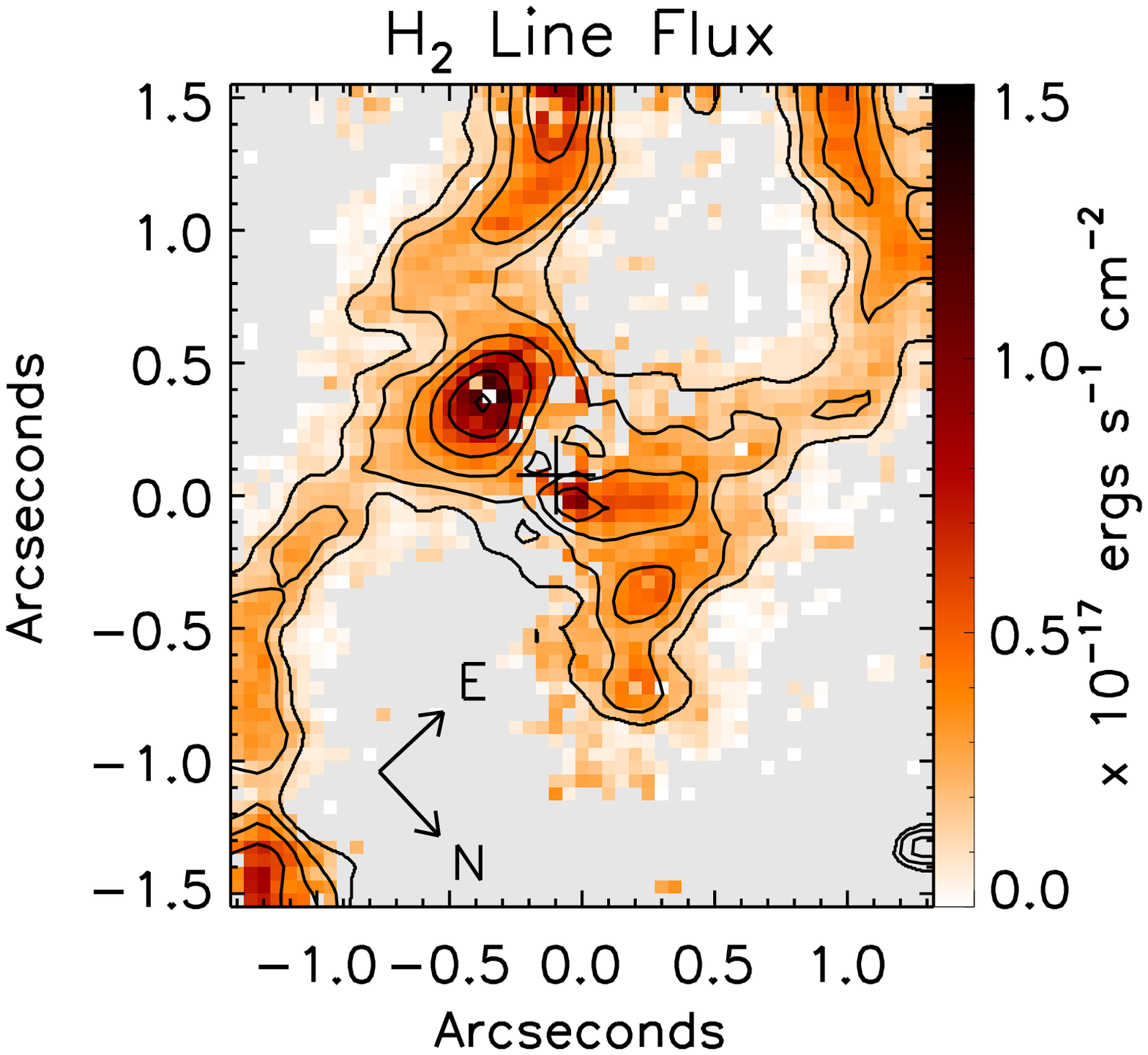} \\

\caption{Top, middle, and bottom rows display centroid velocity, FWHM, and integrated flux maps of H$_{2} ~ 2.12 ~ \mu$m
emission-line profiles, respectively. Black contours represent integrated, continuum-subtracted H$_{2}$ flux images. 
Continuum centroid is depicted by a cross.}

\label{fig:H2maps}
\end{figure}

\begin{figure}
\centering
\includegraphics[width=0.46\textwidth]{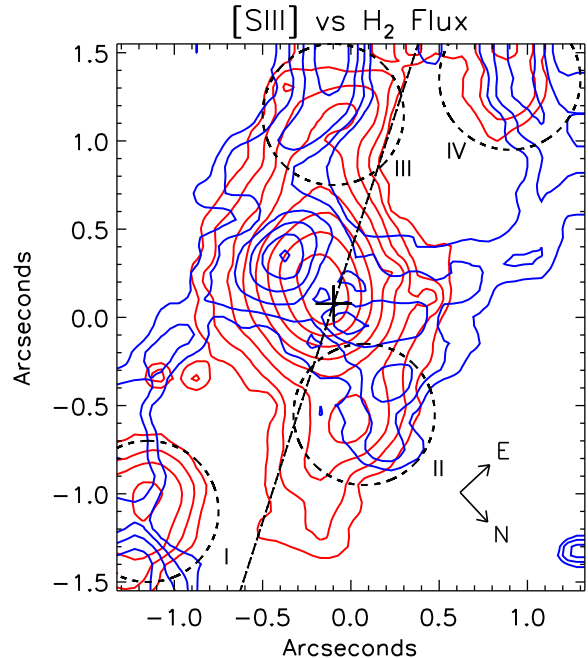}
\vspace{-0.75cm}

\caption{Stacked contour maps of continuum-subtracted [S~III] 0.95$\mu$m (red) and H$_2$ 2.12$\mu$m (blue) 
emission in the overlapping field of view in NIFS Z- and K-band observations. Outer contours represent S/N of 30$\sigma$ and 1$\sigma$. 
The black cross depicts the continuum centroid for both cubes. The dashed line represents the position angle for the projected axis of the NLR, 
PA $= 128\degree$. Dashed circles surround regions of entwined ionized and molecular gas along the edge of the NLR morphology.}

\label{fig:compare}
\end{figure}

\begin{figure*}
\centering

\includegraphics[width=0.38\textwidth]{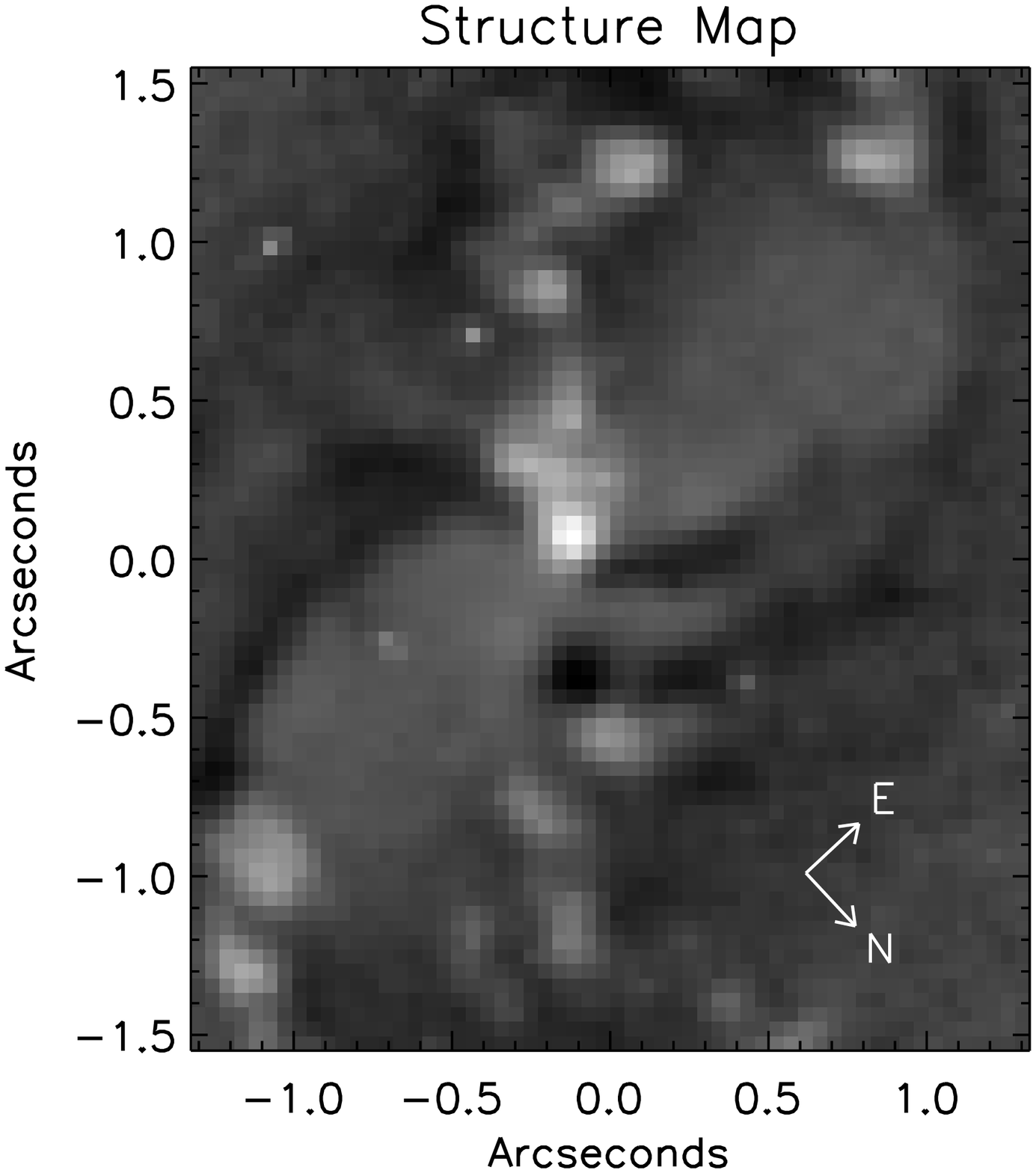} \hspace{-1.63cm}
\includegraphics[width=0.38\textwidth]{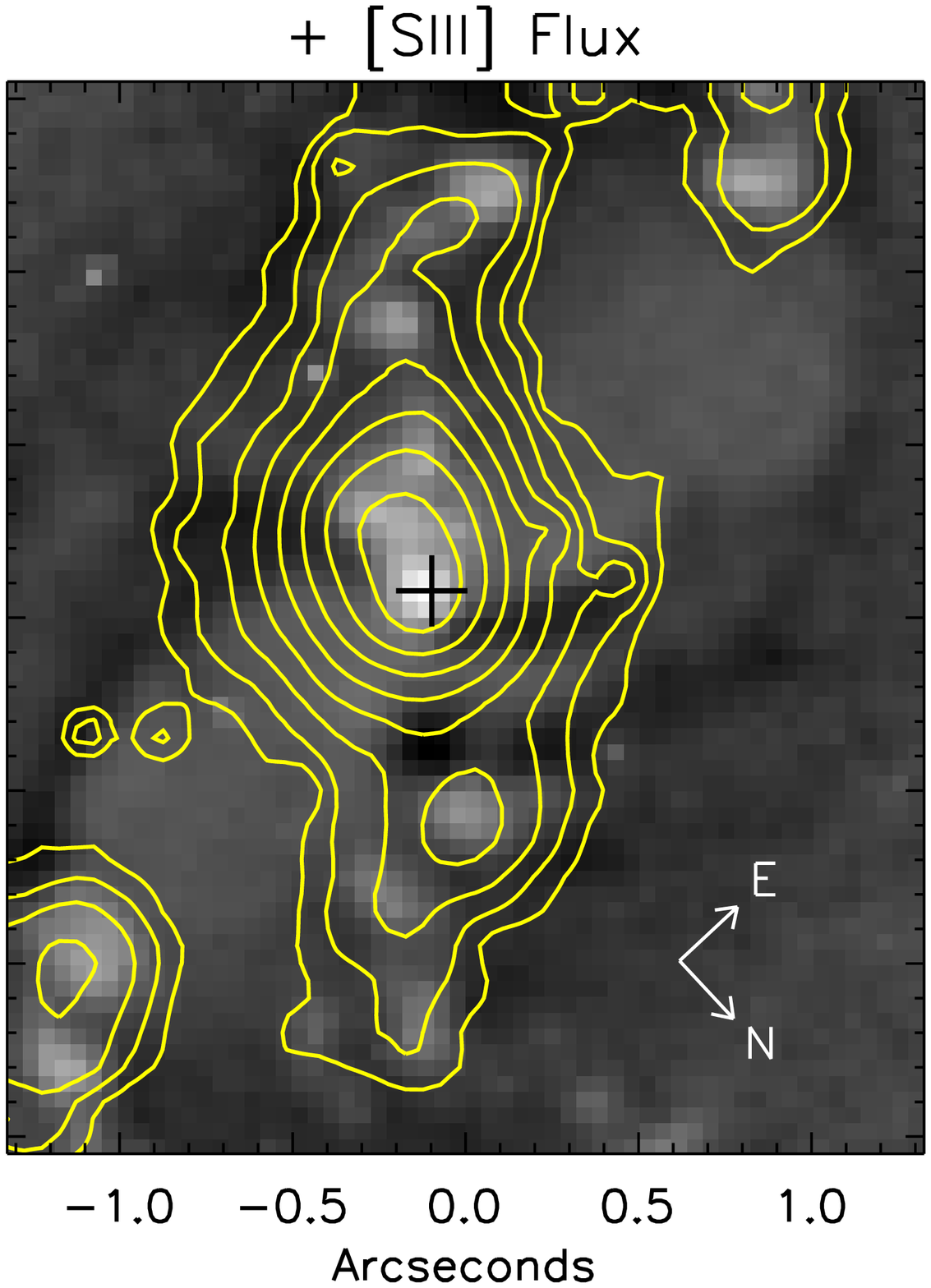}\hspace{-1.53cm}
\includegraphics[width=0.38\textwidth]{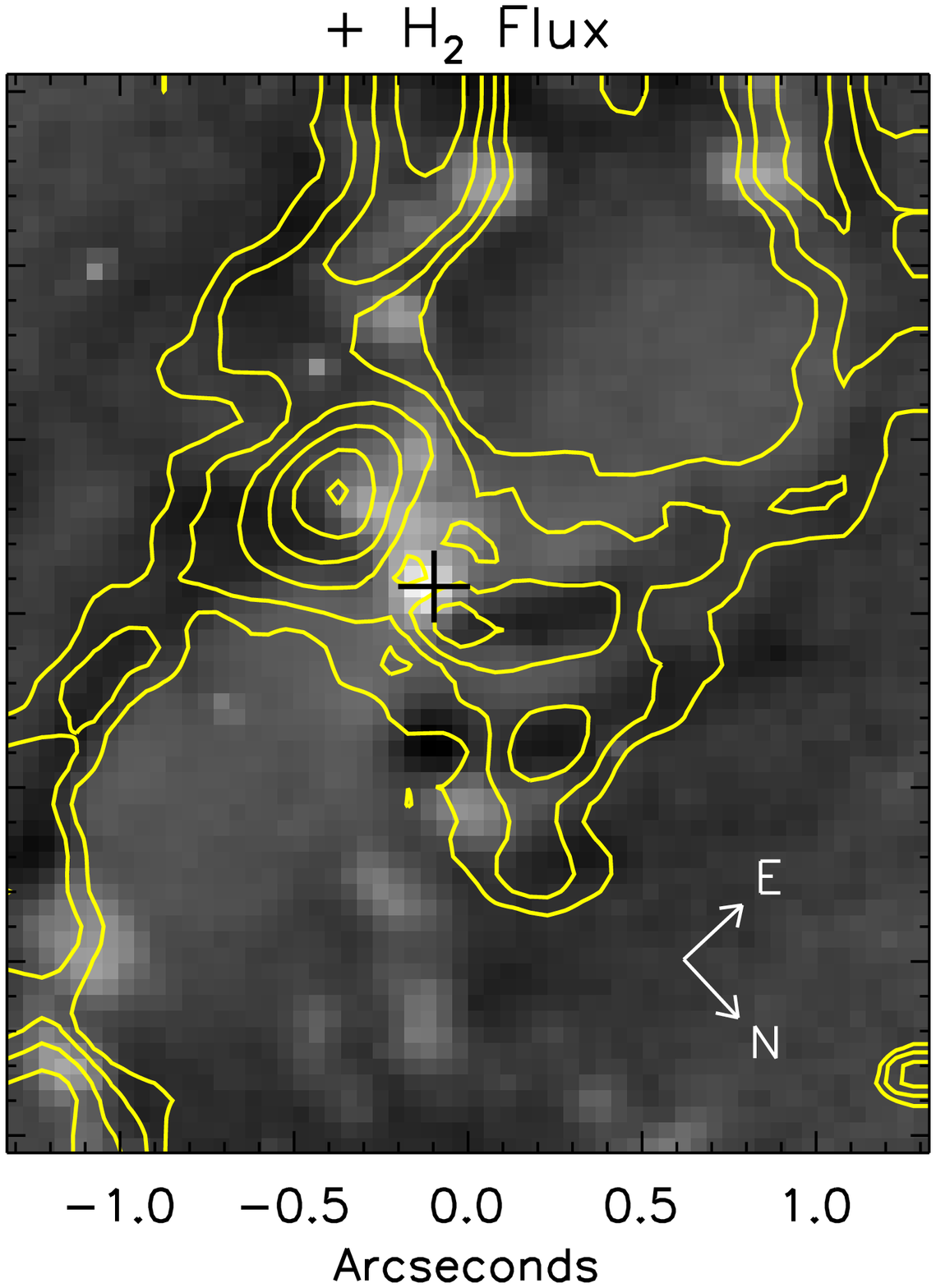} \\

\caption{Left: Structure map from Figure \ref{fig:structure}, and Paper I, cropped to the NIFS K$_{l}$-band FOV. Center: Continuum-subtracted [S~sIII] 
0.95$\mu$m contours mapped to the same structure map, where contours align with knots of emission seen in optical NLR imaging. 
Right: Continuum-subtracted H$_2$ 2.12$\mu$m contours mapped to the same structure map, where contours align with dark, dusty 
absorption features. Contour levels are identical to those in Figure \ref{fig:compare}.} 

\label{fig:struc_flux}
\end{figure*}

\section{Gemini NIFS : Nuclear Ionized and Molecular Gas Kinematics}
\label{sec:gaskin}

Gas fluxes and kinematics in our NIFS observations were determined by fitting emission lines with Gaussians in an automated routine, described in depth 
in Appendix A. Typical fits to the [S~III] and H$_2$ emission-line profiles are shown in Figures \ref{fig:SIIIflux} and \ref{fig:H2flux}. A majority of the spectra with 
detected emission lines contained single-component lines for both [SIII] and H$_2$, which were fit with a single Gaussian, however regions near the nucleus 
contained two- and three-component [S~III] emission lines, either as individual peaked lines or a single peaked line with asymmetric wings, which were fit with 
multiple Gaussians.  

Fit parameters for both emission lines were used to calculate their observed velocity, FWHM, and integrated flux, mapped in 
Figures \ref{fig:SIIImaps} and \ref{fig:H2maps}. The Doppler shifted velocity for each emission-line component is given in the 
rest frame of the galaxy using rest wavelengths of 9533.2 \AA~and 21218.3\AA~for [S~III] and H$_2$ respectively. 

We found Z-band [S~III] emission lines to contain up to three components and sorted components between [S~III] maps by FWHM. Of the 
measurable line-component parameters, binning by FWHM is the most successful way to trace kinematics in specific knots of gas across 
several spaxels. Emission-line components were fit into narrow-, mid-, and broad-width component maps, with the narrowest and second-narrowest 
components always being placed in the narrow- and medium-width component maps. No thresholds were placed on widths when assigning specific
components to maps (i.e.~an emission line fit with only one component, such as knot A in Figure \ref{fig:SIIIflux}, would have that component placed in the 
narrow-component map regardless of width). Component blending is likely observed in regions near borders 
between fits with different numbers of components. Particularly, redshifted components observed in the narrow-component of the 
[S~III] measurements east of the nucleus are likely a combination of the two-component fits observed closer toward the 
nucleus, as a jump in line width is observed at the border between single and double component fits. As seen through the lower flux 
values for these measurements, lower signal-to-noise ratios of these single-component lines are the likely culprit in this discrepancy.

\begin{figure*}
\centering

\hspace{1.0cm}\includegraphics[width=0.9\textwidth]{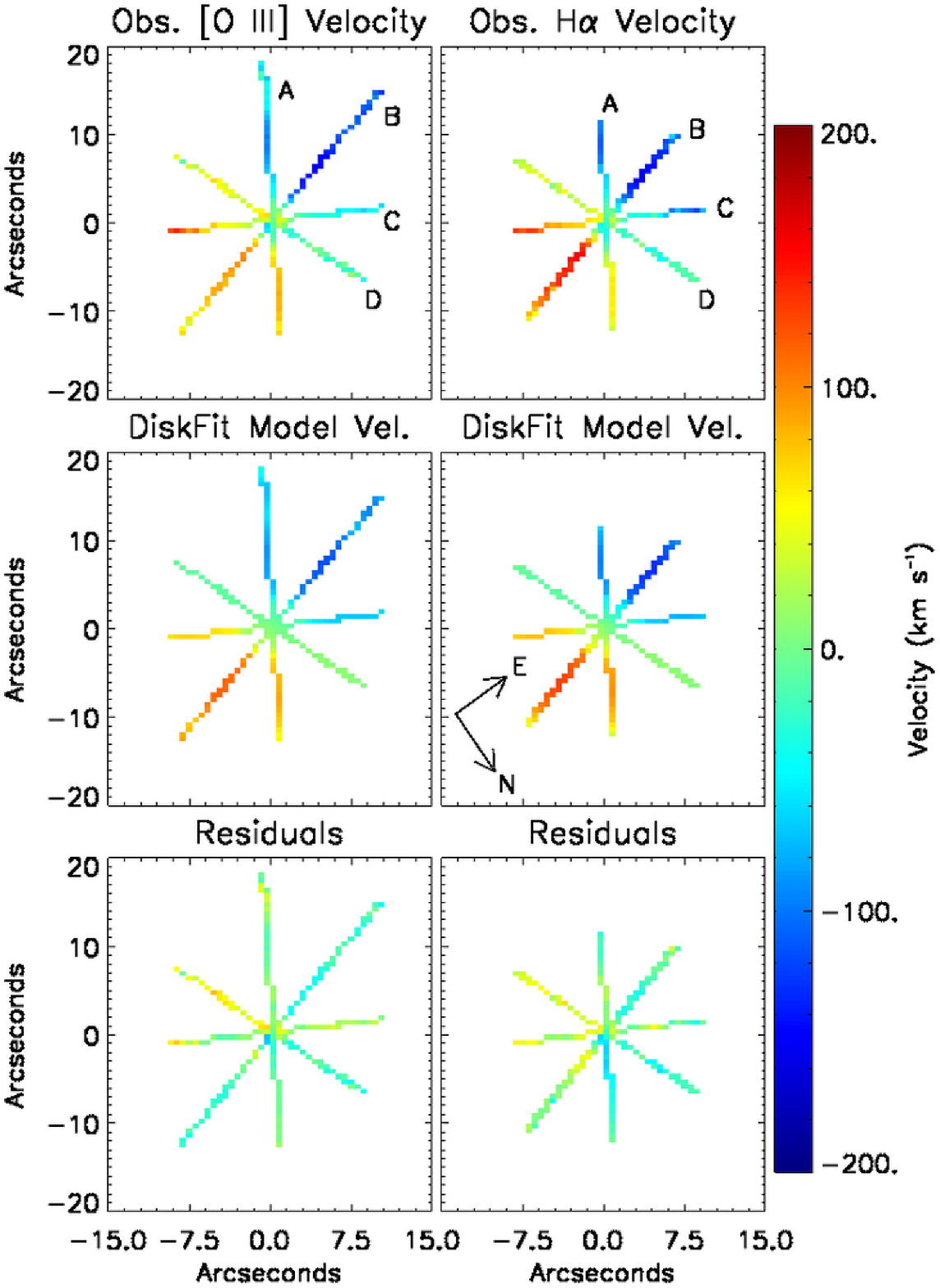}

\caption{Top: Pseudo-IFU centroid velocity map of [O~III] $\lambda$5007 and H$\alpha ~ \lambda$6563 kinematics in Mrk 573 generated from 
several long-slit observations from APO/DIS at varying position angles. Long-slits A, B, C, and D have position angles of 140$\degree$, 103$\degree$, 
55$\degree$, and 8$\degree$, respectively. Mid: DiskFit rotation models based on the observed kinematics. Bottom: Residuals between models and data.} 

\label{fig:bigmaps}
\end{figure*}

\begin{figure*}
\centering
\includegraphics[width=0.49\textwidth]{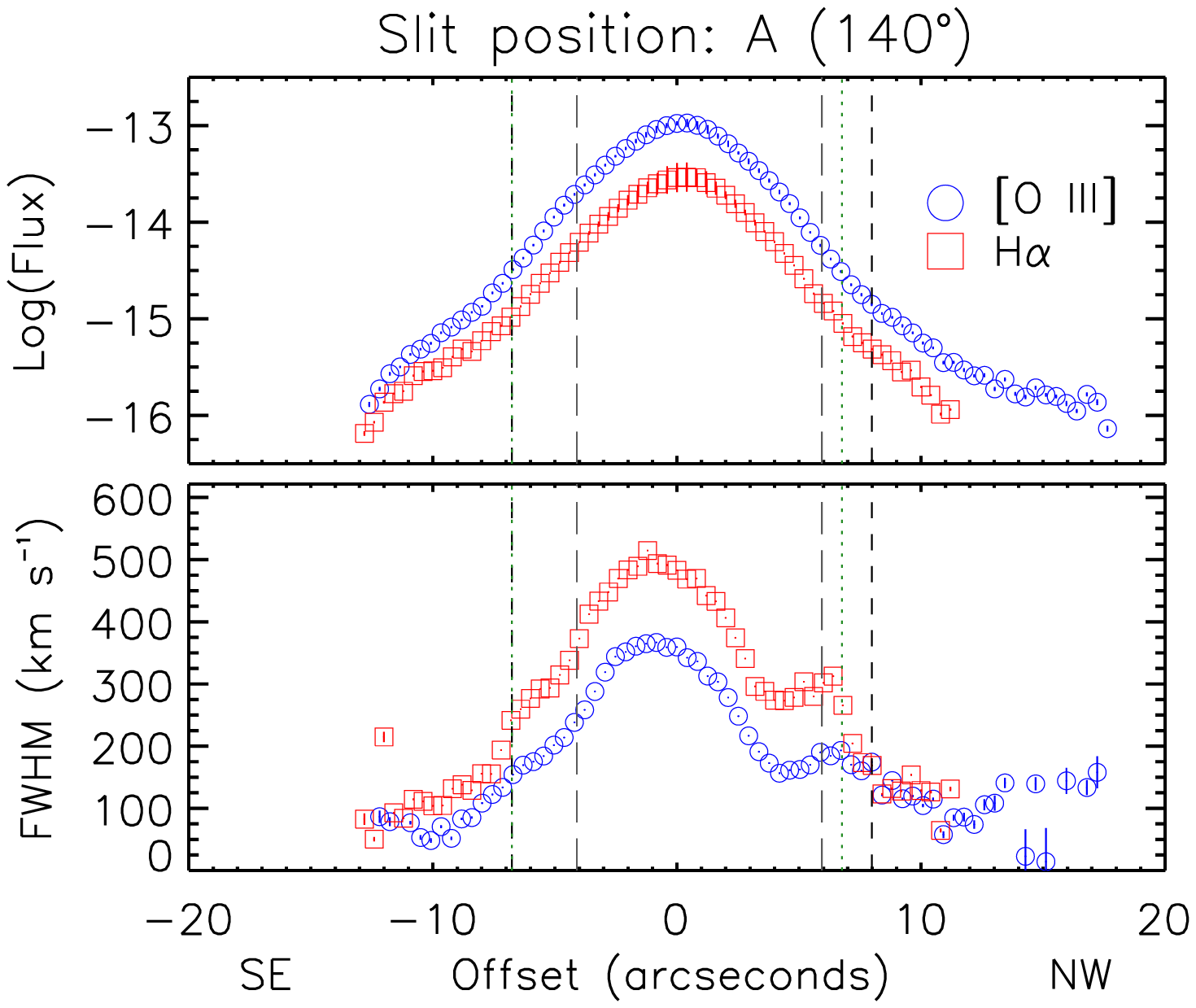}
\includegraphics[width=0.49\textwidth]{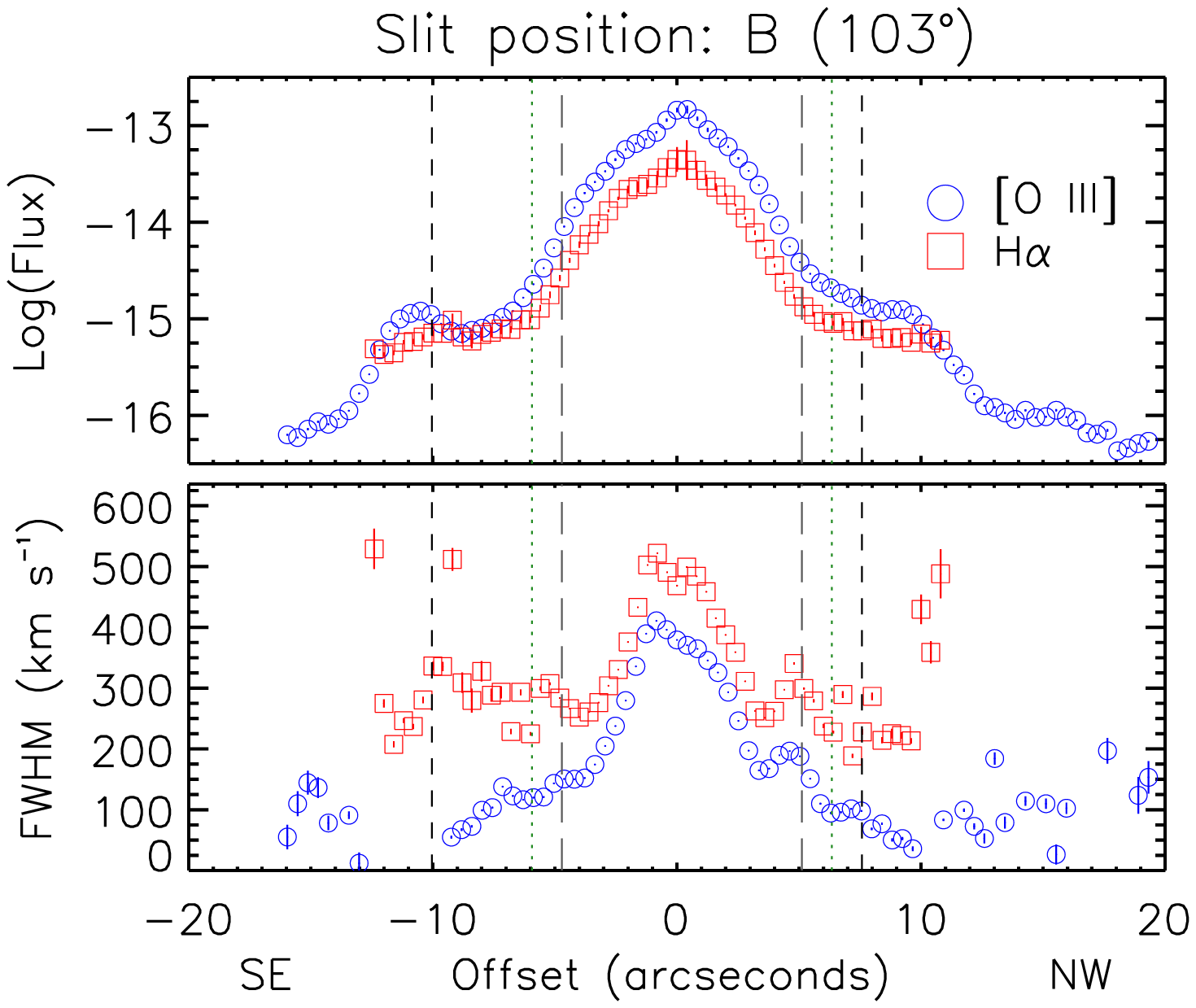}\\
\includegraphics[width=0.49\textwidth]{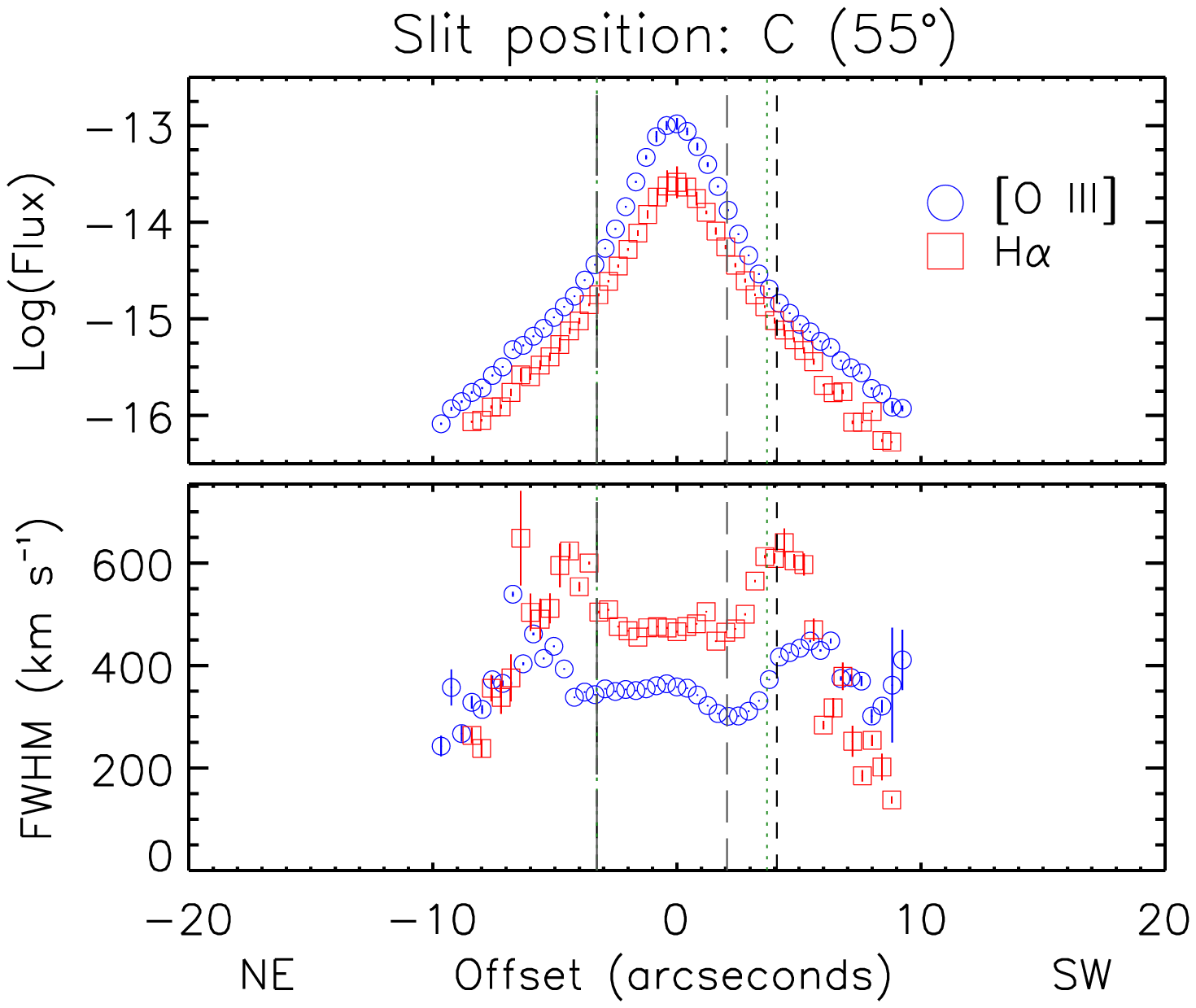}
\includegraphics[width=0.49\textwidth]{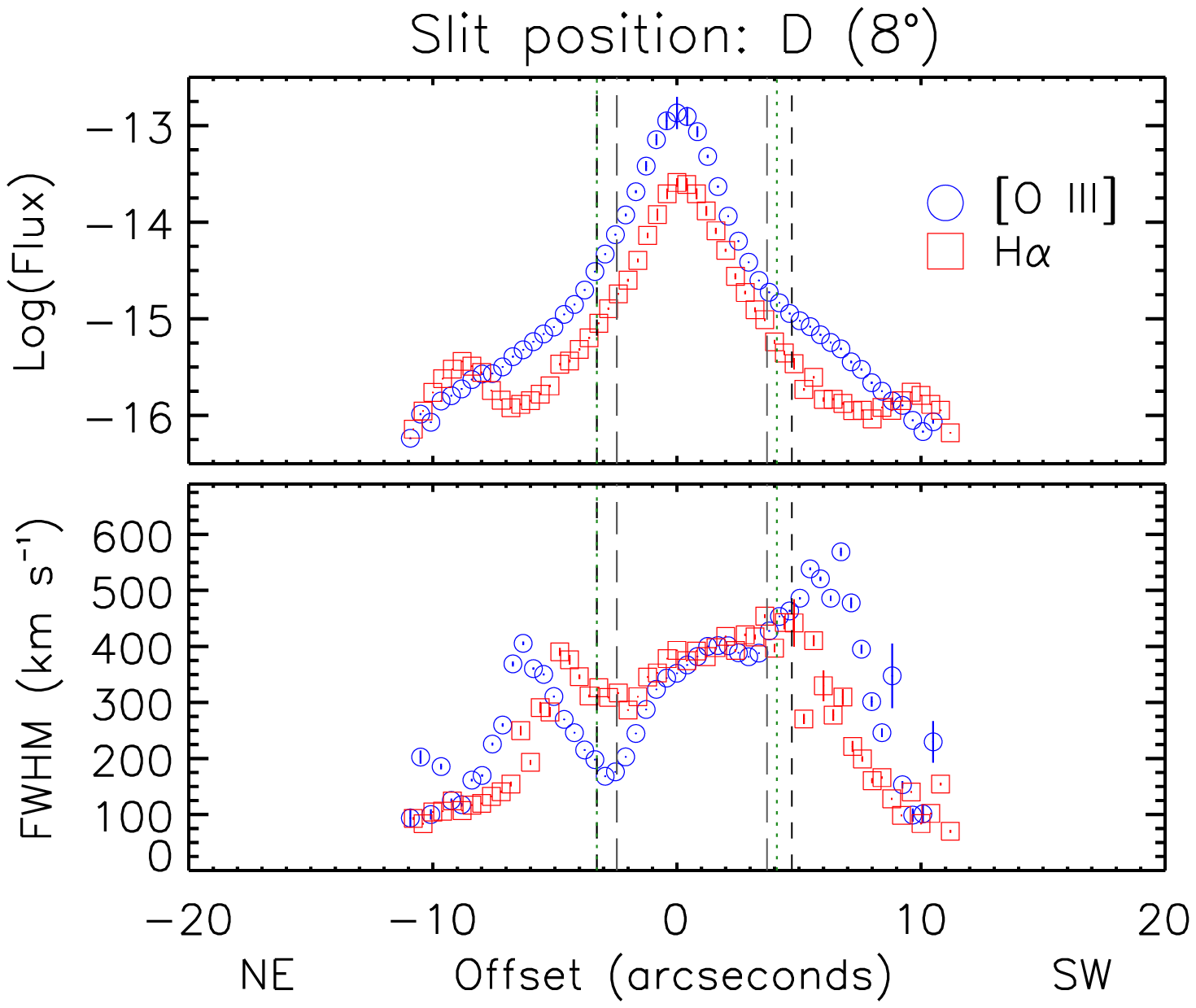}

\caption{Flux distributions and FWHM measurements for [O~III] $\lambda$5007 and H$\alpha$ emission-lines along each APO DIS long-slit 
observation. Maximum radial distances for [O~I] $\lambda$6300, [N~II] $\lambda$6584, and [S~II] $\lambda\lambda$ 6716,6731 emission-line 
measurements required for BPT diagnostics are plotted as gray, black, and green dashes, respectively. Positive offsets correspond 
to measurements located above the continuum centroid on the CCD, which is oriented at the position angle of the observation. } 

\label{fig:disfwhmflux}
\end{figure*}

\begin{figure*}
\centering

\includegraphics[width=0.49\textwidth]{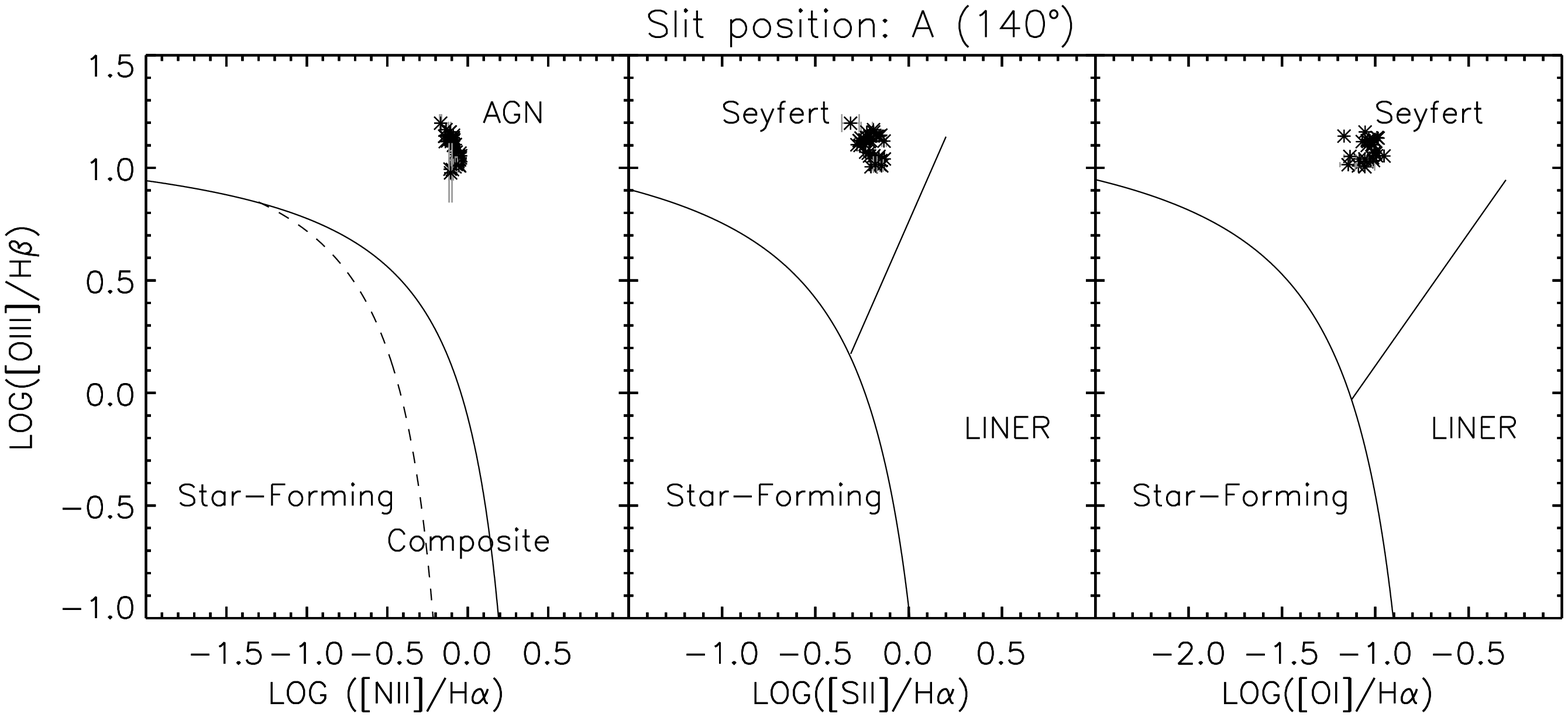}
\includegraphics[width=0.49\textwidth]{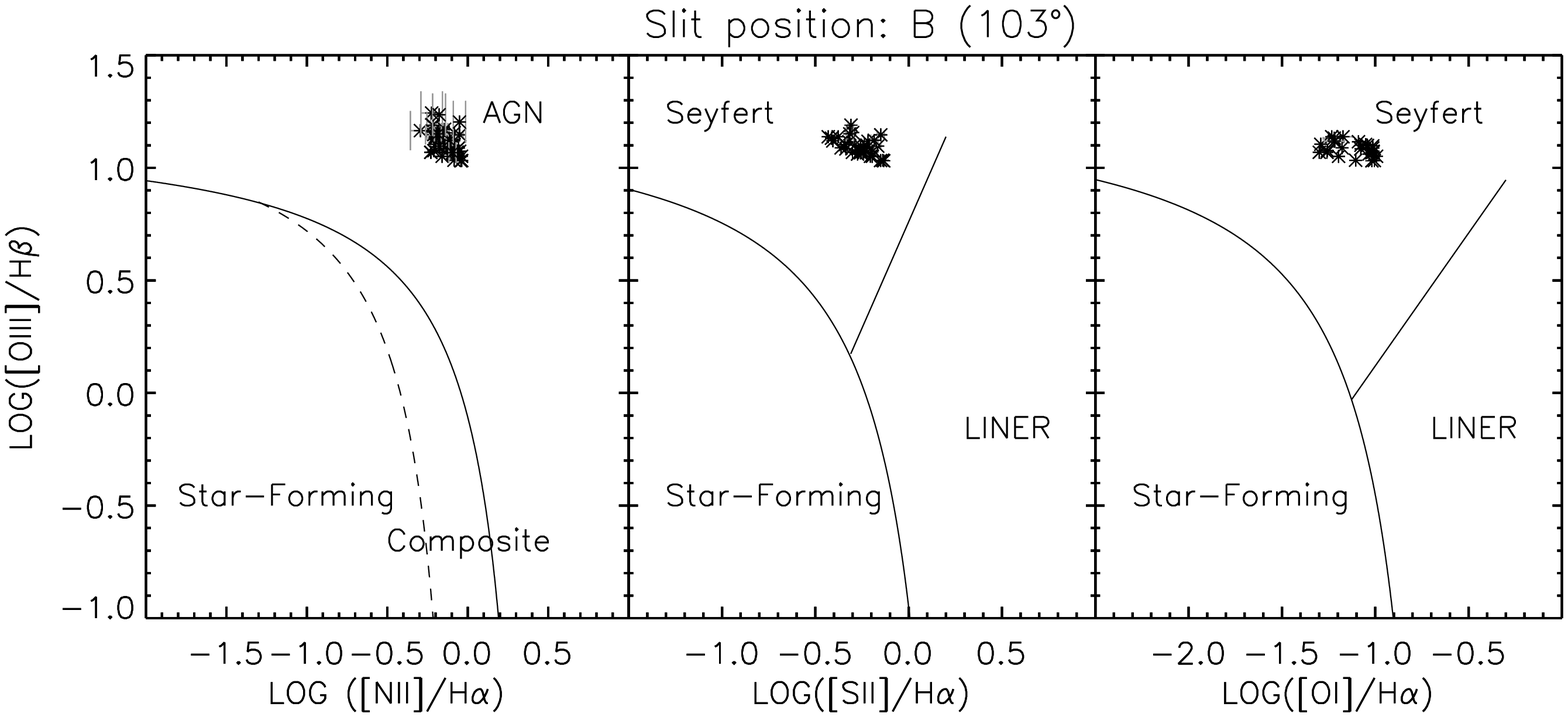}\\
\includegraphics[width=0.49\textwidth]{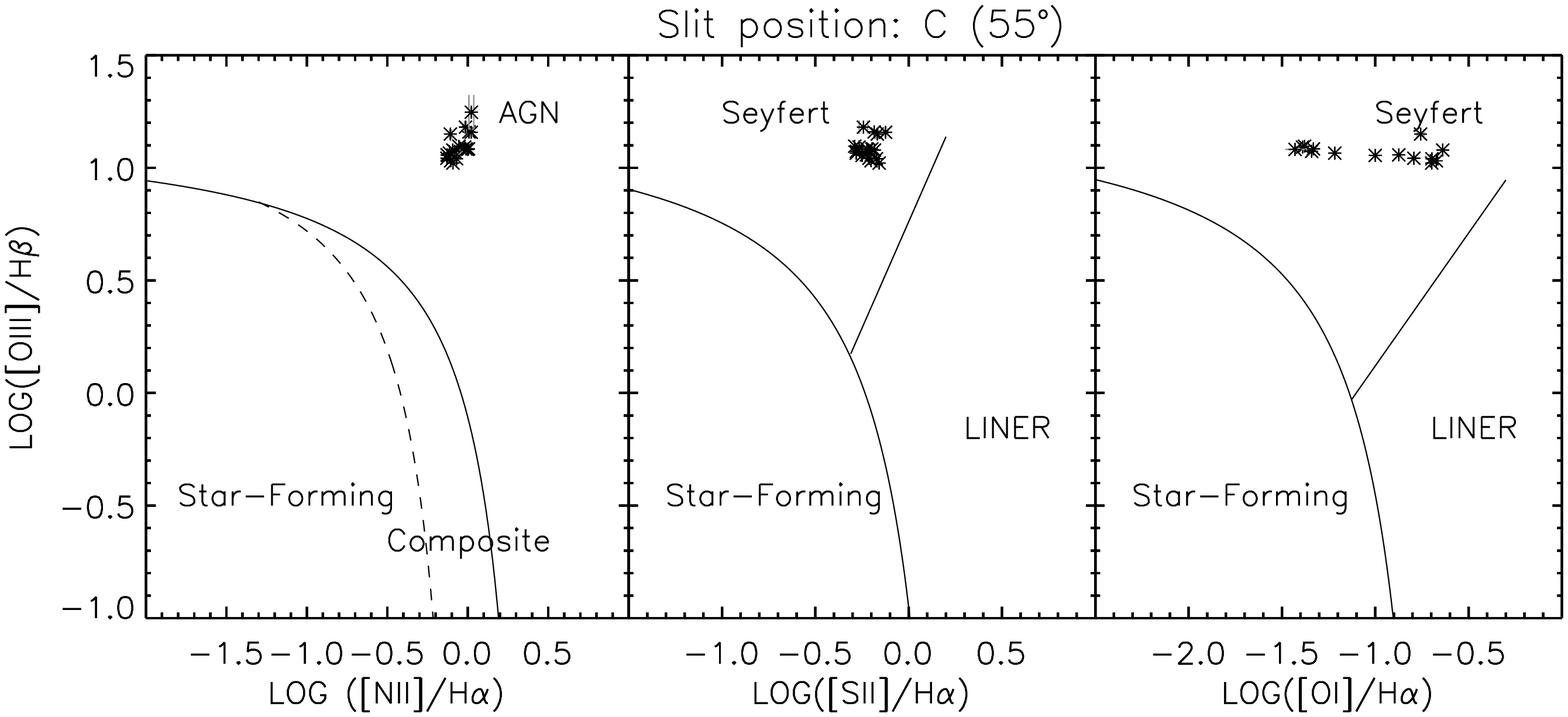}
\includegraphics[width=0.49\textwidth]{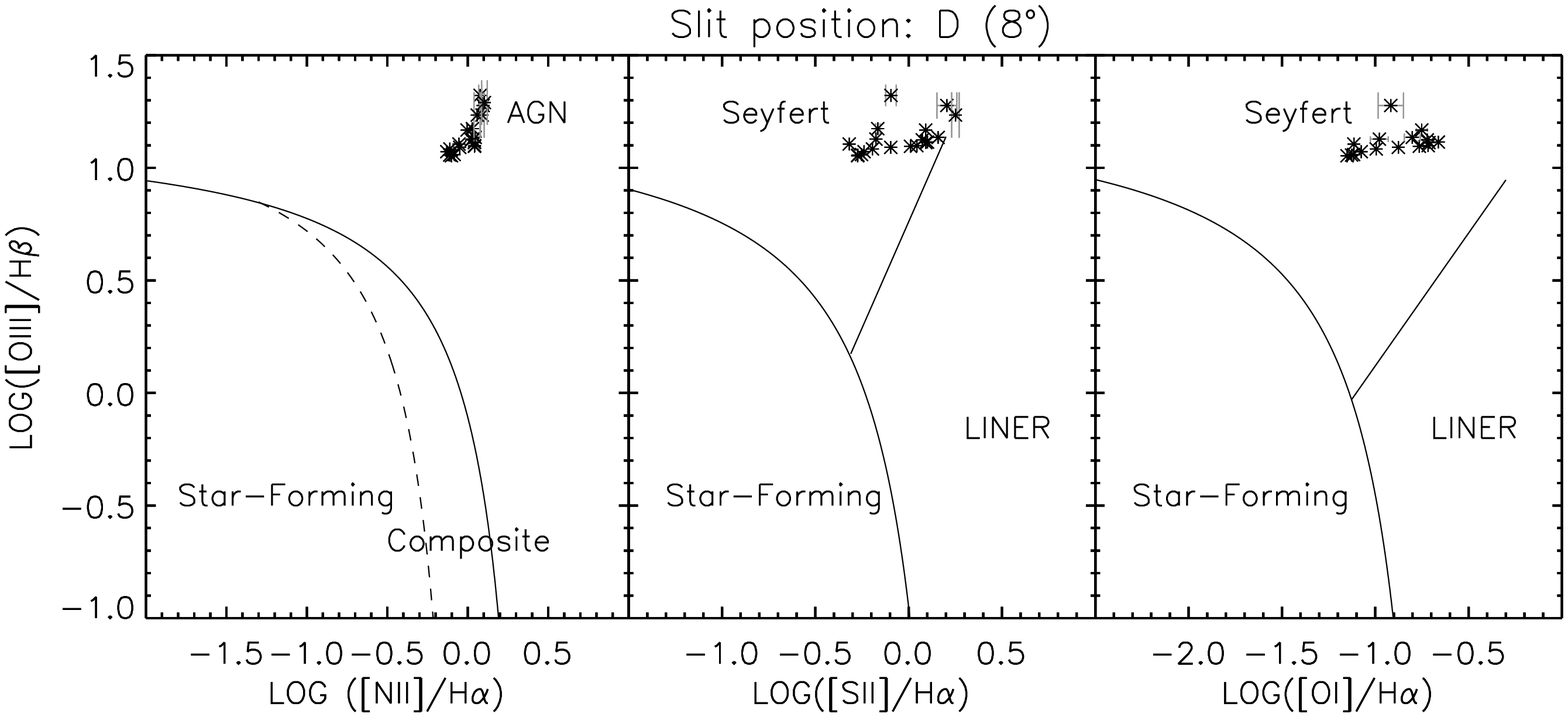}

\caption{BPT diagnostic diagrams for each long-slit APO DIS observation with measurable line fluxes, identifing the source of ionization in the 
observed gas as an AGN/Seyfert.} 

\label{fig:BPT}
\end{figure*}

Overlaying the flux map of the [S~III] emission from Figure \ref{fig:SIIIflux} onto the [S~III] kinematics, and comparing the morphology to 
what is seen in the larger structure map in Figure \ref{fig:structure}, the observed NLR kinematics can be largely credited to two types of structures.
The first structure is a near-linear feature extending from the nucleus to the southeast and northwest to radii $\sim$ 500 pc ($\sim$1.5$"$), along an 
approximate position angle of $\sim$131$^{\circ}$ and reaching peak radial velocities of $\sim 300$ km s$^{-1}$. This feature is not solid, but 
consists of several, aligned knots of emission. Kinematic measurements over the nucleus and bright knots adjacent to the nucleus reveal 
that the line emission in these regions is generally composed of a bright, narrow (200 km s$^{-1}$), near-systemic component and a 
fainter, broad (800 km s$^{-1}$), offset component, visible in profiles N and B-G in Figure \ref{fig:SIIIflux}. Three component fits to the 
[SIII] emission line are observed in a very small region 0.4$"$ ($\sim$140 pc) southeast of the nucleus (profile D in Figure \ref{fig:SIIIflux}), 
where a second, bright component exists at redshifted velocities $\sim 300$ km s$^{-1}$ with some components requiring a FWHM 
$\sim$2000 km s$^{-1}$. Along the linear set of knots, redshifted and blueshifted velocities peak at different distances from the nucleus, 
redshifted velocities in the southeast peak at approximately 0.7$"$ ($\sim$240 pc) and blueshifted velocities in the northwest peak at 
approximately 1.3$"$ ($\sim$450 pc), near emission-line knots C/D and G respectively. These amplitudes and directions of the radial velocities 
in this component agree with those from our {\it HST}/STIS long-slit measurements of [O~III] in the inner +/- 1$"$ of Mrk 573 (Paper I).

The remaining kinematics can be attributed to the arc features seen in Figure \ref{fig:structure}. Arcs to the northwest of the nucleus are 
largely outside of our field of view, with only the closest portion of the inner arc visible 1.6$"$ ($\sim$550 pc) west of the nucleus (profile 
H in Figure \ref{fig:SIIIflux}). Arcs to the southeast of the nucleus are largely contained within the combined FOV of the Z-band observations. 
The inner southeast arc, located at radii ranging between $\sim$ 1.6$"$ and 2.2$"$ (550 pc - 750 pc), contains both redshifted and 
blueshifted velocities. The central and southern portions of the inner arc have redshifted gas velocities of $\sim$100 km s$^{-1}$ 
(profile A in Figure \ref{fig:SIIIflux}), similar to the velocities in the adjacent linear feature. Ionized gas in the northern portion of the 
arc has blueshifted velocities also of $\sim$100 km s$^{-1}$ (profile I in Figure \ref{fig:SIIIflux}). Due to lower intensity, line emission 
detection in the outer arc (r $\sim$ 2.5$"$ - 3.3$"$, or $\sim$0.99 - 1.30 kpc) is less successful. Comparing the location of the arc in 
Figure \ref{fig:structure} with successful measurements in the NIFS FOV, the brightest, northern portion of the outer arc shows 
blueshifted velocities similar to the adjacent blueshifted knot. Emission-line measurements for all arc features consisted largely of 
single Gaussian measurements. The amplitudes and directions of the radial velocities in the arcs are similar to those found in the {\it HST} 
long-slit spectra at distances $>$ 1$"$ (Paper I).

Figure \ref{fig:SIIImaps} also incorporates archive radio observations from VLA, from which we can compare its morphology with the 
morphologies of the ionized gas and determine what effect the radio jet may have on the kinematics of the NLR gas. Knots of radio emission 
are aligned with the ionized linear feature observed in [SIII], however, their association to one another is unclear. Radio knots do not overlap 
with centers of emission-line clouds seen in the IR, and exist both radially interior and exterior to the peak redshifted velocities and are co-
located with peak blueshifted velocities. These findings are in agreement with similar studies of NGC 1068 and NGC 4151, where there was a lack 
of connection between radio jet and ionized gas flows \citep{Das05,Das06}. Therefore, as IR cloud positions and velocities are not correlated with the radio knots, 
we do not expect that the clouds in the linear feature are radially driven by the radio jet,  

K$_{l}$-band H$_2$ emission lines typically contain a single component, shown in Figure \ref{fig:H2maps}. Morphologically, the H$_2$ gas is 
different than what is observed in [S~III]. Initially, the molecular gas appears to be similar to a figure eight. Velocities on either side 
of the $``$eight$"$ are roughly symmetric, but opposite, with the southwest half being redshifted and the northeast half being blueshifted generally 
to velocities $\sim$100 km s$^{-1}$. Peak redshifted and blueshifted velocities $>$150 km s$^{-1}$ reside near the peak velocity positions 
observed in [S~III]. Additionally, the northern, blueshifted filament exhibits redshifted velocities near the continuum centroid of the system.

Comparing the morphologies and kinematics of the [S~III] and H$_2$ gas, we find that they are complimentary. Figure \ref{fig:compare} plots 
contours of the [S~III] and H$_{2}$ integrated flux over one another, from which we can compare the locations of the brightest knots in both 
gases. This comparison shows that the gasses are not co-located, but that the ionized [S~III] gas is located interior to the H$_{2}$ gas with 
respect to the NLR axis. Evidence for this is highlighted in four regions labeled in Figure \ref{fig:compare}: I) The west molecular lane has a divot where [SIII] knot resides. II) 
The north H$_{2}$ filament rakes out along the northwest [SIII] filament and ends where ionized-gas emission knots are present. III) the 
southeast linear section also has H$_{2}$ clouds at further radial distances than the [S~III] clouds. IV) The east molecular lane exists at a 
further radial distance than the adjacent [SIII] arc. Comparing kinematics between the ionized and molecular gasses, both near the arcs 
and the linear filament, velocities of both gasses are comparable where their projected positions overlap, such that velocities in the molecular 
gas begin to accelerate to higher velocities near the linear ionized gas feature. 

Figure \ref{fig:struc_flux} compares the combined ionized and molecular gas morphology to what is observed in the inner region of Figure 
\ref{fig:structure}. Here, we see that the ionized gas is located in the same knots of gas that emit in the optical, while molecular gas is co-
located with the inner dust lane morphology, with emission directly north / south of the AGN indicating a possible fueling flow to the AGN. 
Assuming a typical biconical NLR geometry, like the bicone projected onto the host disk in Paper I, these knots of [SIII] and H$_{2}$ gas appear to be 
interwoven near the edges of the NLR. This is consistent with our findings in Paper I, as the inner surfaces of spiral arms, which house the 
illuminated molecular gas, are likely becoming ionized when they are inside the illumination cone, forming the arc structures observed in imaging. This also 
suggests that the high velocity linear components of ionized gas do not originate from the nucleus, but are instead an ionized portion of rotating 
spiral arms that is accelerated in situ away from the nucleus as it enters the NLR at small radii. Additional examples of in situ acceleration have 
already been identified in this AGN, as {\it HST}/STIS observations in Paper I had previously shown velocity gradients to exist across the flux 
peaks of each ionized-gas arc in the NLR where the gas on the interior side of the arm is being ablated off the main structure. 

Combining the kinematics of both the ionized and molecular gas therefore creates a continuous story where the spatially resolved NLR is 
largely the biconical illumination of gas residing in the rotating host disk. As molecular-gas possessing spiral arms pass into the NLR, radiation 
flooding out from the central source illuminates and ionizes the radial interior of all structure within the volume of the bicone. In turn, this 
explanation suggests that a majority of the NLR gas originates in the pre-existing fueling flow and was ionized after the AGN turned on and is 
now being driven out by the AGN. 

\section{APO DIS: Extended Ionized Gas Kinematics}

While IFU observations allow for excellent analysis of the few, inner arcseconds of Mrk 573, we are also able to observe kinematics of 
ionized gas in Mrk 573 at larger radial distances using the long-slit observations obtained with APO DIS. For each of the four slit positions, 
we employed a similar line-fitting procedure to the routine discussed in Section 3 to fit Gaussians to [O III] $\lambda$5007 and H$\alpha$, in 
the blue and red images respectively, to characterize the kinematics of the ionized gas out to distances $>$3.5 kpc. Resultant velocities from our 
measurements are shown in Figure \ref{fig:bigmaps}, with flux and FWHM parameters along each slit plotted in Figure \ref{fig:disfwhmflux}. 
Gaussian profiles were also fit to H$\beta$, [O~I] $\lambda$6300, [N~II] $\lambda$6584, and [S~II] $\lambda\lambda$ 6716,6731 to distinguish the 
ionization mechanism of the gas throughout our observations. The extent of flux measurements for each emission-line is shown in Figure \ref{fig:disfwhmflux}. 
Narrow-line ratio diagnostics, i.e. BPT diagrams  \citep{Bal81,Vei87,Kew06}, in Figure \ref{fig:BPT} compare [OIII]/H$\beta$ ratios to [NII]/H$\alpha$, [SII]/H$\alpha$, 
and [OI]/H$\alpha$ ratios, which illustrate that Seyfert/AGN ionization dominates out to distances $>$ 2 kpc in most directions. 

Ionized gas kinematics in these datasets appear to be largely rotational, with the kinematic major axis being sampled in observations along 
slit B (PA $= 103\degree$) in Figure \ref{fig:bigmaps}. Velocities within a radius of 1.5$"$ do not follow the rotation pattern, which is to be expected as this gas 
experiences high velocity outflows as observed in the NIFS data. Bumps in H$\alpha$ emission at 10$"$ in Slit D (PA $= 8\degree$) are due to star formation from 
spiral arms coming off of a large stellar bar, as observed by \citet{Pog95,Afa96}.

\section{Comparison With Rotation}

\begin{figure*}
\centering

\includegraphics[width=0.45\textwidth]{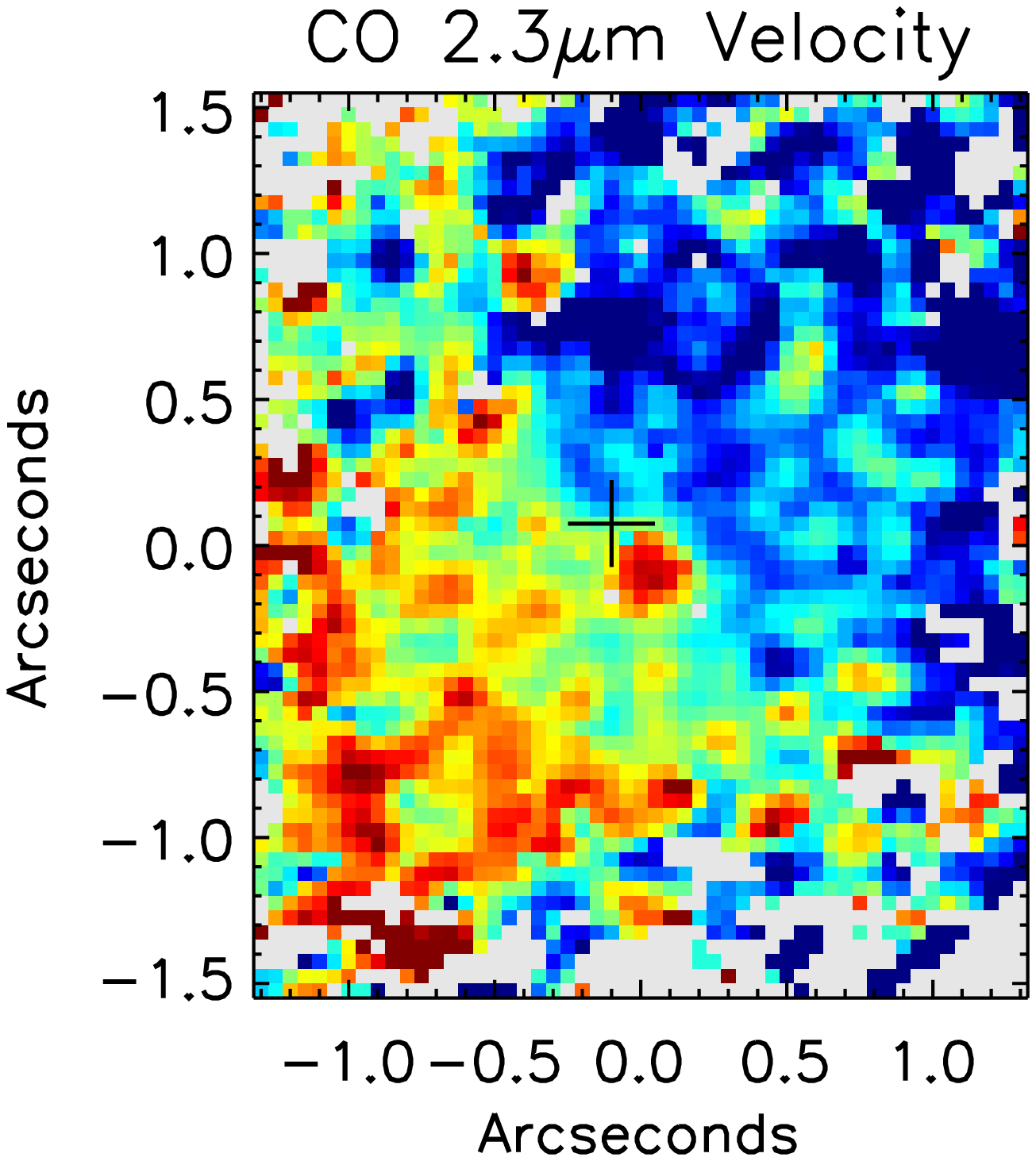} \hspace{-3.5cm}
\includegraphics[width=0.45\textwidth]{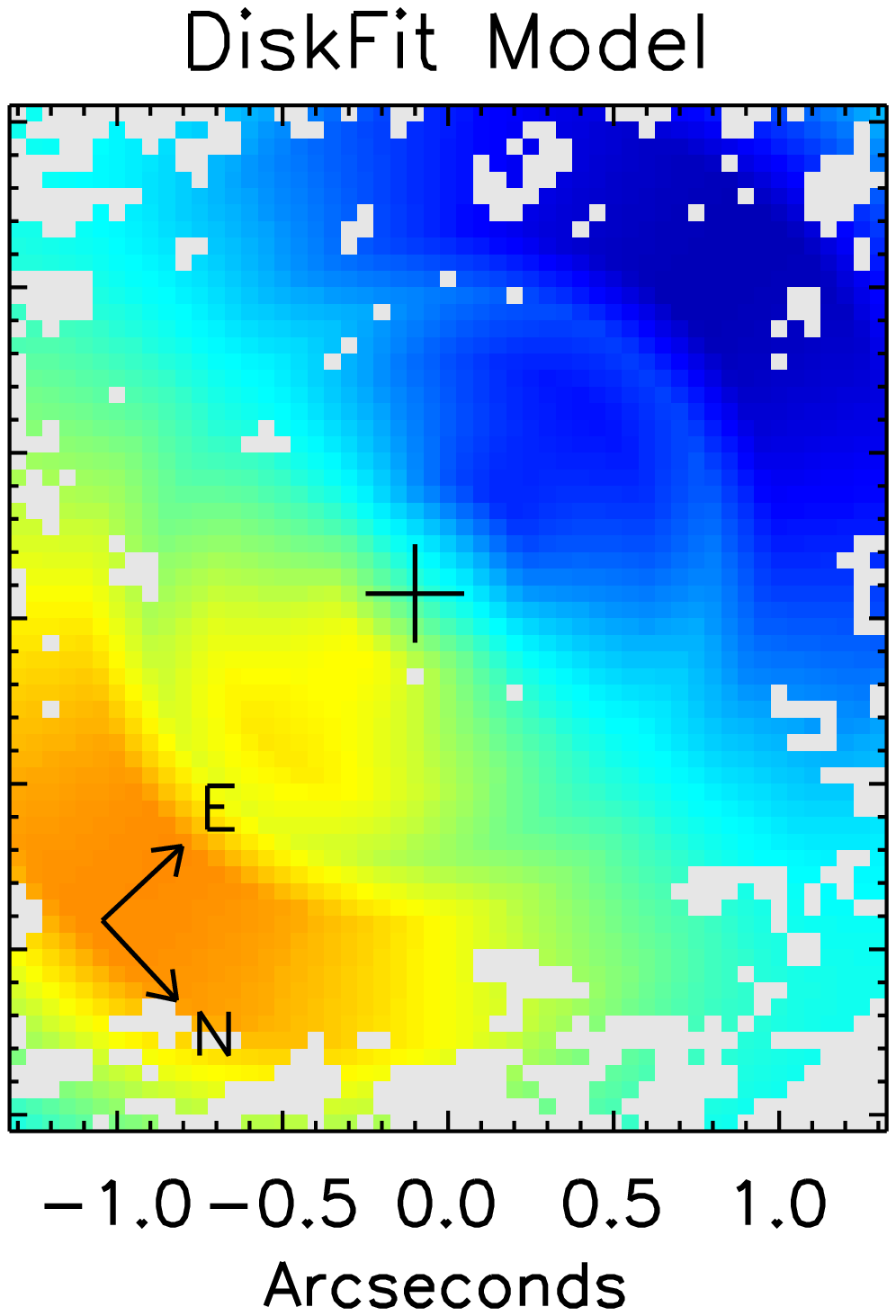}\hspace{-3.5cm}
\includegraphics[width=0.45\textwidth]{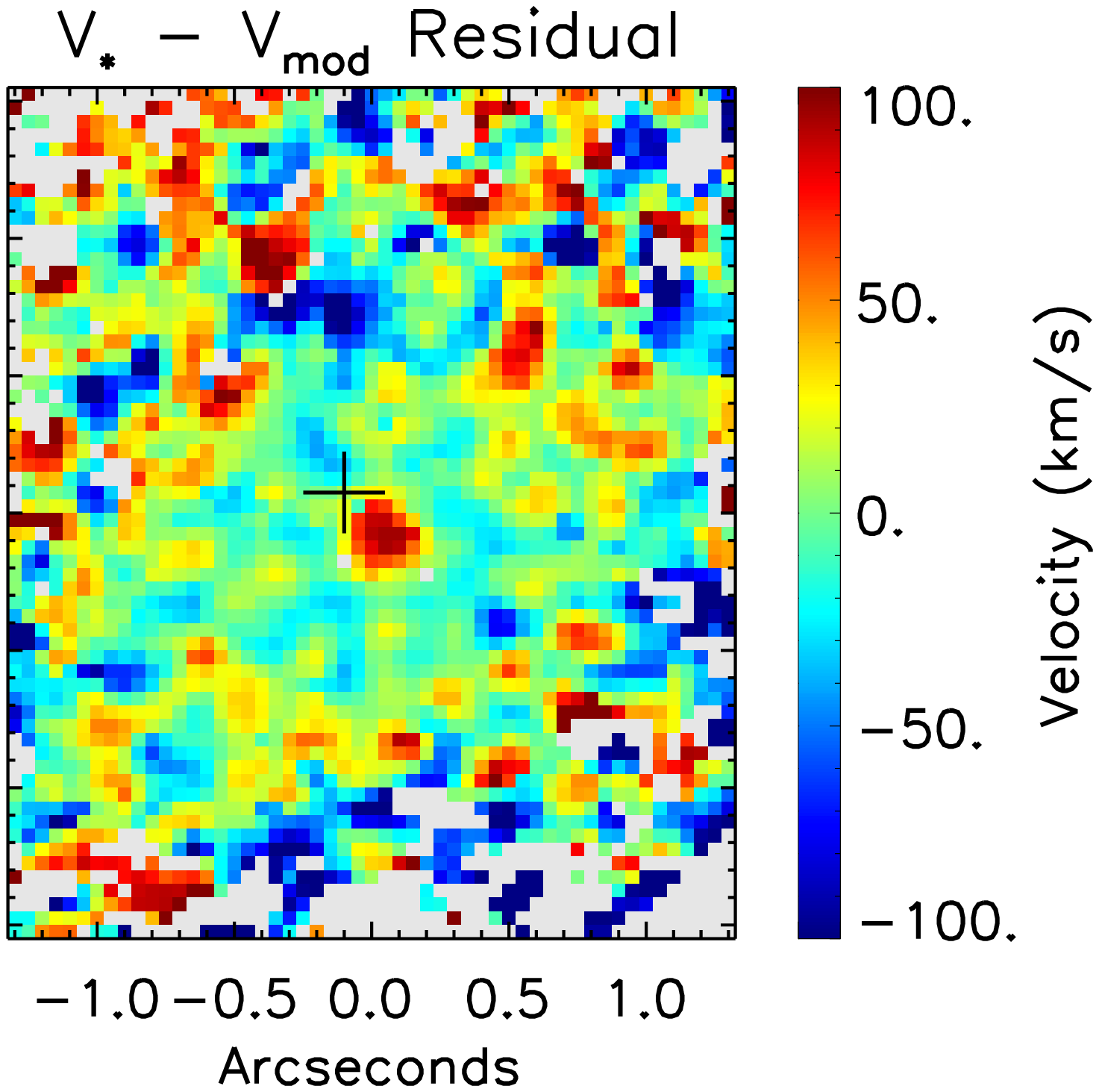}\\
\includegraphics[width=0.45\textwidth]{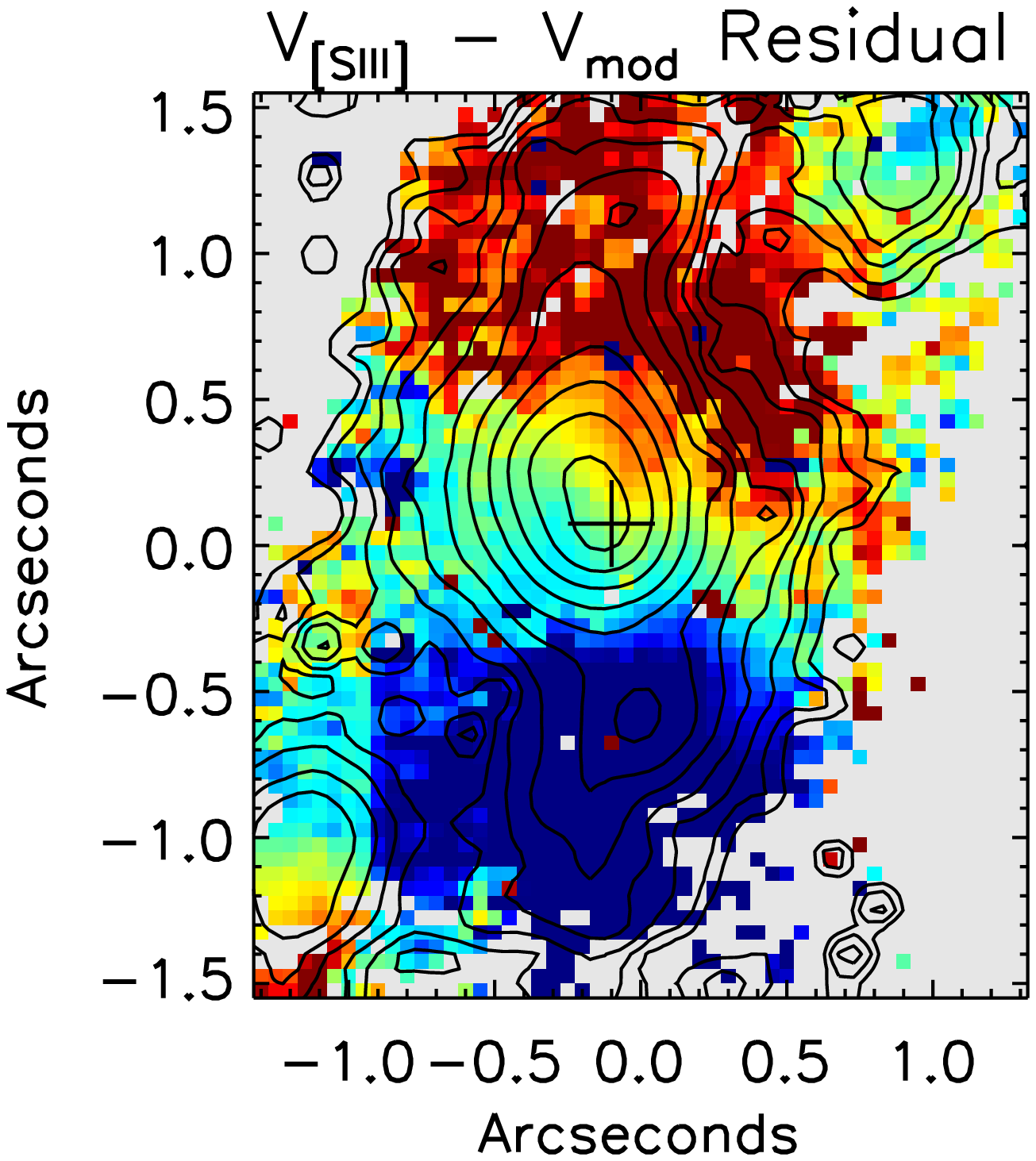} \hspace{-3.5cm}
\includegraphics[width=0.45\textwidth]{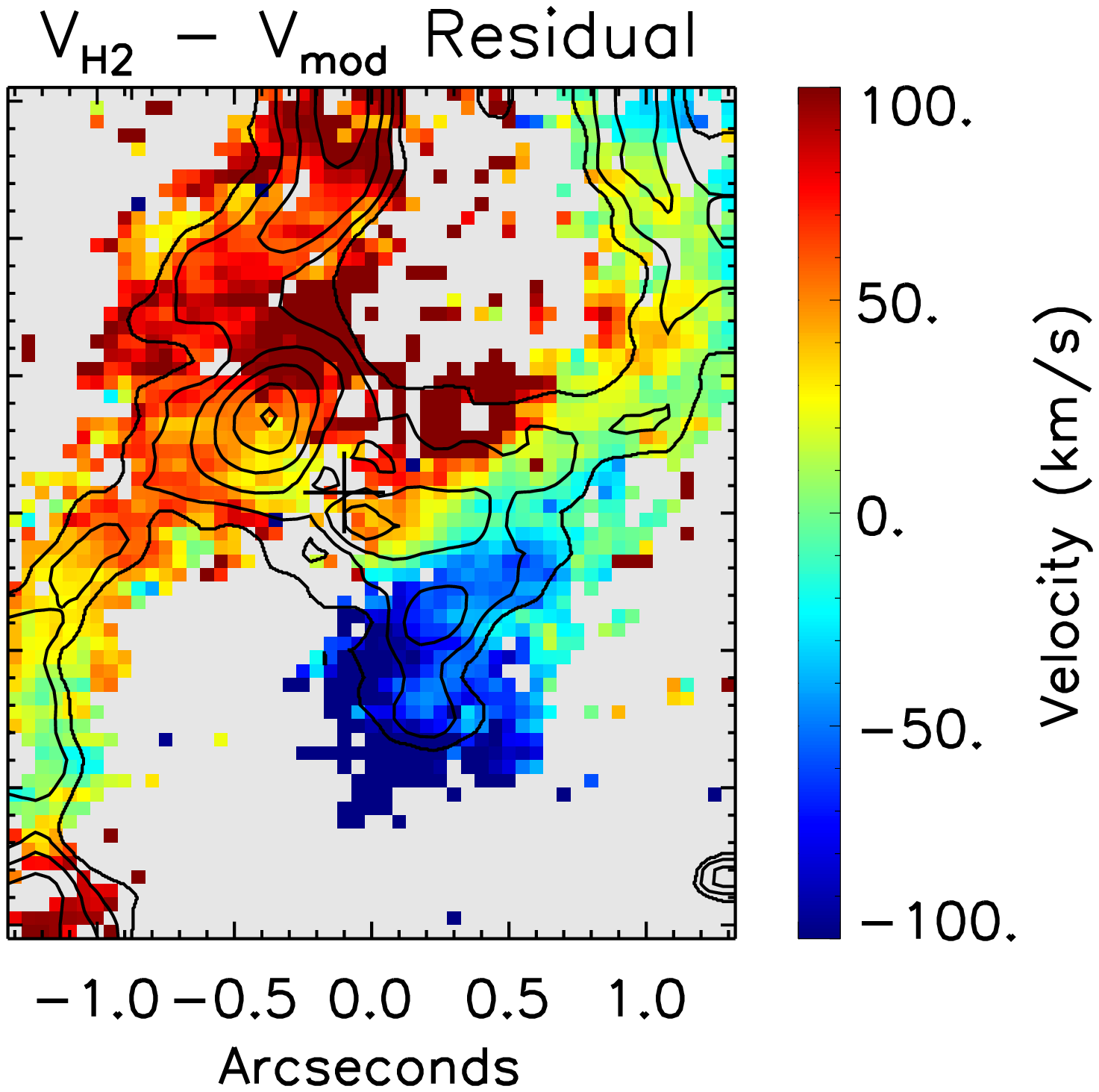} \hspace{-1.25cm}\\

\caption{Top Left: Observed stellar velocity field. Top Center: DiskFit rotating disk model derived from stellar kinematics. Top Right: Residual map 
of model stellar kinematics subtracted from observed stellar kinematics. Bottom Left: Residual map of model stellar kinematics subtracted from 
extended, narrow-component [S~III] kinematics. Bottom Right: Residual map of modeled stellar kinematics subtracted from H$_2$ kinematics. 
Continuum centroid is depicted by a cross.} 

\label{fig:stellar}
\end{figure*}

In order to determine where the observed ionized and molecular gas kinematics are affiliated with rotation, we can compare their 
velocities to stellar kinematics in the host disk. We measured stellar kinematics in the NIFS data using the penalized pixel-fitting (PPXF) method of 
Cappellari \& Emsellem (2004). We fit $^{12}$CO$_{\lambda2.29}$, $^{12}$CO$_{\lambda2.32}$, and $^{13}$CO$_{\lambda2.34} ~ \mu$m 
stellar absorption lines within the K-band FOV following the procedure described in \citet{Rif08}. Stellar templates of 60 early-type 
stars \citep{Wing09} were used to obtain the stellar line-of-sight velocity distribution at each position. The observed stellar velocities are 
shown in the left panel of Figure \ref{fig:stellar}. White regions in the velocity map correspond to positions where the signal-to-noise ratio 
in the CO bands was not high enough to allow good fits. Several spaxels near the nucleus of the galaxy contain spectra that could not be 
properly fit due to the dilution of the CO absorptions by non-stellar continuum emission. Kinematics depict a rotation pattern with blueshifts 
to the east and redshifts to the west of the nucleus, a maximum velocity of $\sim$200 km s$^{-1}$, and a major axis along the east/west 
direction. 

To characterize the rotation of the stellar kinematics within the host disk, we used DiskFit \citep{Spe07,Sel10,Kuz12}, a publicly available code 
that fits non-parametric models to a given velocity field. We applied the rotation model to the NIFS stellar kinematics using initial host major axis position 
angle and ellipticity parameters based on isophote measurements from previous I-band photometric analysis \citep{Sch00}. Within a radius of 
5$"$, which contains the entirety of our NIFS K-band FOV, the position angle and average ellipticity ($e = 1-b/a$) of the system are 
approximately 97$\degree$ and 0.1 respectively. The resultant model and residuals are shown in Figure \ref{fig:stellar}. We note that despite 
significant localized residuals due to large uncertainties in the stellar absorption measurements, the rotation model provides a reasonable global 
fit to the stellar kinematics.

We are able to analyze the source of the gas kinematics in the NIFS nuclear FOV by comparing them to our stellar rotation model. The 
bottom panels of Figure \ref{fig:stellar} map the residual velocity difference between the [SIII] / H$_2$ gas and stellar kinematics. 
Ionized gas arcs and molecular gas lanes east and west of the nucleus largely agree with rotation, as the residuals between gas and stars 
in these locations are $<$ 25 km s$^{-1}$. Molecular gas is in rotation until it reaches close proximity to the ionization cones emitted from the 
central engine, where it exhibits kinematics similar to the ionized [SIII] gas in the linear filament. Here, both gasses have velocities largely 
offset from rotation, traveling outward from the nucleus. The northeast molecular hydrogen gas depicts this prominently, with gas being driven 
away in the redshifted direction before it approaches the nucleus and is driven away in the blueshifted direction after it passes the nucleus. 
Therefore, comparing our measurements of stellar and ionized/molecular gas kinematics of the inner, nuclear region, we find that the observed 
gas follows a rotation pattern until it enters the NLR. 

\begin{figure}
\centering

\includegraphics[width=0.49\textwidth]{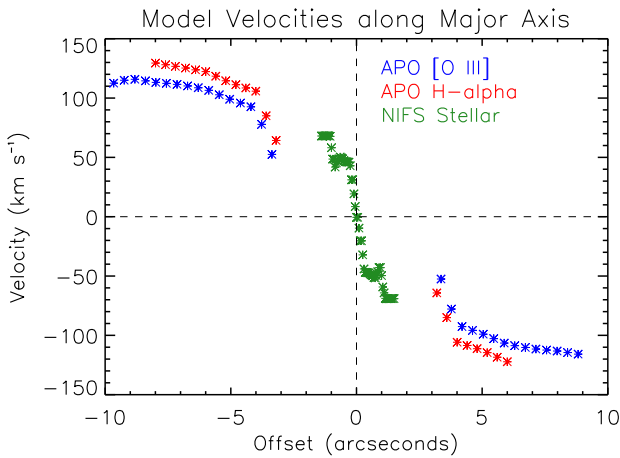}

\caption{Comparison of rotation model radial velocities near the major axis of Mrk 573 (DIS Slit position B). DIS and NIFS observations are cropped at offsets $< 3.0"$ and $> 1.5"$ respectively.} 

\label{fig:rot_compare}
\end{figure}

We can also compare the stellar kinematics observed with NIFS to the large-scale ionized gas kinematics in the host galaxy through 
measurements obtained from our APO DIS observations. We again used DiskFit to characterize the rotation of the extended gas kinematics 
within the host disk, applying a rotation model to the DIS [O~III] and H$\alpha$ kinematics. As DIS observations extend to radii greater than 
5$"$, we employ host parameters determined via isophote fits of the greater host disk morphology as observed in the SDSS image in Figure 
\ref{fig:structure}. Using the ellipse IRAF task, the position angle and ellipticity for the outer disk were measured as 92$\degree$ and 0.275, respectively, 
similar to those found from the inner disk in the NIFS data. From these large-scale, long-slit measurements we find that the kinematics at radii $>$ 5$"$ 
again largely follow a rotation pattern, even in regions that are aligned along the NLR axis, which provides evidence that gas can be illuminated inside 
the NLR at large radii which does not exhibit outflow kinematics. We see that the large radius ionized gas kinematics largely agree with the stellar 
kinematics at smaller radii by comparing velocities obtained from our stellar and ionized gas kinematic models, as shown in Figure \ref{fig:rot_compare}. 
In this figure, we plot the kinematics of DIS ionized gas outside 3$"$ to avoid kinematic contamination from outflows, and the NIFS stellar kinematics 
inside 1.5$"$ to avoid low S/N measurements near the edge of the FOV, along the position angle of Slit B. We find both datasets to be in general 
agreement and representative of a typical disk rotation curve, supporting the notion that rotation dominates any ionized gas kinematics at distances 
greater than 5$"$ ($\sim$1.75 kpc).

With the additional information gained from the NIFS and DIS spectra, it is clear that a majority of the observed kinematics do in fact agree 
with a rotation model. We find that the maximum observed velocity in the large-
scale APO observations to be $\sim$130 km s$^{-1}$. With the outer host disk having an inclination of $\sim43\degree$, determined via 
our ellipse fitting, deprojecting the observed maximum velocity would result in a rotational velocity $\sim$190 km s$^{-1}$. The remaining puzzle lies in 
explaining the deviations from the rotation curve, observed as the high velocity linear feature and, to a lesser extent, the velocity gradients 
observed across the emission-line arcs noted in Paper I. Assuming that these deviations are correlated, an immediate guess to the culprit 
would be radiative driving, as the interaction between AGN radiation and pre-existing host material has already been made evident. 


\section{Comparison with Radiative Driving}

Here, we compare the radiative acceleration to the gravitational deceleration experienced by the gas at a given 
radius to determine if radiative driving could produce the observed non-rotating kinematics inside the NLR at the observed radii. 

\subsection{Gravitational Deceleration}

\begin{figure*}
\centering

\includegraphics[width=0.99\textwidth]{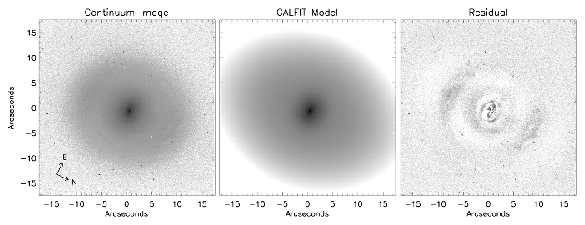} 

\caption{{\it Left:} {\it HST} WFPC2/PC F814W continuum image of Mrk 573. {\it Center:} Best fit galaxy decomposition model (3 components) for Mrk 573. {\it Right:} Residuals between image and model.}

\label{fig:galfit}
\end{figure*}

\begin{figure}
\centering

\includegraphics[width=0.45\textwidth]{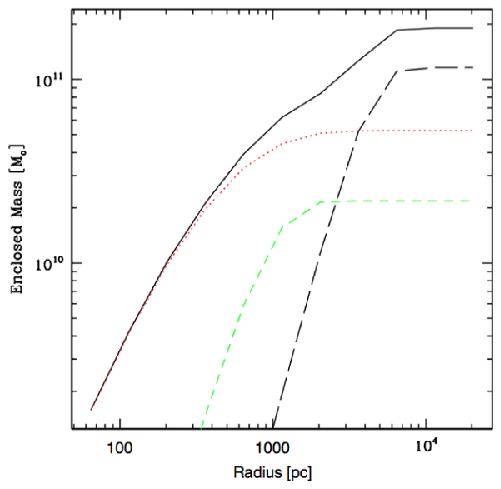} 

\caption{Mass distribution profiles for each component in our model. Red, green, black dashed, and black solid lines represent inner, intermediate, and outer components and the sum of the 3 components, respectively. Our radial mass distribution is calculated using the expressions from Terzi{\'c} \& Graham (2005) assuming a mass-to-light ratio of 5.}

\label{fig:mass}
\end{figure}

Gravitational deceleration was measured by determining the enclosed mass as a function of radius. Using a method similar to the one used 
by \citet{Das07} for NGC 1068, {\it HST} WFPC2/PC F814W imaging of this galaxy was decomposed using GALFIT version 
3.0.5 \citep{Pen02,Pen10} in order to measure enclosed mass of Mrk 573. We find that the best fitting model is composed of three S\'ersic 
components with parameters described in Table \ref{tab:galfit}. The original image, GALFIT model, and resulting residual map, 
are presented in Figure \ref{fig:galfit}. Component 1, the innermost one, can be identified as a disk, which is coplanar with the outer disk and ring such that 
position angle and inclination properties agree with our previous measurements and has a morphology similar to that described 
by \citet{Mar01,Lai02}. Component 2 has a very small deviation from circular symmetry,which is consistent with a bulge, but given the 
S\'ersic index n=0.58 it is better classified as a pseudobulge \citep{Fis08}. Component 3 is elongated similar to component 1, due to illuminated 
spiral arms north and south of the nucleus, and classified as a bar by \citet{Lai02}.

The radial mass distribution of these three components (Figure \ref{fig:mass}) was calculated using the expressions from Terzi\'c \& Graham (2005). 
The S\'ersic profile is given by the following expression \citep{Pen10}:

\begin{table}[h!]
  
  \caption{GalFit Model Results}
  \begin{threeparttable}
  \centering
  \label{tab:galfit}
  \begin{tabular}{ccccccc}
    \toprule
	Comp. & I (mag) & R$_e$ (pc) & n    & b/a  &  PA (deg) & f  \\	
    \midrule
	1         & 14.29   &  365       & 1.79 & 0.64 & 94.7     & 0.21  \\
	2         & 14.26   &  670       & 0.58 & 0.97 & 88.0     & 0.20  \\
	3         & 13.13   & 2880       & 0.39 & 0.78 &  1.7     & 0.59  \\
    \bottomrule
  
  \end{tabular}
  
	\begin{tablenotes}
	\item Col. (1) indicates the S\'ersic component; Col. (2) gives the integrated I band
	magnitude; Col. (3) Effective radius; Col. (4) gives the S\'ersic index; 
	Cols. (5) and (6) give the axial ratio and position angle of the component;
	Col. (7) gives the fraction of the integrated flux from each component. 
	\end{tablenotes}
	\end{threeparttable}
\end{table}

\begin{equation}
\Sigma(r) = \Sigma_e exp [ -\kappa ((\frac{r}{r_e})^{1/n} - 1)]
\end{equation}

where $\Sigma(r)$ is the surface brightness, $\Sigma_e$ is the surface brightness at the effective radius, $\kappa$ is a constant that 
depends on $n$, the index of the profile, and $r_e$ is the effective radius.

The value $\Sigma_e$ is calculated using the equation:
\begin{equation}
F_{tot} = 2 \pi  r_e^2  \Sigma_e e^\kappa n \kappa^{-2n} \Gamma(2n) q/R(C_0,m)
\end{equation}

where $\Gamma$ is the gamma function, $q=b/a$ is the axial ratio of the S\'ersic component and $R(C_0,m)$ represents deviations from 
a perfect ellipse \citep{Pen10}. This term has a value of the order of unity and will be disregarded in our calculations.

Following Equation~4 in \citet{Ter05} we have that the mass density of a S\'ersic component is given by the following expressions:

\begin{equation}
\rho(r) = \rho_0 (\frac{r}{r_e})^{-p} e^{\kappa} e^{(-\kappa (\frac{r}{r_e})^{1/n})}
\end{equation}

\begin{equation}
p = 1 - \frac{0.6097}{n} + \frac{0.05563}{n^2}
\end{equation}

\begin{equation}
\rho_0 = \frac{M}{L} \Sigma_e \kappa^{n(1-p)}  \frac{\Gamma(2n)}{(2 r_e \Gamma(n(3-p))} 
\end{equation}

where $\frac{M}{L}$ is the mass to light ratio, assumed to be 5. Notice that due to a difference in notation between \citet{Pen10} and 
\citet{Ter05}, the expression for $\rho(r)$ has an addition $e^{\kappa}$ term. Also, the expression for $p$ corresponds to $n$ values in the 
range $0.6<n<10$. Two of our components are slightly outside this range, but we do not expect a large deviation in the results by using this 
expression.

Finally, we have from Equation~A2 in \citet{Ter05} that the mass profile is given by:

\begin{equation}
M(r) = 4 \pi \rho_0 r_e^3 n \kappa^{n(p-3)} \gamma(n(3-p),Z)
\end{equation}

where $\gamma(n(3-p),Z)$ is the incomplete gamma function and $Z$ is given by $Z = \kappa (\frac{r}{r_e})^{1/n} $. Using 
Equation 10 of \citet{Ter05}, we can calculate the enclosed mass at a given radius and thus determine the gravitational deceleration at said radius. 
This is critical in determining where gas can be radiatively accelerated, as shown below.

\subsection{Radiative Acceleration}

\begin{figure}
\centering

\includegraphics[width=0.35\textwidth,angle=90]{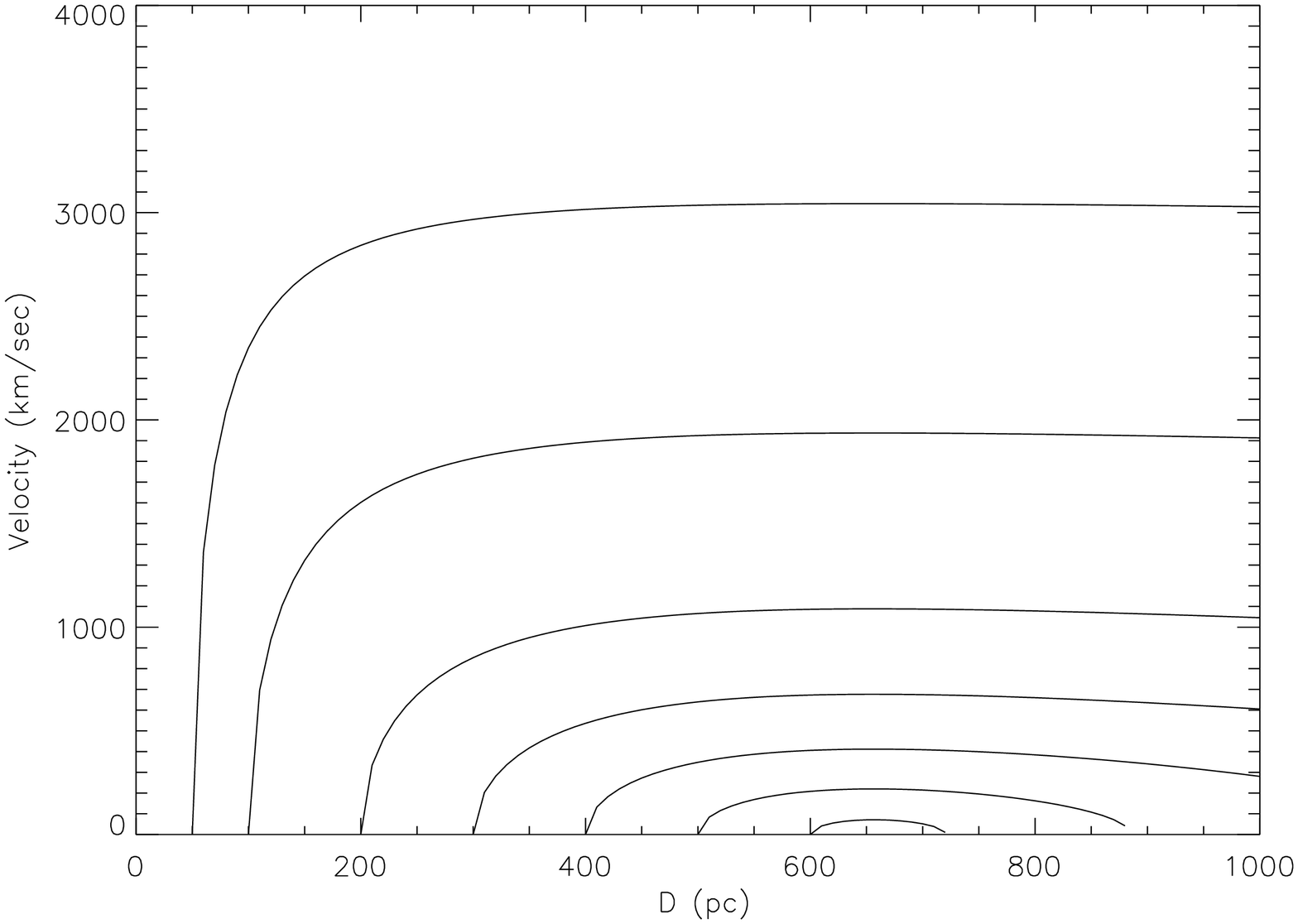}\\
\includegraphics[width=0.35\textwidth,angle=90]{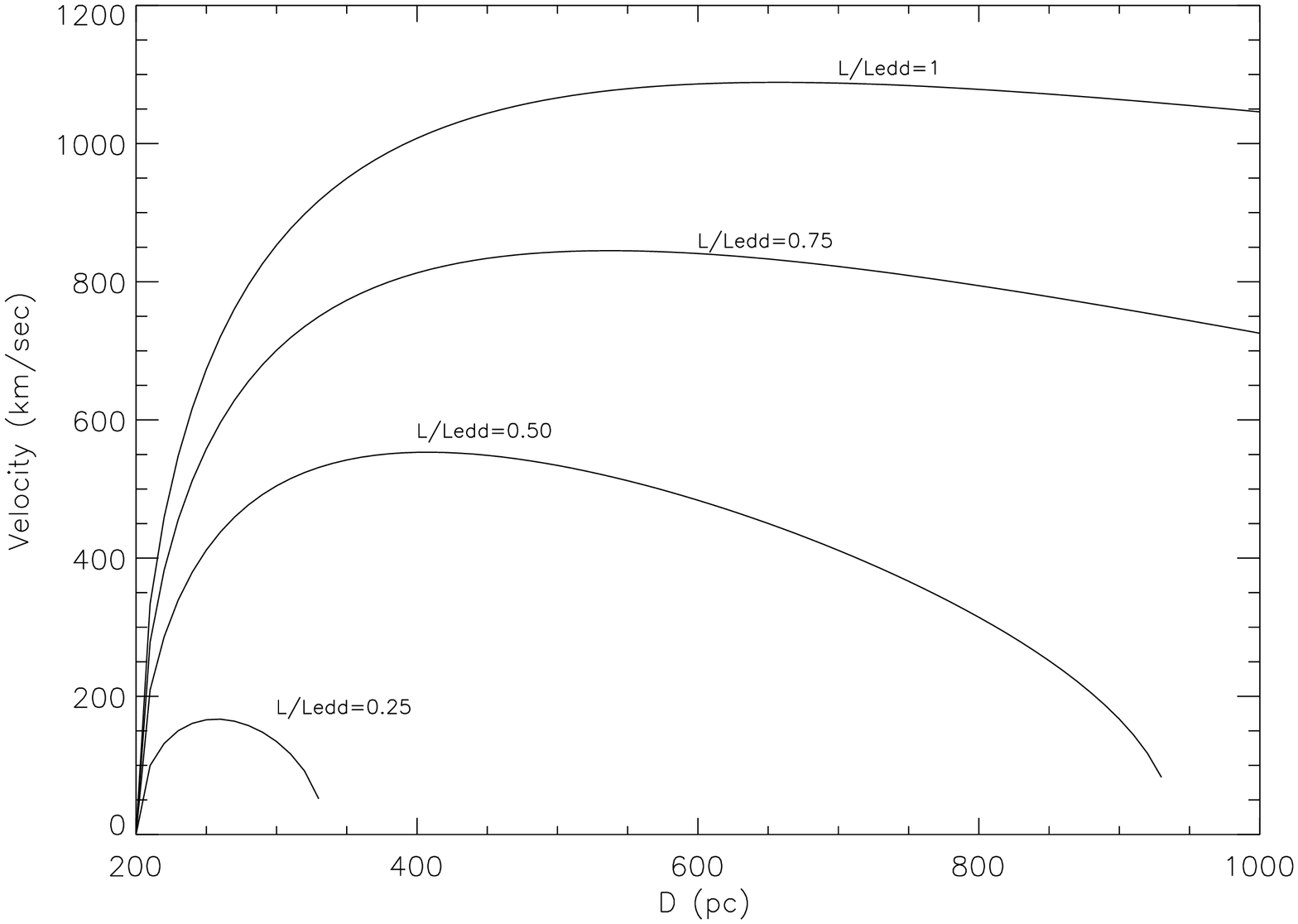}\\

\caption{{\it Top}: Velocity profiles for various launch radii (all generated assuming $\mathcal{M}$ $=$ 3300), in the absence
of interaction with an ambient medium. Based on these
results, radiatively accelerated gas can escape the inner $\sim$ kpc if launched from $D < 500$ pc.
 {\it Bottom}: Velocity profiles for a lauch radius of 200 pc, over a range in $L/L_{\rm edd}$. From this distance, if $L/L_{\rm edd} \leq 0.5$,
 radiatively accelerated gas would not reach a distance of 1 kpc.} 

\label{fig:cloudy}
\end{figure}

In order to determine whether the emission-line gas can be radiatively
accelerated in situ, we used the radiation-gravity formalism detailed in
\citet{Das07}. Assuming an azimuthally symmetric distribution, velocity as a function of radial
distance, $v(r)$, in units of km s$^{-1}$ and pc, can be written as:

\begin{equation}
v(r) = \sqrt{\int_{r_{1}}^{r}\big[6840L_{44}\frac{\mathcal{M}}{r^{2}} - 
8.6\times10^{-3} \frac{M(r)}{r^{2}}\big]dr},
\end{equation}

where $L_{44}$ is the bolometric luminosity, $L_{bol}$ in units of 10$^{44}$ ergs s$^{-1}$,
$\mathcal{M}$ is the Force Multiplier, or ratio of the total photo-absorption
cross-section to the Thomson cross-section, $M(r)$ is the enclosed mass at the
distance $r$, determined from the radial mass distribution described above,
and $r_{1}$ is the launch radius of the gas.

In \citet{Kra09} we derived a value for $L_{bol} = 10^{45.4}$ 
ergs s$^{-1}$. In order to determine $\mathcal{M}$, we generated photo-ionization
models with Cloudy 13.03 \citep{Ferl13} and the ionizing continuum derived in Kraemer et al.
Models with log$U \approx -2.5$ predict that S$^{+2}$ is the peak ionization state for sulfur,
and therefore can be used to constrain the physical conditions in the [S~III] emission-line gas. 
At this ionization, Cloudy predicts $\mathcal{M}$  $=$ 3300 at the ionized face of an illuminated 
slab and we use this value to solve for $v(r)$. Calculating $M(r)$ required solving an
incomplete Gamma function, hence we determined it at 10 pc intervals over a
range of $10~{\rm pc} < r < 1~{\rm kpc}$. We derived an expression for the enclosed mass as a function of $r$,
in each 10~pc interval from $r_{1}$ to $r_{2}$, using a powerlaw of the form 
$M(r) = M(r_{1}) \times (\frac{r_{1}}{r_{2}})^{\beta}$. We were then able to solve for $v(r)$ analytically, 
by integrating within each interval.     

In calculating $v(r)$, we address two points regarding mass outflow in Mrk 573.
First, can the emission-line gas be radiatively accelerated in
situ? In the top panel of Figure \ref{fig:cloudy}, we show $v(r)$ for several different values of $r_{1}$, 
which illustrates that in situ radiative acceleration is possible (out to $\sim$0.5
kpc) in this object. The bottom panel plots the effect of different Eddington ratios ($L/L_{edd}$, with Mrk 573 having a ratio of $L/L_{edd} \sim 1$) on a cloud 
launched from 200 pc. This illustrates that AGN as luminous as 0.5 $L_{edd}$ cannot successfully launch outflows from a distance of 200 pc.
Second, given the deprojected radial distances and velocities of individual emission-line knots,
can we determine the radial distances at which they originated? 
Using the major axis position angle of the inner disk of 95$\degree$ and a maximum ellipticity of the inner 5$"$ from 
\citet{Sch00}, $e = .18$, we calculate the portion of the host disk containing the NLR knots near PA $=128\degree$ 
to be inclined $\sim17\degree$ out of the plane of the sky. Observed velocities and distances, deprojected 
velocities and distances and resulting origin distances, model velocities, and distances traveled for emission line components 
in each knot are given in Table \ref{tab:knots}. Knot emission lines without a modeled velocity and travel distance originate 
at a distance too small to calculate using Equation 11, i.e. the clouds originate at a distance less than 10 pc from 
their current position.

Using our radiative acceleration model, we find that many of the observed emission-line clouds originate locally, i.e. within 17 parsecs, or 1 spatial pixel,
of their observed position. Clouds which originate further from their observed position, Mid-width components in 
Knots B and C and Narrow- and Mid-width components in Knots F and G, have travel distances that can place their origin 
position at knot flux peaks radially interior to their position. Mid-width components in Knots B and C travel distances of $\sim$0.25$"$ and $\sim$0.2$"$, 
placing their origin points near flux peaks west and northwest of the current positions, respectively. Narrow- and Mid-width components in Knots F have travel
distances of $\sim$0.18$"$ and $\sim$0.15$"$, placing their origin points near Knot E. Narrow- and Mid-width components in Knots G have 
travel distances of $\sim$0.78$"$ and $\sim$0.31$"$, placing their origin points near Knots E and F, respectively. Except for the mid-width component of Knot G, 
all kinematic components can be shown to originate in bright knots of ionized gas located near molecular gas lanes outside the NLR (see Figure \ref{fig:struc_flux}), 
consistent with the explanation that the radiatively driven NLR gas kinematics are due to in situ acceleration of gas originating in the host disk outside of the AGN.

\begin{table*}[h!]
  \centering
  \caption{Mrk 573 [S~III] Knot Kinematic Properties for Pure Radiative Acceleration / Gravitational Deceleration}
  \label{tab:knots}
  \begin{tabular}{lcccccccc}
    \toprule
    Knot & Component
& Projected V       &   True V & Projected D  
& True D & Origin D &
Modeled V   & Travel D \\   
       & &   (km s$^{-1}$) &  (km s$^{-1}$)  
&  (pc)         & (pc)     &  (pc)              
& (km s$^{-1}$) &  (pc)
\\
    \midrule
        A   & Narrow & -16
& -56 & 602
& 628 & 615
& -56 &
13 \\
        B        & Narrow
& 8 & 28
& 397 & 414
& 414 & ---
& ---
\\
                  & Mid
& 126 & 441
& 397 & 414
& 330 & 456
& 84
\\
        C        & Narrow
& 47 & 165
& 294 & 307
& 307 & ---
& ---
\\
                  & Mid
& 182 & 637
& 294 & 307
& 240 & 652
& 67
\\
        D       & Narrow
& 6 & 21
& 120 & 125
& 125 & ---
& ---
\\
                  & Mid
& 260 & 910
& 120 & 125
& 115 & 981
& 10
\\
                  & Wide
& 285 & 997
& 120 & 125
& 115 & 981
& 10
\\
        E        & Narrow
& -138 & -483
& 229 & 239
& 225 & -466
& 14
\\
                  & Mid
& -25 & -88
& 229 & 239
& 239 & ---
& ---
\\
        F        & Narrow
& -168 & -588
& 311 & 324
& 260 & -567
& 64
\\
                  & Mid
& -152 & -532
& 311 & 324
& 270 & -512
& 54
\\
        G       & Narrow
& -359 & -1257
& 428 & 446
& 170 & -1231
& 276
\\
                  & Mid
& -137 & -480
& 428 & 446
& 335 & -482
& 111
\\                
    \bottomrule
  \end{tabular}
\end{table*}

\section{Discussion}
\begin{figure*}
\centering

\includegraphics[width=0.98\textwidth]{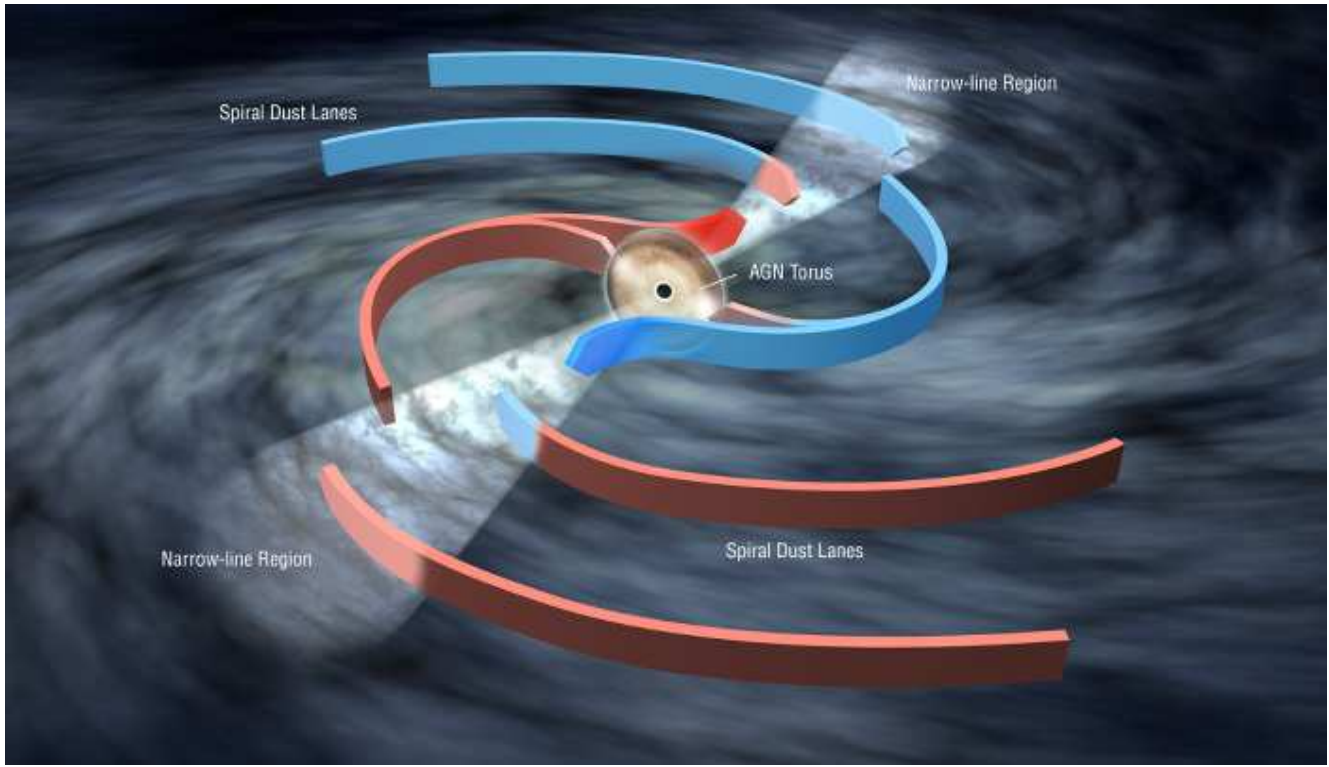}

\caption{Cartoon interpretation of the AGN feedback process occurring in Mrk 573. Host disk gas initially rotates in the galaxy plane. After the AGN turns on, 
gas in spiral arms enters the NLR and is ionized. Gas located at small radii ($<$ 750 pc) is radiatively accelerated away from the nucleus as outflows. Gas located at larger 
radii is ionized but not driven away from the nucleus and remains in rotation.} 

\label{fig:cartoon}
\end{figure*}

From our analysis and modeling of the gas and stellar kinematics in Mrk 573, it is clear that the observed kinematic profiles in this system consist of both 
outflows and rotation. Gas outside the NLR ionizing bicone and at large distances from the AGN contain velocities that are consistent with rotation derived 
from stellar kinematics. Gas inside and immediately adjacent to the NLR at small radii have high-velocity kinematics which deviate from 
rotation and appear to travel radially from the central engine. From our work, it can now be shown that these two sets of kinematics may 
be attributed to one process: the interaction between ionizing radiation from the AGN and its host disk (Figure \ref{fig:cartoon}). Gas in the disk is originally in rotation 
as the AGN turns on and releases ionizing radiation from the central engine into the host disk material. This interaction between AGN radiation and 
host material ionizes gas in the arms, creating the spatially-resolved NLR morphology we observe. At small radii from the AGN, the 
ionized gas can experience enough radiative acceleration to be driven outward from the nucleus, producing kinematics we interpret as 
AGN outflows that are required for bulge evacuation scenarios. At increasingly large radii, gas ionized by the AGN experiences less 
radiative acceleration from the AGN photons and more gravitational attraction from the bulge mass enclosed at that radius, preventing 
further evacuation and preserving the original structure and kinematics of the preexisting gas lanes. Therefore, returning to our hypotheses 
from Paper I, the kinematics observed at $r > 2"$ in {\it HST}/STIS observations are predominantly due to rotation with localized in situ 
acceleration of spiral arm gas. 

Spatially resolved observations of NLR kinematics in other recent works offer interpretations of AGN feedback and NLR / disk interactions 
similar to our own. \citet{Mul11} attributed AGN coronal line kinematics to a combination of rotation and outflows. \citet{Dav14} studied H$_2$ 
and stellar kinematics in several nearby Seyferts and found the molecular gas was often in rotation, except when located adjacent to outflows 
located within the NLR. \citet{Len15} found the kinematics of the extended NLR gas in NGC 1386 to be in rotation, suggesting that the morphology 
results from photoionization of material within the host disk. Recent studies on the extent of AGN feedback by  \citet{Kar16} and \citet{Vil16} have also 
found the size of outflows versus the entire morphological extent of the NLR to be relatively small. Additionally, comparing accretion rates versus 
mass outflow rates for several AGN \citep{Sto10,Cre15,Din15}, outflow rates are often larger by orders of magnitude. If the outflowing material 
originated largely in the nucleus, the AGN duty cycle would be extremely short and the central engine would frequently experience accretion disk 
depletion. Having the gas reservoir exist outside the accretion disk alleviates this problem and is consistent with our conclusion.

With knowledge of the host disk inclination and position angle, we can deproject the maximum radius of the NLR outflows and derive the 
extent of the feedback process. Using the major axis position angle and ellipticity of the inner 5$"$ from 
\citet{Sch00}, the maximum deprojected radii of our measurements along the NLR axis of 128$\degree$ increase by only 7$\%$, and the maximum radial 
extent of radiatively driven gas is $\sim$ 750 pc. As the resolved NLR is the largest structure that depicts the interaction between 
this AGN and its host galaxy, this is the maximum distance at which the AGN can impose negative feedback. With AGN feedback 
unable to fully evacuate a bulge housing an AGN radiating near Eddington, such as Mrk 573, it becomes unclear if bulge clearing negative 
feedback processes are successful in the local universe. Therefore, for negative feedback via AGN outflows to be successful, the size of the 
bulge in radio-quiet AGN at the time it is evacuated in quenching scenarios must be much smaller than typical kpc-scale bulge radii currently 
observed in nearby galaxies, while still maintaining an AGN radiating near Eddington.

From Figures \ref{fig:disfwhmflux}, \ref{fig:stellar}, and \ref{fig:rot_compare}, Mrk 573 has a projected rotating velocity of $\sim$ 50 km s$^{-1}$ near the 
maximum radial outflow distance of 750 pc along the major axis (Slit B). Deprojecting this to a true velocity of 115 km s$^{-1}$, the time required to 
rotate the host disk once at this radius would be $\sim 4 \times 10^7$ years. During each rotational lap, the gas in the host disk would experience 
radiative driving by the AGN at two separate epochs, once for each intersection with a cone of ionizing radiation emitted by the AGN. Therefore, 
gas in the host disk experiences radiative driving every $2 \times10^7$ years. If we assume that the NLR orientation is static and that its intersection 
with the host disk intersection currently impacts approximately half of the host disk volume at radii $>$ 750 pc, then each epoch of radiative driving 
experienced by material in the host disk lasts $\sim 10\times10^6$ years. Assuming an AGN duty cycle of $10^8$ years (galaxy lifetime 
$\sim 10^{10}$ years, 1$\%$ of galaxies are currently active), then the host disk rotates twice during an active period. Cumulatively, the total time spent inside 
the ionizing bicone of the NLR for any parcel of gas at radii $<$ 750pc is then $4 \times 10^7$ years.

By measuring the amount of hydrogen gas in the NIFS FOV, we can form an estimate for the total amount of gas within that a radius 
that would need to be evacuated via radiative driving. Per \citet{Maz13}, the cold molecular gas mass can be estimated as:

\begin{equation}
\frac{M_{cold}}{M_{\astrosun}} \approx 1174 \times (\frac{L_{H_{2}\lambda2.1218}}{L_{\astrosun}})
\end{equation}

where $L_{H_{2}\lambda2.1218}$ is the luminosity of the H$_2$ line. Using an integrated flux of $F_{H_{2}\lambda2.1218} = 2.74 \times 
10^{-15}$ erg s$^{-1}$ cm$^{-2}$ and a distance of D = 70.55 Mpc, we calculate an H$_2$ luminosity of $L_{H_{2}\lambda2.1218} = 
$2.74$\times 10^{-15}$ erg s$^{-1}$ and a cold molecular gas mass of M$_{cold} \sim 5\times10^{8}$M$_{\astrosun}$. As that is the 
currently observed mass, we assume double the mass as an estimate for the original gas mass of 10$^9$ M$_{\astrosun}$. Therefore, with an 
initial mass estimate of 10$^9$ M$_{\astrosun}$ and a cumulative timescale of radiative driving of 4.8$\times 10^{7}$ years, the required mass 
outflow rate to evacuate gas within a radius of 750 pc would be $\sim$25 M$_{\astrosun}$ yr$^{-1}$. This may not be an unreasonable mass outflow rate; 
we have previously estimated a peak outflow rate of $\sim$3 M$_{\astrosun}$ yr$^{-1}$ for NGC 4151 \citep{Sto10,Cre15}, which has a much 
lower luminosity and Eddington ratio. In a future paper (Revalski et al., in prep), we will give the resolved mass outflow rate in the NLR of Mrk 573 for comparison.




As the gas being radially driven from the nucleus is likely more important in understanding the impact of the AGN on its host, we suggest to 
define this gas as the true NLR of Mrk 573. Therefore, assuming a symmetric extent west of the nucleus, radiatively driven gas along the NLR axis 
extends to projected distances of $\sim$ 2$"$ ($\sim$700 pc). Ionized gas exterior to the NLR, with an orderly velocity field characteristic of normal 
galactic rotation, should reside in the Extended Narrow Line Region (ENLR), as originally defined by \citet{Ung87}. In Mrk 573, we detect the ENLR out 
to projected distances greater than 6 kpc. As such, future projects analyzing spatially-resolved kinematics of AGN-ionized emission-lines should clarify 
the extent of the NLR vs the ENLR, as the scale between the two regions in Mrk 573, R$_{NLR}$/R$_{ENLR}$, is no greater than 0.12.

It is notable that, if all gas kinematics can be attributed to material approximately within the disk, a high-velocity, leading-arm host disk in Mrk 573 is no 
longer viable. Rotating gas is observed to be blueshifted to the east of the nucleus and redshifted to the west, yet radial outflows exhibit 
redshifts to the southeast and blueshifts to the northwest. The only orientation that can produce this combination of kinematic signatures 
places the north edge of the disk as the nearest edge (contrary to our claim in Paper I). Moving the north edge out of the plane of the sky results in the disk rotating in the 
clockwise direction, making the host disk a traditional trailing-arm system. Our previous model from Paper I created a biconical outflow 
model that required a host-disk inclined in the opposite direction, with the southeast side coming out of the plane of the sky, to satisfy 
both kinematic and morphological parameters. Thus, as illustrated by our studies of Mrk 573, radial velocity maps from IFU observations are likely to provide 
much better constraints on the outflow geometries than previous long-slit observations at similar angular resolutions.

\section{Conclusions}
\label{sec:conclusions}

We observed the Seyfert 2 AGN Mrk 573 with Gemini NIFS and APO DIS and obtained ionized gas, molecular gas, and stellar kinematic maps 
surrounding its nucleus and ionized gas in the extended host disk. Our main conclusions are as follows:

1) Flux distributions of [S~III] and H$_2$ in Gemini NIFS observations are vastly different yet complementary. [S~III] emission mimics that 
observed in the optical NLR (e.g., in [O~III]); arcs of emission and a linear nuclear feature which initially appear to be kinematically unrelated. However, 
H$_2$ emission reveals arcs of gas that lie outside the NLR bicone, which connect the ionized gas features. 

2) The velocity field of the ionized and molecular arcs in NIFS observations and large scale ionized gas kinematics in APO/DIS observations 
show signatures of rotation as observed in NIFS stellar kinematics. Therefore, the kinematics of these features are credited to rotation. 
Detecting such a continuous morphology in material inside/outside of the NLR, which shows signs of rotation throughout, suggests that the 
NLR kinematics and morphology in Mrk 573 can largely be attributed to material originating in the rotating host disk. 

3) Given that the kinematics are largely rotation, deviations from the expected rotation curve exist along the axis of the projected NLR at 
radii $r < 750$ pc. We can explain these deviations as radiative, in situ acceleration of material residing 
in the host disk with different kinematic profiles in the NLR (i.e. outflow versus rotation) existing as a function of radius.

4) Radiatively driven gas in the NLR only extends to distances of $\sim$750 pc from the nucleus and not the entire length of the combined 
NLR/ENLR, which suggests that AGN outflows in this Seyfert galaxy may have a much smaller range of impact than was previously expected. 

From the evidence seen in Mrk 573, it is likely that the NLR is still biconical in geometry, however the majority of the NLR volume is filled with 
ionizing radiation from the central source that is illuminating material that already exists in the host galaxy environment, providing the 
spatially-resolved morphology we observe.  Should the kinematic explanation of the NLR in Mrk 573 prove to be applicable to a majority of 
AGN, outflows may be more prevalent than previously thought \citep{Fis13,Fis14} but likely do not extend far enough to clear the host bulge. 
Incorporating molecular gas and stellar populations in the kinematic analysis of AGN feeding and feedback is vital to understanding the interaction 
between the central engine and host disk. As such, the Near InfraRed Spectrograph (NIRSpec) on the {\it James Webb Space Telescope} will likely be a boon to observing 
spatially resolved feeding and feedback processes in AGN, allowing for observations at a greater sensitivity out to higher redshifts than current, ground-based 
instruments. 

\acknowledgments

TCF was supported by an appointment to the NASA Postdoctoral Program at the NASA Goddard Space Flight Center, 
administered by Universities Space Research Association under contract with NASA.
This study was based on observations obtained at the Gemini Observatory
(processed using the Gemini IRAF package), which
is operated by the Association of Universities for Research
in Astronomy, Inc., under a cooperative agreement with the
NSF on behalf of the Gemini partnership: the National Science
Foundation (United States), the National Research Council
(Canada), CONICYT (Chile), the Australian Research
Council (Australia), Ministerio da Ciencia, Tecnologia e Inovacao (Brazil) and 
Ministerio de Ciencia, Tecnologia e Innovacion Productiva (Argentina).


\bibliographystyle{apj}             
\bibliography{apj-jour,mrk573_kin}       

\appendix
\section{IFU Spectral Analysis}
\label{sec:fitting}

We have devised a new fitting technique for IFU observations that allows us to determine the number of meaningful kinematic components for 
each emission line based on the fits. This process employed Bayesian model selection, described below, as this technique is well suited to extracting 
individual velocities from blended lines.

Model-fitting estimates the most probable set of model parameters, $\mathbf{\Phi}$ of a model $\mathbf{M_i}$ in comparison to the given data, $\mathbf{D}$. In our case, $\mathbf{\Phi} =$ Gaussian parameters $\mu$ (centroid), $\sigma$ (dispersion), and $H$ (peak height). To determine $\mathbf{\Phi}$, one maximizes the posterior probability 
$p(\mathbf{\Phi}|\mathbf{D},\mathbf{M_{i}})$:

\begin{equation}
p(\mathbf{\Phi}|\mathbf{D},\mathbf{M_{i}}) =\frac{p(\mathbf{D}|\mathbf{\Phi},\mathbf{M_{i}})p(\mathbf{\Phi}|\mathbf{M_{i}})}{p(\mathbf{D}|\mathbf{M_{i}})} ,
\end{equation}

\noindent where $p(\mathbf{D}|\mathbf{\Phi},\mathbf{M_{i}})$ is the likelihood of the model parameters, $p(\mathbf{\Phi}|\mathbf{M_{i}})$ is the 
prior probability of the parameters, and $p(\mathbf{D}|\mathbf{M_{i}})$ is the marginal likelihood, or Bayesian evidence, whose role is to 
normalize the posterior probability. For model $\mathbf{M_i}$, the evidence is constant. Therefore, to determine $\mathbf{\Phi}$, it is sufficient to 
maximize the numerator of Equation 1, the likelihood (or $\chi^{2}$) under prior constraints. However, in order to compare individual line-
component models, the ratio of the evidences is required. The ratio of the probabilities of two models $\mathbf{M_{1}}$ and $\mathbf{M_{2}}$, 
given the data $\mathbf{D}$, can be expressed as:

\begin{equation}
\frac{p(\mathbf{M_{1}}|\mathbf{D})}{p(\mathbf{M_{2}}|\mathbf{D})} =\frac{p(\mathbf{M_{1}})}{p(\mathbf{M_{2}})}\frac{p(\mathbf{D}|\mathbf{M_{1}})}{p(\mathbf{D}|\mathbf{M_{2}})} ,
\end{equation}

\noindent where $p(\mathbf{M_{i}})$ is the a priori probability of model $\mathbf{M_{i}}$. Without a preference for a specific model, $
(p(\mathbf{M_{1}})/p(\mathbf{M_{2}}) = 1)$, the posterior odds ratio $R$ becomes a ratio between the two evidences where $Z(\mathbf{M_{i}}) 
= p(\mathbf{D}|\mathbf{M_{i}})$. The logarithm of the ratio of evidences provides a guide to what constitutes a significant difference between 
models:

\begin{equation}
\Delta ln R = ln\left [\frac{p(\mathbf{M_{1}}|\mathbf{D})}{p(\mathbf{M_{0}}|\mathbf{D})} \right] = ln\left [\frac{Z_{1}}{Z_{0}} \right] , 
\end{equation}

\noindent where a $|\Delta ln R| > 5$ is used in our measurements as strong evidence that the more complex model is superior, per \citet{Fer11}. For a given data set, 
the evidence $Z(\mathbf{M_{i}})$ for model $\mathbf{M_{i}}$ is defined as the marginalized likelihood:

\begin{equation}
Z(\mathbf{M_{i}}) = \int_{\Phi_{1}} ... \int_{\Phi_{n}} p(\mathbf{D}|\Phi,\mathbf{M})p(\Phi|\mathbf{M})d\Phi_{1}...d\Phi_{n} , 
\end{equation}

Our model-fitting code employs the Importance Nested Sampling algorithm as implemented in the MultiNest library
\citep{Fer08,Fer09,Fer13,Buc14} to compute the logarithm of the evidence, $lnZ$, for each model. 

\begin{figure}[htp]
\centering
\includegraphics[width=0.24\textwidth]{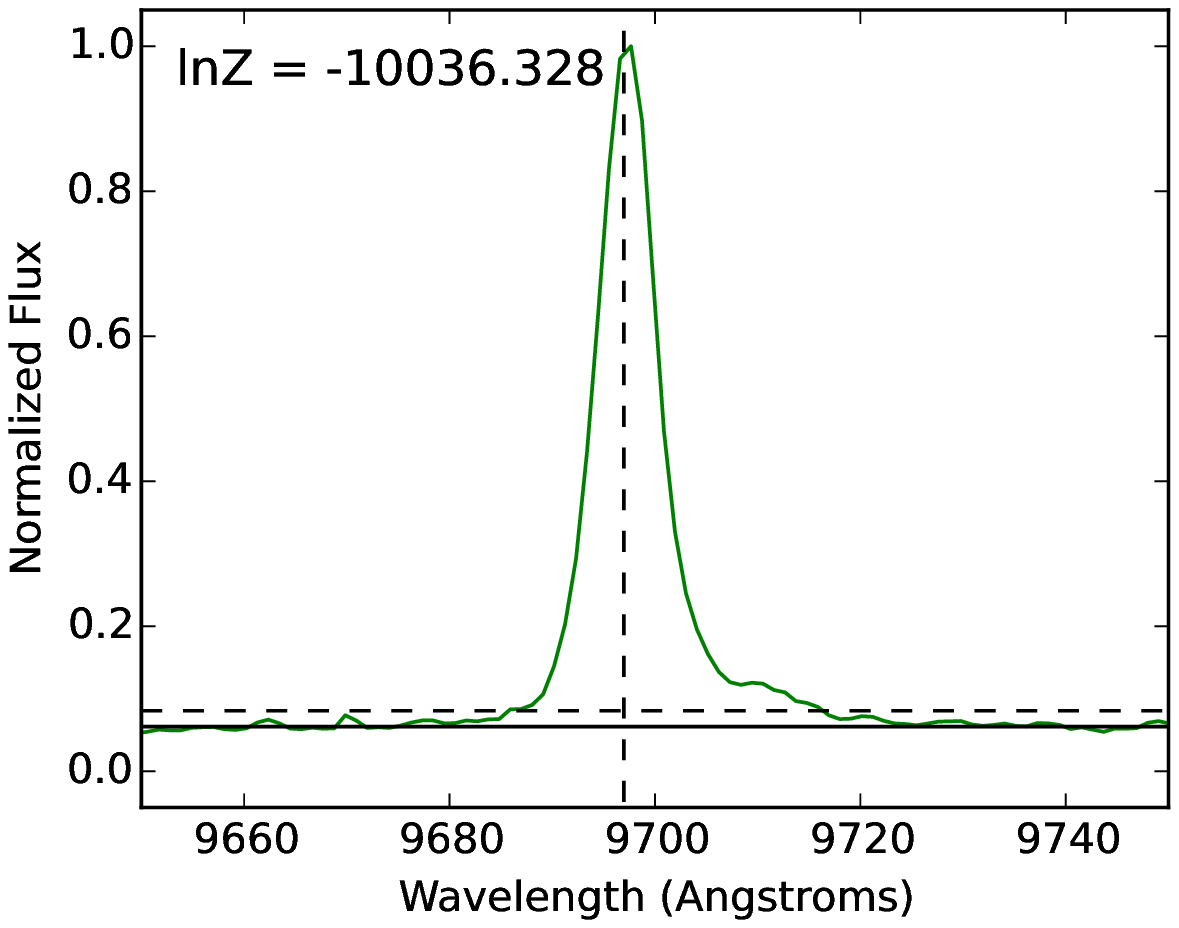}
\includegraphics[width=0.24\textwidth]{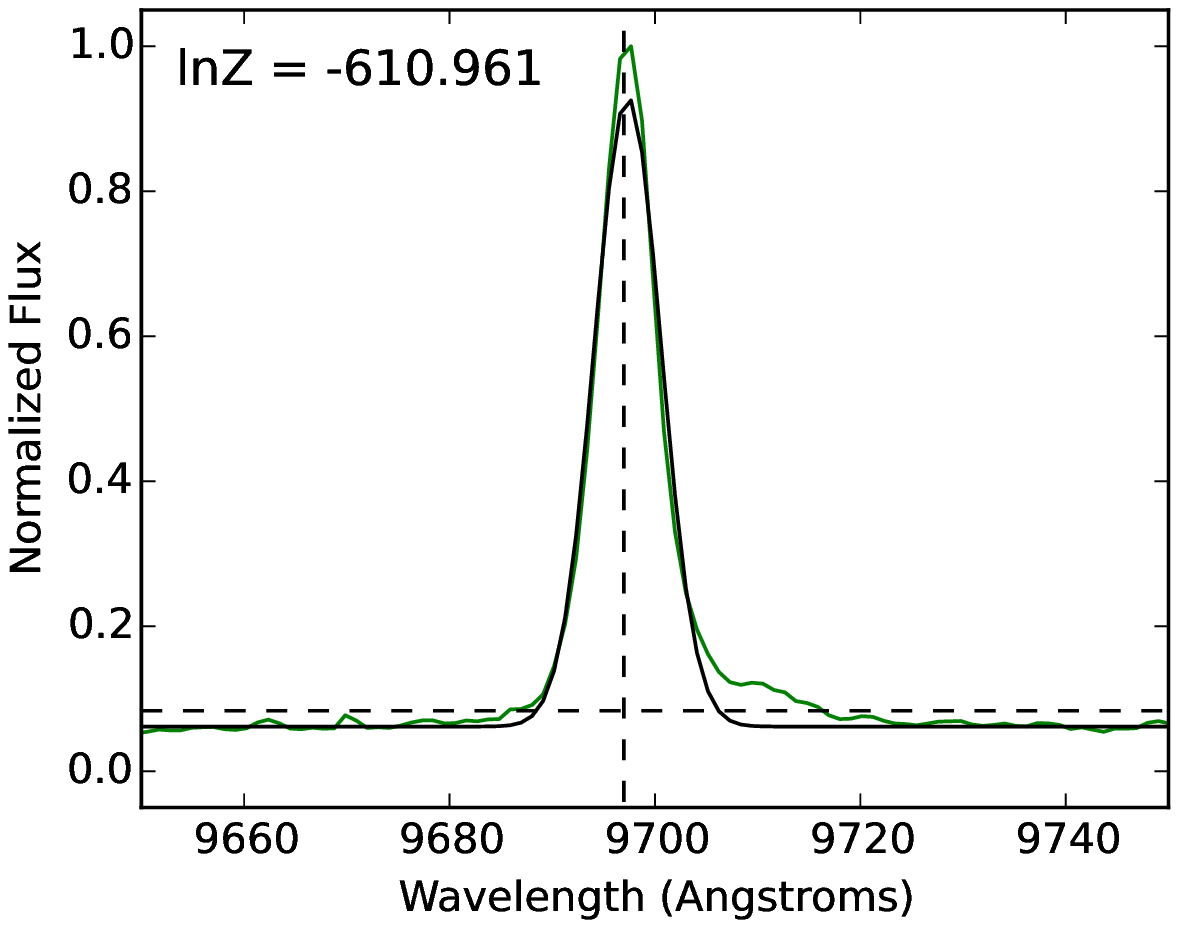}
\includegraphics[width=0.24\textwidth]{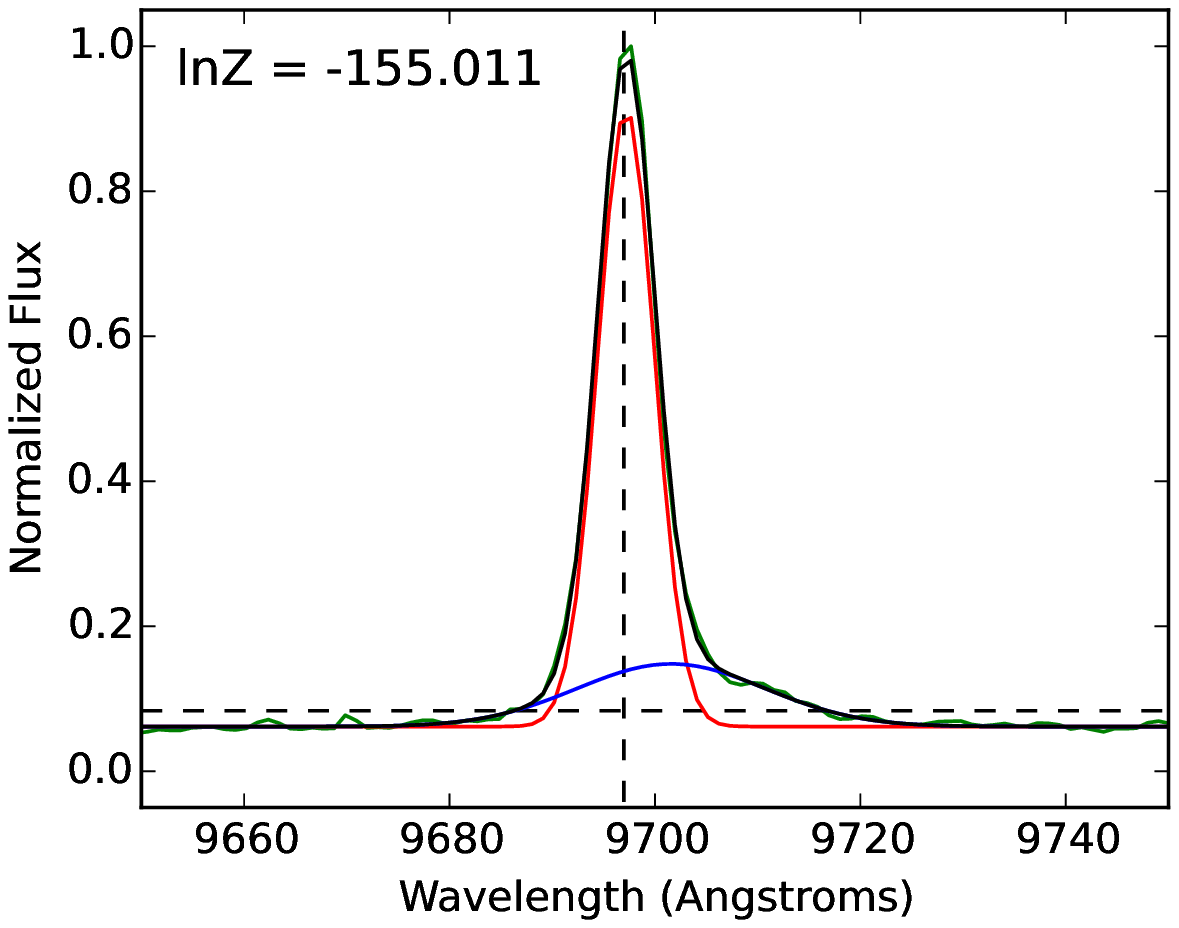}
\includegraphics[width=0.24\textwidth]{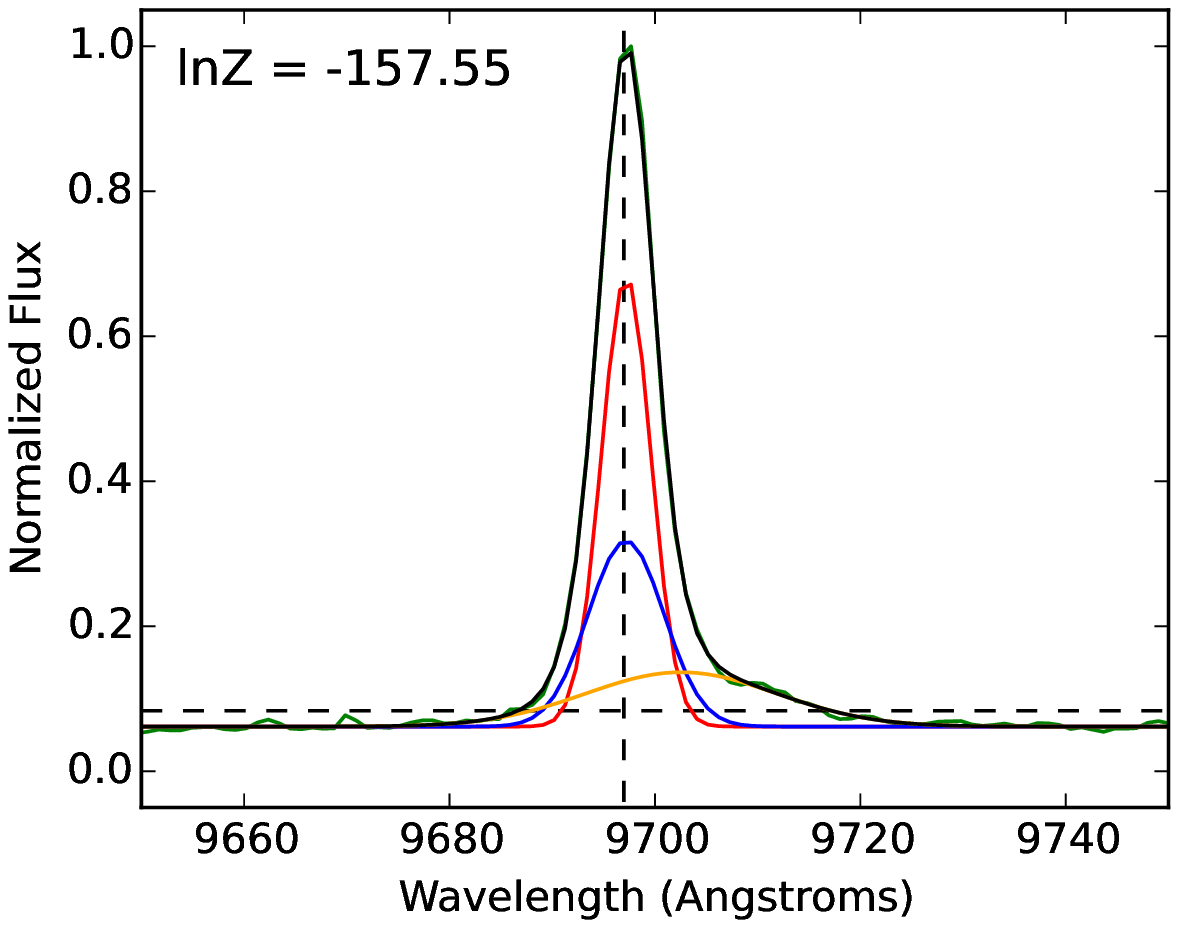}
\caption{[S~III] $\lambda 0.9533 \mu m$ emission-line component fitting example over the continuum peak in Mrk 573 (Point N in 
Figure \ref{fig:SIIIflux}). Green line represents NIFS spectral data. Solid black line represents the 
total model. Red, blue, and orange lines represent individual Gaussians sorted by width, narrowest to widest. 
Vertical dashed black line represents the [S~III] $\lambda 0.9533 \mu m$ wavelength at systemic velocity. Horizontal dashed 
black line represents the $3\sigma$ continuum-flux lower limit for Gaussians in our fitting. {\it Top:} 0- and 1-component model fits. 
{\it Bottom:} 2- and 3-component model fits. The Bayesian evidence, or marginal likelihood, is listed for each model. 
Using a $|\Delta ln R|  = |ln(Z_{1}/Z_{0})| > 5$ filter, the 2-component model fit of this emission line is most probable.} 
\label{fig:fits}
\end{figure}

Our overall procedure, as illustrated in Figure \ref{fig:fits}, is therefore as follows. Models are run for zero components ($\mathbf{M_{0}}$; 
continuum) and one component ($\mathbf{M_{1}}$; continuum plus Gaussian). If $\mathbf{M_{1}}$ is favored over $\mathbf{M_{0}}$, the 
data is analyzed with a two-component model ($\mathbf{M_{2}}$, continuum plus two Gaussians), with the process repeating until the more 
complex model is no longer favored. We utilize Bayesian model selection in our automated fitting as reduced-$\chi^2$ fitting is not suited to 
truly assess the relative probabilities of models because the decrease in $\chi^2$ observed when introducing a more complex model could be due 
to either the presence of another line component present in the profile or simply to over-fitting. 

Priors in our models are selected based on physical considerations. The centroid position ($\mu$) for each Gaussian was limited 
to a 50\AA~range around each emission line that contained the entirety of the line profile throughout the datacube. 
Gaussian standard deviation ($\sigma$) ranged from a minimum width determined by the spectral resolution in each band to 
an artificial limit of 15\AA~(Z-band FWHM $\sim$1100 km s$^{-1}$; K-band FWHM $\sim$ 500 km s$^{-1}$ ). Gaussian height ($H$)
was restricted to a minimum value of 3 times the standard deviation of the continuum ($\sigma_{c}$) and given a virtually unbound maximum 
height restriction of 3$\sigma_{c} \times$ 10$^8$.

\end{document}